\documentclass{article}

\usepackage{iclr2027_conference,times}
\usepackage[nopatch=footnote]{microtype}  
\usepackage{xurl}
\usepackage{algorithm}
\usepackage{algorithmic}
\usepackage{hyperref}

\usepackage{amsmath,amsfonts,bm}

\def\eqref#1{equation~\ref{#1}}

\def\1{\bm{1}}

\DeclareMathAlphabet{\mathsfit}{\encodingdefault}{\sfdefault}{m}{sl}
\SetMathAlphabet{\mathsfit}{bold}{\encodingdefault}{\sfdefault}{bx}{n}

\usepackage{listings}
\usepackage{fvextra}
\usepackage{float}
\usepackage{needspace}
\usepackage{nicematrix}

\usepackage[utf8]{inputenc}
\usepackage{lipsum}

\usepackage[T1]{fontenc}

\usepackage{etoolbox}
\newbool{includeappendix}
\setbool{includeappendix}{true}

\ifdefined\isoverfull
\else
\fi

\usepackage{subcaption}

\definecolor{my-full-blue}{HTML}{1F77B4}

\definecolor{my-full-orange}{HTML}{FF7F0E}
\definecolor{my-extra-orange}{HTML}{7C4514}

\definecolor{my-full-green}{HTML}{2CA02C}

\definecolor{my-full-red}{HTML}{d62728}

\definecolor{my-full-purple}{HTML}{9467bd}

\colorlet{my-blue}{my-full-blue!30}
\colorlet{my-orange}{my-full-orange!30}
\colorlet{my-green}{my-full-green!90}
\colorlet{my-red}{my-full-red!90}
\colorlet{my-purple}{my-full-purple!30}

\definecolor{mygreen}{HTML}{B5E3B5}
\definecolor{mylightblue}{HTML}{B0DBFF}

\definecolor{darkblue}{RGB}{34,49,63}
\definecolor{lightblue}{RGB}{238,244,249}
\definecolor{accentblue}{RGB}{88,139,202}
\definecolor{lightgreen}{RGB}{220,245,230}
\definecolor{lightred}{RGB}{250,225,225}
\definecolor{darkgreen}{RGB}{40,167,69}
\definecolor{darkred}{RGB}{220,53,69}
\definecolor{textgray}{RGB}{120,120,120}
\definecolor{lightgray}{RGB}{240,240,240}
\definecolor{accentorange}{RGB}{127,17,144}
\definecolor{gray2}{HTML}{FCFCFC}

\usepackage{listings}

\usepackage{textcomp}

\usepackage[scaled=0.8]{beramono}

\definecolor{ckeyword}{HTML}{7F0055}
\definecolor{ccomment}{HTML}{3F7F5F}
\definecolor{cstring}{HTML}{2A0099}

\lstdefinestyle{numbers}{
	numbers=left,
	framexleftmargin=20pt,
	numberstyle=\tiny,
	firstnumber=auto,
	numbersep=1em,
	xleftmargin=2em
}

\lstdefinestyle{layout}{
	frame=none,
	captionpos=b,
}

\lstdefinestyle{comment-style}{
	morecomment=[l]//,
	morecomment=[s]{/*}{*/},
	commentstyle={\color{ccomment}\itshape},
}

\lstdefinestyle{string-style}{
	showstringspaces=false,%
}

\lstdefinestyle{keyword-style}{
	keywordstyle={\ttfamily\bfseries},
	morekeywords={
		function,
		constructor,
		int,
		bool,
		return,
		returns,
		uint
	},
	morekeywords = [2]{},
	keywordstyle = [2]{\text},
	sensitive=true,
}

\lstdefinestyle{input-encoding}{
	inputencoding=utf8,
	extendedchars=true,
	literate=
	{ℝ}{$\reals$}1%
	{→}{$\rightarrow$}1%
	{–}{-}1%
	{—}{---}3%
	{≠}{$\neq$}1%
	{§}{\S}1%
	{…}{...}3%
	{∈}{$\in$}1%
	{⟺}{$\Leftrightarrow$}1%
	{α}{$\alpha$}1%
	{β}{$\beta$}1%
	{λ}{$\lambda$}1%
	{θ}{$\theta$}1%
	{ϕ}{$\phi$}1%
}

\lstdefinestyle{escaping}{
	escapeinside={(*@}{@*)},
	mathescape=true
}

\lstdefinestyle{default-style}{
	basicstyle=\fontencoding{T1}\ttfamily\footnotesize,
	style=numbers,
	style=layout,
	style=comment-style,
	style=string-style,
	style=keyword-style,
	style=input-encoding,
	style=escaping,
	tabsize=2,
	upquote=true
}

\lstdefinelanguage{BASIC}{
	language=C++,
	style=default-style
}[keywords,comments,strings]%

\lstdefinelanguage{JavaScript}{
morekeywords=[1]{break, continue, delete, else, for, function, if, in,
new, return, this, typeof, var, void, while, with, const, let},
morekeywords=[2]{false, null, true, boolean, number, undefined, string,
Array, Boolean, Date, Math, Number, String, Object},
morekeywords=[3]{eval, parseFloat, escape, unescape},
sensitive,
morecomment=[s]{/*}{*/},
morecomment=[l]//,
morecomment=[s]{/**}{*/},
morestring=[b]',
morestring=[b]"
}[keywords, comments, strings]
\definecolor{delim}{RGB}{20,105,176}
\definecolor{numb}{RGB}{106, 109, 32}
\definecolor{string}{rgb}{0.64,0.08,0.08}

\lstdefinelanguage{json}{
    showspaces=false,
    showtabs=false,
    breaklines=true,
    postbreak=\raisebox{0ex}[0ex][0ex]{\ensuremath{\color{gray}\hookrightarrow\space}},
    upquote=true,
    morestring=[b]",
    morecomment=[l]//,
    stringstyle=\color{string},
    literate=
     *{0}{{{\color{numb}0}}}{1}
      {1}{{{\color{numb}1}}}{1}
      {2}{{{\color{numb}2}}}{1}
      {3}{{{\color{numb}3}}}{1}
      {4}{{{\color{numb}4}}}{1}
      {5}{{{\color{numb}5}}}{1}
      {6}{{{\color{numb}6}}}{1}
      {7}{{{\color{numb}7}}}{1}
      {8}{{{\color{numb}8}}}{1}
      {9}{{{\color{numb}9}}}{1}
      {\{}{{{\color{delim}{\{}}}}{1}
      {\}}{{{\color{delim}{\}}}}}{1}
      {[}{{{\color{delim}{[}}}}{1}
      {]}{{{\color{delim}{]}}}}{1},
}
\definecolor{dkgreen}{rgb}{0,0.6,0}
\definecolor{dred}{rgb}{0.545,0,0}
\definecolor{dblue}{rgb}{0,0,0.545}
\definecolor{lgrey}{rgb}{0.9,0.9,0.9}
\definecolor{gray}{rgb}{0.4,0.4,0.4}
\definecolor{darkblue}{rgb}{0.0,0.0,0.6}
\lstdefinelanguage{cpp}{
      breaklines=true,               
      postbreak=\raisebox{0ex}[0ex][0ex]{\ensuremath{\color{gray}\hookrightarrow\space}},
      deletekeywords={...},          
      escapeinside={\%*}{*)},                  
      language=C++,                
      keywordstyle=\color{purple},  
      morekeywords={string,float}, 
      identifierstyle=\color{black},
      stringstyle=\color{blue},      
      showspaces=false,               
      showstringspaces=false,        
      showtabs=false,                
      tabsize=5,                     
    }

\usepackage[english]{babel}
\definecolor{darkpastelblue}{HTML}{0279AF}
\hypersetup{breaklinks=true,
      colorlinks,
      linkcolor=darkpastelblue,
      citecolor=darkpastelblue,
      urlcolor=darkpastelblue,
      filecolor=darkpastelblue}
\usepackage{color,soul}
\usepackage{ bbold }
\usepackage{multirow}
\usepackage{multicol}
\usepackage{caption}
\usepackage{tabularx,booktabs,xltabular,colortbl}
\usepackage{graphicx}
\usepackage{tabu}
\usepackage{array}
\usepackage{siunitx}
\usepackage{fontawesome5}
\usepackage{stackengine}
\usepackage{svg}
\usepackage{mathtools}
\usepackage{xspace}
\usepackage{wrapfig}
\makeatletter
\let\WF@finale@orig\WF@finale
\def\WF@finale{\WF@finale@orig
  \ifnum\@listdepth>\z@ \parshape\@ne\@totalleftmargin\linewidth \fi}
\makeatother
\usepackage{makecell}
\usepackage{placeins}
\usepackage{float}
\usepackage{pifont}
\usepackage{svg}
\usepackage[most]{tcolorbox}
\usepackage{array}
\usepackage{dsfont}
\usepackage{enumitem}
\usepackage[utf8]{inputenc}
\usepackage{textcomp}
\usepackage{amsthm}
\usepackage{amssymb,amsmath}

\newcolumntype{x}[2]{S[table-format=#1.#2,table-auto-round]}

\usepackage{tikz}

\usetikzlibrary{arrows}
\usetikzlibrary{automata}
\usetikzlibrary{calc}
\usetikzlibrary{backgrounds}
\usetikzlibrary{decorations.markings}
\usetikzlibrary{decorations.pathmorphing}
\usetikzlibrary{decorations.pathreplacing}
\usetikzlibrary{fit}
\usetikzlibrary{patterns}
\usetikzlibrary{positioning}
\usetikzlibrary{shadows}
\usetikzlibrary{shapes}
\usetikzlibrary{shapes.geometric}
\usetikzlibrary{arrows.meta}
\usetikzlibrary{shadows.blur}

\definecolor{blue}{HTML}{347bc6}
\definecolor{green-underline}{HTML}{2de12c}
\definecolor{yellow-underline}{HTML}{ffd700}

\tikzstyle{block} = [
    minimum width=1cm,
    minimum height=0.5cm,
    align=center,
    fill=gray!30,
    rounded corners=5pt,
]

\tikzstyle{arrow} = [
	-{Latex[length=1.4mm, width=1.4mm]},
	draw=blue,
]

\tikzstyle{dashedline} = [
    draw=blue,
    dashed
]

\tikzstyle{line} = [
    draw=blue
]

\definecolor{lightblue}{RGB}{173,216,230}

\usepackage{pgfplots}
\pgfplotsset{compat=1.18}
\usepackage[capitalize,noabbrev]{cleveref}
\AtBeginDocument{
\crefname{appendix}{Appendix}{Appendices}
}

\usepackage{pifont}

\definecolor{pastelgreen}{HTML}{059C05}
\definecolor{pastelred}{HTML}{FF7373}

\usepackage{setspace}

\renewcommand{\algorithmicrequire}{\textbf{Input:}}
\renewcommand{\algorithmicensure}{\textbf{Output:}}
\renewcommand{\algorithmicfunction}{\textbf{function}}
\newcommand{\ttt}[1]{\text{\texttt{#1}}}

\newcommand{\eemph}[1]{\textbf{#1}\hspace{1em}\ignorespaces}

\usepackage{bm}

\usepackage{scalerel}

\newcommand{\Hsquare}{%
  \text{\kern2\scriptspace\fboxsep=-.2pt\fbox{\rule{0pt}{1ex}\rule{1ex}{0pt}}\kern2\scriptspace}%
}

\def\epsilon{\ensuremath{\varepsilon}}

\def\presuper#1#2%
  {\mathop{}%
   \mathopen{\vphantom{#2}}^{#1}%
   \kern-\scriptspace%
   #2}
\def\surround#1#2#3%
  {\mathop{}%
   \mathopen{\vphantom{#2}}^{#1}%
   \kern-3\scriptspace%
   #2%
   \kern4\scriptspace%
   \mathopen{\vphantom{#2}}^{#3}%
   }

\newcounter{algoline}[algorithm]

\newcommand{\bench}{PROBE\xspace}
\newcommand{\benchlong}{\textbf{P}roactive \textbf{R}epository-wide \textbf{O}pen-world \textbf{B}ug \textbf{E}valuation\xspace}

\newcommand{\musespark}{Muse Spark 1.3\xspace}
\newcommand{\luna}{GPT-5.6 Luna\xspace}
\newcommand{\sol}{GPT-5.6 Sol\xspace}
\newcommand{\terra}{GPT-5.6 Terra\xspace}
\newcommand{\gptfivefour}{GPT-5.4\xspace}
\newcommand{\opus}{Claude Opus 5\xspace}
\newcommand{\sonnet}{Claude Sonnet 5\xspace}
\newcommand{\glmflash}{GLM 5.3 Flash\xspace}
\newcommand{\deepseekflash}{DeepSeek V4.1 Flash\xspace}
\newcommand{\geminiflash}{Gemini 3.5 Flash\xspace}

\definecolor{modelbg}{RGB}{66,66,66}
\definecolor{systemborder}{RGB}{255,140,0}
\definecolor{userborder}{RGB}{70,130,180}
\definecolor{assistantborder}{RGB}{82,183,136}
\definecolor{toolborder}{RGB}{143,129,246}
\definecolor{resultborder}{RGB}{175,175,175}

\tcbuselibrary{skins, breakable}

\newtcolorbox{promptbox}[2]{
  enhanced,
  breakable,
  before skip=3mm,
  after skip=3mm,
  colback=#2!3,
  colframe=#2,
  boxrule=1pt,
  arc=2mm,
  left=2mm,
  right=2mm,
  top=1.5mm,
  bottom=-2mm,
  fonttitle=\sffamily\bfseries,
  coltitle=black,
  title=#1,
  attach boxed title to top left={yshift=-2mm, xshift=3mm},
  boxed title style={colback=#2, arc=1.5mm, boxrule=0pt},
}

\newtcbox{\modelpill}{
  on line,
  colback=modelbg,
  colframe=modelbg,
  coltext=white,
  fontupper=\sffamily\bfseries\footnotesize,
  arc=1.5mm,
  boxrule=0pt,
  left=1.5mm, right=1.5mm, top=0.6mm, bottom=0.6mm,
}

\newcommand{\promptheading}[3]{%
  \needspace{6\baselineskip}%
  \subsubsection{#1}%
  \label{#2}%
  \noindent\modelpill{#3}\hspace{0.6em}%
}

\newcommand{\benchNumRepos}{20}
\newcommand{\benchNumTargets}{2,457}

\newcommand{\benchRepoRecallW}{0.75}
\newcommand{\benchRepoPrecisionW}{0.92}
\newcommand{\benchRepoBugsRecallRho}{-0.27}

\newcommand{\fpAuditSample}{100}
\newcommand{\fpAuditRejected}{209}
\newcommand{\fpAuditRejectionRate}{8.5\%}
\newcommand{\fpAuditSampleRejected}{17}

\newcommand{\fpAuditHumanRejected}{12}
\newcommand{\fpAuditBothRejected}{6}

\newcommand{\fpAuditKappa}{0.29}
\newcommand{\dedupNumFindings}{307}
\newcommand{\dedupNumRepos}{3}
\newcommand{\dedupNumGold}{67}
\newcommand{\dedupNumPairs}{17,177}

\newcommand{\dedupFiredRate}{2.7\%}

\newcommand{\dedupNumDropped}{16,708}

\newcommand{\dedupNumRuns}{5}
\newcommand{\dedupPrecisionMean}{98.8\%}

\newcommand{\dedupRecallMean}{97.6\%}

\newcommand{\dedupTPMin}{65}
\newcommand{\dedupTPMax}{66}

\newcommand{\dedupFPMax}{1}

\newcommand{\dedupCensusPrecision}{98.5\%}
\newcommand{\dedupCensusRecall}{98.5\%}
\newcommand{\dedupAgreementMin}{99.8\%}
\newcommand{\dedupKappaMin}{0.991}
\newcommand{\dedupCostUSD}{\$94}
\newcommand{\validityRuns}{nine}
\newcommand{\validitySample}{100}

\newcommand{\validityCost}{\$772}
\newcommand{\validityCostPerFinding}{\$0.86}
\newcommand{\validityPooledRate}{85\%}

\newcommand{\validityMatchedRate}{96\%}

\newcommand{\validityUnmatchedRate}{84\%}

\newcommand{\validityHumanGraded}{180}
\newcommand{\validityBothDecided}{131}
\newcommand{\validityAgreement}{93\%}
\newcommand{\validityKappa}{0.77}
\newcommand{\taxCorpus}{5,416}
\newcommand{\taxCorpusHuman}{2,444}
\newcommand{\taxCorpusAgentFound}{3,205}
\newcommand{\taxCorpusBoth}{233}
\newcommand{\taxInductionSample}{2,546}
\newcommand{\taxMinCategories}{25}
\newcommand{\taxMaxCategories}{40}
\newcommand{\taxInducedCategories}{38}
\newcommand{\taxCategories}{33}
\newcommand{\taxMergeFloor}{50}
\newcommand{\taxBatchSize}{50}
\newcommand{\taxAgentRuns}{14}

\newcommand{\taxTriggerHumanInput}{74.3\%}
\newcommand{\taxTriggerAgentsInput}{83.9\%}
\newcommand{\taxTriggerHumanEnvironment}{10.8\%}
\newcommand{\taxTriggerAgentsEnvironment}{4.7\%}
\newcommand{\taxTriggerHumanInspection}{3.8\%}
\newcommand{\taxTriggerAgentsInspection}{0.7\%}

\usepackage[capitalize]{cleveref}

\crefformat{section}{\S#2#1#3}

\crefrangeformat{section}{\S#3#1#4\crefrangeconjunction\S#5#2#6}

\crefmultiformat{section}{\S#2#1#3}{\crefpairconjunction\S#2#1#3}{\crefmiddleconjunction\S#2#1#3}{\creflastconjunction\S#2#1#3}

\newcommand{\crefrangeconjunction}{--}

\crefname{listing}{Lst.}{listings}
\crefname{line}{Line}{Lines}
\crefname{appendix}{App.}{App.}

\newcommand{\app}[1]{%
	\ifbool{includeappendix}{\cref{#1}}{the appendix}%
}
\newcommand{\App}[1]{%
	\ifbool{includeappendix}{\cref{#1}}{The appendix}%
}

\newcommand{\faHuggingface}{\raisebox{-0.15em}{\includegraphics[height=1em]{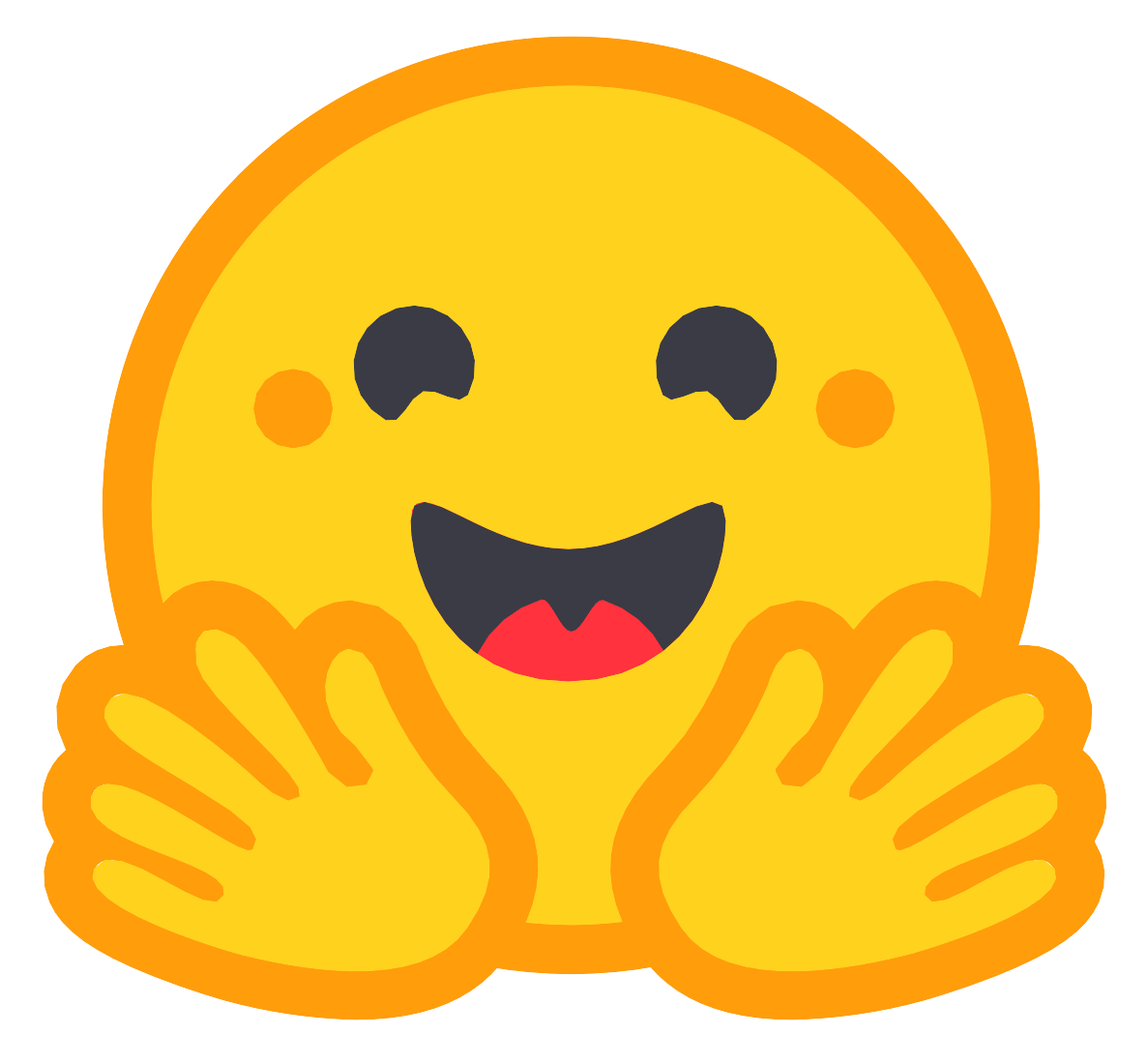}}}

\title{PROBE: Frontier Coding Agents Find \\Different Bugs Than Maintainers Fix}

\author{
Marco Paina$^1$,
Mark Niklas Müller$^1$,
Niels Mündler-Sasahara$^2$,
Martin Vechev$^2$ \\
$^1$ LogicStar.ai, $^2$ ETH Zurich \\
\texttt{\{\href{mailto:marco@logicstar.ai}{\textcolor{black}{marco}}, \href{mailto:mark@logicstar.ai}{\textcolor{black}{mark}}\}@logicstar.ai}
}

\iclrfinalcopy

\hypersetup{
  pdftitle={PROBE: Frontier Coding Agents Find Different Bugs Than Maintainers Fix},
  pdfauthor={Marco Paina, Mark Niklas Müller, Niels Mündler-Sasahara, Martin Vechev}
}

\begin{document}

\maketitle
\lhead{Preprint}

\vspace{-2.2em}
\begin{center}\small
  \definecolor{preprintlinkblue}{rgb}{0,0,1}
  \hypersetup{urlcolor=preprintlinkblue}
  \faGlobe\enspace\href{https://probebench.com}{\texttt{probebench.com}}\hspace{2em}
  \faHuggingface\enspace\href{https://huggingface.co/datasets/LogicStar/probe}{\texttt{LogicStar/probe}}
\end{center}

\begin{abstract}
Proactive bug discovery is an important software-engineering problem. Even though coding agents are already widely deployed in this domain, their capabilities remain understudied. In particular, the challenging task of finding all bugs that human maintainers would repair across an entire repository has not been studied yet. To close this gap, we introduce \bench, a benchmark consisting of all 2,457 bugs fixed over the course of 18 months across 20 diverse GitHub repositories, together with a method for matching candidate findings to this partial reference set.
We evaluate coding agents across frontier models and find they all achieve a recall of below $21\%$ and a precision of less than $9\%$ with respect to the reference set. While stronger models increase recall substantially from $8\%$ to $17\%$, they do not improve precision. Surprisingly, despite the low precision, LLM and human evaluators still judge $\validityPooledRate$ of findings valid. Crucially, we find that found bug distributions are similar across agents but differ substantially from those of bugs fixed by maintainers. The practical relevance of bug discovery and the low recall of frontier models make this benchmark a valuable resource for both model users and developers.
\end{abstract}

\makeatletter
\@startsection{section}{1}{\z@}{-0.5ex}{0.3ex}{\large\sc\raggedright}{Introduction}
\makeatother
\label{sec:introduction}

Coding agents have dramatically increased the pace at which code is being produced \citep{agentsinthewild,octoverse2025}, outpacing human review \citep{he2026aiwritesfasterhumans,xu2026aiassistedprogrammingdecreasesproductivity}. While coding agents have been shown to reliably fix bugs reported by humans \citep{swebench,swebenchpro}, it is an open question whether they can also find such meaningful bugs autonomously, a capability receiving increasing commercial attention \citep{codexsecurity,glasswing,bugbot,cubic}. Such a capability would ease the human burden of code review and application testing.

Prior work answers this question only incompletely, pointing the agent to a proposed patch or files known to contain a flaw \citep{swrbench,activeswe}, or focusing on security vulnerabilities \citep{vuleval,repoaudit,iris}, often ones that can be reliably exploited \citep{cybench,cybergym,bountybench}. Whether agents can find the bugs that maintainers would fix across a whole repository remains unexplored. Answering this question is hard, as the full set of bugs that a repository contains is unknown \citep{bugdatasets}. 

\begin{figure}[t]
    \centering
    \includegraphics[width=\linewidth,trim=0 0 0 10,clip]{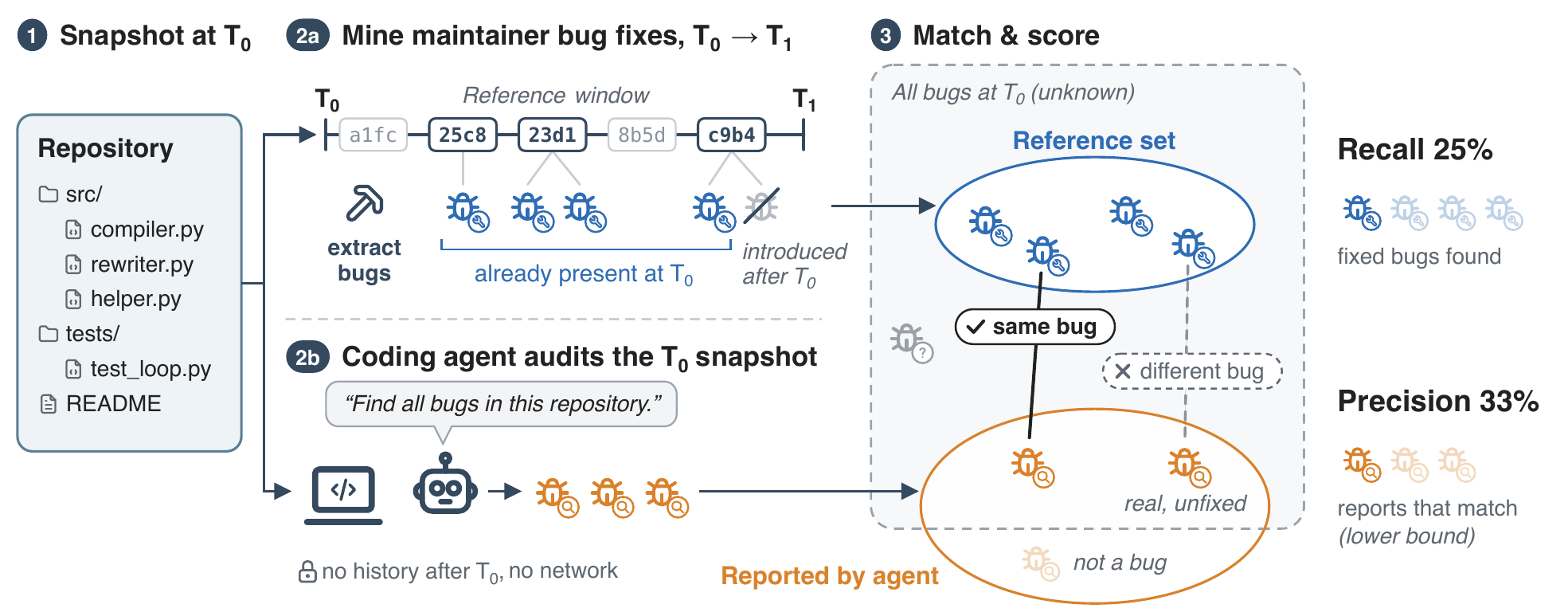}
    \vspace{-7mm}
    \caption{Overview of \bench. (1) We snapshot a repository at $T_0$. (2a) We mine the maintainers' commits in the reference window $T_0 \to T_1$, extract all bugs they fix, and keep those already present at $T_0$ as the reference set. (2b) The bug finder we are benchmarking audits the $T_0$ snapshot. (3) Reported bugs are matched against the reference set to compute recall and precision.}
    \label{fig:overview}
    \vspace{-6mm}
\end{figure}

However, we observe that maintainers reveal which bugs matter to them by fixing them over time. We build \bench on this signal (\cref{fig:overview}): we snapshot 20 Python, JavaScript and TypeScript repositories at the start of 2025 and collect the 2,457 bugs present in the snapshot that maintainers fixed over the following 18 months. An LLM-based matcher scores findings against these bugs; on expert-labeled pairs of findings, it reaches 99\% precision and 98\% recall. 

Asked to audit a whole repository, coding agents report only a handful of findings, recovering under 1\% of the reference bugs. Prior bug finding agents \citep{hound,agenticpbt} fare little better, achieving at most 3.2\% recall. We therefore split each repository into small sets of files and audit each in a separate agent session, raising recall roughly tenfold. 

Even with this harness, coding agents find few of the bugs maintainers fix: across ten models, the best configuration recovers only 20.8\% of the reference bugs, and none exceeds 8.1\% precision. Stronger models and higher reasoning effort raise recall mainly by reporting more findings (Pearson $r = 0.94$), while precision barely changes.
Low precision does not mean that findings are incorrect. Since maintainers keep fixing snapshot bugs after the reference window, even a perfect bug finder would reach only an estimated 26--47\% precision. Indeed, an independent panel of LLM judges, validated against a human grader, confirms \validityUnmatchedRate{} of the findings that do not match the reference set as real bugs. Thus, agent-found bugs are generally real, but different from those fixed by maintainers.

In addition, we find that all agents report similar bug types. It appears that their reported bugs are sampled from a distribution that differs substantially from that of maintainer-fixed bugs: agents over-report local correctness errors such as wrong arithmetic or comparisons (19\% of agent findings vs.\ 10\% of maintainer fixes), while maintainers more often fix incomplete input handling, such as a missing branch for a supported input variant (8\% vs.\ 18\%). This extends to individual bugs: within the same bug type, \luna runs overlap with each other a median of 6.3 times more than with maintainer fixes.


In conclusion, while agents find real bugs, they are mostly not the ones maintainers fix. Our benchmark \bench{} makes this gap measurable.

\paragraph{Key Contributions} Our three key contributions in this work are:
\begin{itemize}[topsep=0pt, itemsep=1pt, parsep=2pt, partopsep=0pt, leftmargin=15pt]
    \item \bench, a benchmark of 2,457 maintainer-fixed bugs across 20 repositories for whole-repository bug finding, built by a validated pipeline applicable to any repository (\cref{sec:method}).
    \item An evaluation of ten frontier models in a new multi-pass audit harness, with independent validation of findings beyond the reference set (\cref{sec:bug-finding,sec:experiments}).
    \item An analysis showing that agents agree with each other but not with maintainers (\cref{sec:different-bugs}).
\end{itemize}

\section{Evaluating Open-Ended Bug Discovery}
\label{sec:method}

\subsection{Benchmark Construction}

\paragraph{Overview}
We provide an overview of our approach, illustrated in \cref{fig:overview}. We collect a diverse set of open-source code repositories and take a snapshot at time $T_0$ (1st of January 2025). We then scrape all commits to the repository until $T_1$ (1st of July 2026); we call this time range the \emph{reference window}. We analyze every commit during this period to extract fixed bugs. For every detected bug, we check whether it was already present at $T_0$, and filter out bugs that were introduced in later commits. The remaining bugs form the \emph{reference set}. This reference set is incomplete, as bugs still unfixed at $T_1$ are missing.
Below, we provide more detail about each construction step and give all prompts for LLM-based steps in \cref{app:prompt}.

\paragraph{Repository Curation}
We manually selected a total of 20 Python, JavaScript, and TypeScript repositories from diverse domains that are actively developed, do not depend on internet access, special hardware or third-party services, did not experience a major refactoring in the reference window, have at most 200k source lines of code at $T_0$, and have a permissive license.

\paragraph{Environment Preparation}
We create historically faithful execution environments by using coding agents to set up Docker images for each repository, pinning the project dependencies to versions available at $T_0$. We ensure that the repository test suite passes in this state and remove all historical revisions, tags and branches.

\begin{wraptable}[12]{r}{0.42\textwidth}
  \vspace{-\baselineskip}
  \centering
  \caption{Commits and candidate bugs remaining after each stage of the bug collection pipeline. $\Delta$ is the change relative to the preceding stage.}
  \label{tab:funnel}
  \vspace{-2mm}
  \small
  \setlength{\tabcolsep}{4pt}
  \begin{tabular}{lrr}
    \toprule
    Stage & Count & $\Delta$ \\
    \midrule
    Commits                  & 13,254 & --        \\
    Bug-Fix Commits          &  4,296 & $-67.6\%$ \\
    Candidate Bugs           &  4,763 & $+10.9\%$ \\
    Anchored Candidates      &  3,345 & $-29.8\%$ \\
    Present Candidates       &  2,501 & $-25.2\%$ \\
    Deduplicated Bugs        &  2,457 & $-1.8\%$  \\
    \bottomrule
  \end{tabular}
  \vspace{-\baselineskip}
\end{wraptable}

\paragraph{Reference Bug Collection}
Considering all commits, we aim to extract all bugs fixed during the reference window using the following six-step pipeline. We note that relying on explicit links to issues has been shown to lead to small and biased datasets \citep{bugdatasets}. Unless otherwise specified, we use \terra (medium) for all steps requiring an LLM. \cref{tab:funnel} shows the number of commits and candidate bugs remaining after each stage. 

\begin{enumerate}[topsep=0pt, itemsep=1pt, parsep=2pt, partopsep=0pt, leftmargin=15pt]
  \item \eemph{Identify Bug-Fix Commits.} We filter all commits on the main branch, treating merges as a single commit, using a combination of deterministic rules and LLM classification to identify bug-fix commits. First, we remove commits made by automated repository maintenance and edits that touch only generated or vendored files. Second, we use an LLM to classify whether the commit corrects any runtime behavior, based on the commit message, the patch, related user issues, and associated pull requests. We explicitly include commits that additionally contain non-bug-fixing changes.
  \item \eemph{Extract Bugs.} From each bug-fix commit we extract all distinct \emph{candidate bugs}, again using an LLM, splitting bugs that do not share a root cause. Each candidate bug contains a brief summary of the incorrect behavior and the key code hunks modified to resolve the bug.
  \item \eemph{Anchor Code Changes.} Since we extract candidate bugs at the time of their fix $T'$ ($T_0 < T'$), the affected code ranges need to be mapped back onto $T_0$. To this end, we first trace file renames and then correct for lines added or removed between $T_0$ and $T'$. Lines that did not yet exist at $T_0$ are dropped. If no line survives, we try to relocate the block by its content or by its enclosing function name; if that fails too, the code was added or moved to a different file after $T_0$ and we remove the candidate bug.
  \item \eemph{Check Bug Presence.} We verify whether a candidate bug was actually present at $T_0$, using a coding agent with access to both the full revision history and the bug description. We additionally tell the agent whether AG-SZZ \citep{ag-szz} traces the lines modified by the fix to commits already present at $T_0$. If the agent establishes that the bug was absent at $T_0$ and only introduced by a later change, we exclude it. This also includes cases where inputs triggering the bug are not supported at $T_0$. For this step, we use high reasoning effort. \label{item:check}
  \item \eemph{Synthesize Bug Reports.} We now synthesize a detailed bug report based on the original description of the incorrect behavior and the repository state at $T_0$. The agent is instructed not to reference the fix in the report.
  \item \eemph{Deduplicate Candidates.} Finally, we merge candidate bugs that describe the same defect, using our LLM-based bug matching, see \cref{sec:evaluation}, under a stricter criterion requiring a shared root cause.\label{item:dedup}
\end{enumerate}

\subsection{Benchmark Overview}
\label{sec:benchmark-overview}

\bench (\benchlong) consists of \benchNumTargets{} distinct bugs across 10 Python, 6 TypeScript, and 4 JavaScript repositories, covering diverse domains such as parsing, rendering, math, and workflow orchestration. We show core statistics in \cref{tab:benchmark-overview} and a per-repository breakdown in \cref{fig:benchmark-repos,tab:benchmark-repos}.
Every bug consists of a description of the incorrect behavior, a primary location in the snapshot at $T_0$ (line ranges, file, and repository), the surrounding code snippet, a severity and a suggested fix (more details in \cref{app:bug-details}).

\begin{table}[t]
\begin{minipage}[t]{0.46\textwidth}
  \vspace{0pt}
  \centering
  \small
  \caption{Per-repository statistics of \bench.}
  \label{tab:benchmark-overview}
  \vspace{-2mm}
\begin{tabular}{@{}lrrr@{}}
\toprule
 & Min & Mean & Max \\
\midrule
\# Files & 17 & 223.4 & 657 \\
\# Lines (k) & 2.1 & 57.0 & 122.9 \\
\# Tests & 182 & 3,144.8 & 11,831 \\
\midrule
\# Commits & 75 & 662.7 & 3,465 \\
\# Bug-fixing commits & 6 & 214.8 & 979 \\
\# Reference bugs & 3 & 122.8 & 512 \\
\bottomrule
\end{tabular}

\end{minipage}\hfill
\begin{minipage}[t]{0.51\textwidth}
  \vspace{0pt}
  \centering
  \small
  \vspace{-2mm}
  \includegraphics[width=\textwidth]{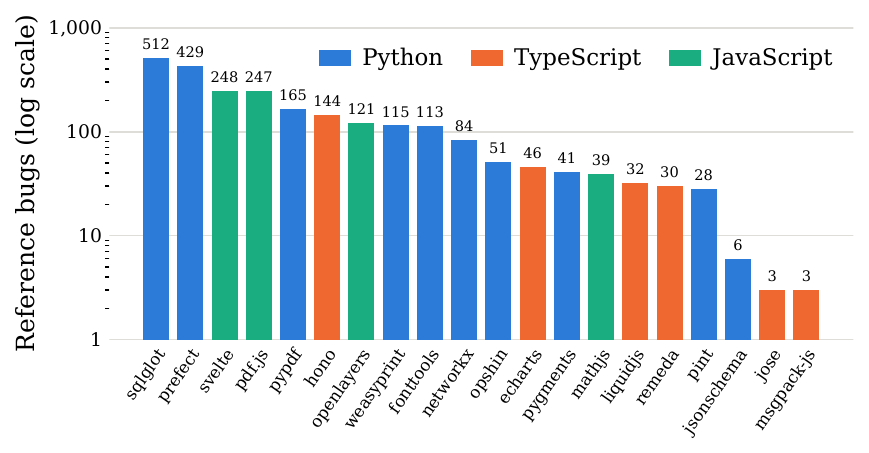}
  \vspace{-8mm}
  \captionof{figure}{Reference bugs per repository.}
  \label{fig:benchmark-repos}
\end{minipage}
\vspace{-4mm}
\end{table}

\paragraph{Benchmark Validation}
We assess the quality of the reference set with a two-stage audit.
First, we review all \benchNumTargets{} reference bugs with a high-effort \terra{} agent with access to the repository history. It categorizes \fpAuditRejected{} bugs (\fpAuditRejectionRate{}) as incorrect (introduced after $T_0$, no defect, not at indicated location, description contradicts code).
Second, we manually label a stratified sample of \fpAuditSample{} bugs, blind to the agent's verdicts. This yields \fpAuditHumanRejected{} rejected bugs, only weakly overlapping (\fpAuditBothRejected{} bugs) with the \fpAuditSampleRejected{} rejected by the agent. As agreement barely exceeds chance ($\kappa = \fpAuditKappa{}$), we keep the whole set.

\paragraph{Total Bug Pool Estimate}
\begin{wrapfigure}[12]{r}{0.43\textwidth}
    \vspace{-\baselineskip}
    \vspace{1mm}
    \centering
    \includegraphics[width=0.98\linewidth,trim=10 0 3 0,clip]{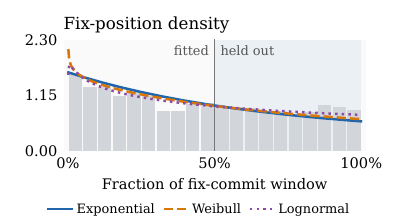}
    \vspace{-3mm}
    \caption{Fix-position density observed (gray bars) and fitted based on the first half of each repository's fix-commit sequence.}
    \label{fig:bug-fixing-holdout}
\end{wrapfigure}
We estimate the total number of bugs humans would eventually fix by modeling the rate at which $T_0$ bugs are fixed. We fit exponential, Weibull, and lognormal distributions and show the resulting fits in \cref{fig:bug-fixing-holdout}. We observe a heavy tail with the lognormal distribution fitting the data best. Fitting the full data, we estimate between 5.2k and 5.9k bugs fixed within 5 years and between 7.2k and 9.4k within 10 years, see \cref{tab:drain-horizons}. As humans tend to discover more relevant bugs earlier, we assume bugs detected in the far future to be irrelevant. Given the observed 2,457 bugs, this implies a ceiling on precision of between 26\% and 47\% for a perfect bug finder that finds exactly the bugs humans would fix within 5 to 10 years.
For a more detailed description of the model and the exact fitted parameters, see \cref{app:bug-fixing-rate}.

\subsection{Evaluation}
\label{sec:evaluation}

\paragraph{Metrics}
We report recall, precision, and F1 score with respect to the reference set. Recall is the proportion of reference bugs that were matched by an agent finding. Precision is the proportion of findings that match reference bugs. 
We additionally report cost for one run in USD at API prices.


\paragraph{Bug Matching}
Reliably matching findings to reference bugs is essential for the validity of our evaluation. We thus propose the following two-step method, which considers all pairs of bugs using their description, code snippet, and line ranges, so a bug finder must provide at least this information. This approach costs about \$5--8 per 1,000 findings.

\begin{enumerate}[topsep=0pt, itemsep=1pt, parsep=2pt, partopsep=0pt, leftmargin=15pt]
  \item \eemph{Rule-based Filter.} We compute six similarity metrics per bug pair: line overlap, cosine similarities of an embedding of the description and an embedding additionally considering the code snippet, and inverse document frequency-weighted Jaccard similarities of the tokenized code snippet, identifiers in the code snippet and description, and called function identifiers. If any metric exceeds a threshold tuned to avoid false negatives with respect to the second stage, we retain the pair. For details on the metrics and thresholds, see \cref{tab:dedup-signals}.
  \item \eemph{LLM-based Judge.} We instruct \geminiflash, given the description, line range, and code snippet for both bugs, to determine whether resolving the underlying root cause for either one of them would also resolve the other. If so, we count them as matching.
\end{enumerate}

We validate this approach on \dedupNumFindings{} findings across \dedupNumRepos{} repositories with \dedupNumGold{} expert-validated matches. Of the \dedupNumPairs{} within-repository pairs, \dedupFiredRate{} pass the rule-based filter, including all matches. All removed pairs were also rejected by the LLM judge in a validation experiment. Across \dedupNumRuns{} runs of the judge, precision averages \dedupPrecisionMean{} and recall \dedupRecallMean{}, with each run identifying \dedupTPMin{} or \dedupTPMax{} of the \dedupNumGold{} matches and marking at most \dedupFPMax{} additional pair as a match. For more details, see \cref{app:dedup}.

\paragraph{Independent Bug Validation}
To assess precision beyond the incomplete reference set, we use a two-tiered panel of agents.
If the first tier (\luna and \geminiflash) disagrees, we refer the finding to a second tier (\opus and \terra; \sol often refused to answer due to cybersecurity guardrails), reporting an inconclusive result if they disagree.
All judges are instructed to produce a proof of concept demonstrating the described bug on a real user input, given the snapshot codebase and the bug report. We report validity rate as the proportion of confirmed findings, ignoring inconclusive results. See \cref{prompt:panel-validate} for the exact instructions.

\section{Bug Finding Harness}
\label{sec:bug-finding}
As we observe vanilla coding agents to achieve $<1\%$ recall when asked to audit a whole repository (\cref{tab:harness_ablation}), we propose a harness boosting recall. To this end, we partition a repository's files, where each partition is reviewed in a separate \emph{pass} with access to the whole repository and a context document. Every model receives identical file partitions and context documents.

\paragraph{Preprocessing}
For every repository, we create two auxiliary documents, shared across all passes: i) a \emph{context document} compiled by a \opus exploration agent (\cref{prompt:explore}), describing the project's purpose, architecture, public API and supported inputs; ii) a function-level call graph of the non-test source files, built with Python's \texttt{ast} module or the TypeScript compiler API.

\paragraph{File Partitioning}
We partition all non-test source files by complexity and size. We score every file by its number of cross-file callers and callees, its size, and its number of functions. The most complex files are assigned their own group, while the rest are grouped by directory, splitting groups that exceed a budget of 60k tokens. Across the 20 repositories, this yields between 5 and 90 passes each, with a total of 740 passes covering 4,468 files. See \cref{app:pass-planning} for more details.

\paragraph{Prompt Construction}
The prompt is constructed from the audit instructions (\cref{prompt:pass-instructions}), the context document, the full target files, a call graph excerpt, and short snippets of callers and callees in other files. The agent is tasked to report each bug in JSON format using the same fields as the reference bugs (see \cref{tab:bug-record}) plus a confidence score. We remove findings violating the schema.

\paragraph{Execution and Isolation}
We run each pass with a timeout of three hours in a separate Docker container with historically faithful dependencies, but without revision history. Network access and the CLIs' web search tools are disabled.

\section{Experimental Evaluation}
\label{sec:experiments}

\paragraph{Experimental Setup}
We evaluate ten models: \luna, Terra and Sol; \gptfivefour; \sonnet and Opus 5 (\cref{app:lane-subset}); \geminiflash; \glmflash; \deepseekflash; and \musespark. We run Anthropic, OpenAI, and Google models in their vendors' coding-agent CLIs and the remaining models using Codex CLI. Unless otherwise specified, we use medium reasoning effort for the OpenAI, Anthropic and Google models and high effort for \glmflash, \deepseekflash and \musespark, evaluate all models with our harness, and run each configuration once for cost reasons.

\subsection{Agents Recover Few Maintainer-Fixed Bugs}

\begin{table}
\centering
\caption{Recall, precision, and F1 score w.r.t.\ the reference set, sorted by F1, with the effort sweeps of \sol and \luna last, from highest to lowest reasoning effort. Cells for \musespark and \luna (medium) are means over 5 and 10 runs, respectively.}
\label{tab:lane_summary}
\footnotesize
\vspace{-2mm}
\sisetup{detect-weight}
\small
\begin{tabular}{llS[table-format=2.1]S[table-format=2.1]S[table-format=2.1]rr}
\toprule
{Model} & {Effort} & {Recall (\%)} & {Precision (\%)} & {F1 (\%)} & {Findings} & {Cost} \\
\midrule
DeepSeek V4.1 Flash & high & 15.2 & \bfseries 8.1 & \bfseries 10.5 & 5,019 & \$73 \\
GLM 5.3 Flash & high & 15.8 & 7.0 & 9.7 & 6,029 & \$120 \\
GPT-5.4 & medium & 11.0 & 6.9 & 8.5 & 4,204 & \$655 \\
GPT-5.6 Terra & medium & 9.1 & 7.6 & 8.3 & 3,093 & \$224 \\
Gemini 3.5 Flash & medium & 9.8 & 6.4 & 7.8 & 3,867 & \$493 \\
Muse Spark 1.3 & high & 15.9 & 4.8 & 7.4 & \textbf{8,372} & \$164 \\
Claude Sonnet 5 & medium & 7.9 & 5.8 & 6.7 & 3,472 & \$421 \\
\cmidrule(lr){1-2}
\multirow{3}{*}{GPT-5.6 Sol} & high & \bfseries 20.8 & 6.7 & 10.2 & 8,285 & \$1,185 \\
 & medium & 16.9 & 6.7 & 9.6 & 6,675 & \$678 \\
 & low & 11.4 & 7.5 & 9.0 & 3,903 & \$276 \\
\cmidrule(lr){1-2}
\multirow{4}{*}{GPT-5.6 Luna} & xhigh & 17.3 & 7.1 & 10.1 & 6,636 & \$131 \\
 & high & 13.0 & 6.9 & 9.0 & 5,017 & \$55 \\
 & medium & 7.6 & 7.0 & 7.3 & 2,774 & \$15 \\
 & low & 5.3 & 6.5 & 5.8 & 2,027 & \textbf{\$8} \\
\bottomrule
\end{tabular}

\vspace{-3mm}
\end{table}

\begin{wrapfigure}[11]{r}{0.45\textwidth}
    \vspace{-\baselineskip}
    \vspace{-1mm}
    \centering
    \includegraphics[width=\linewidth]{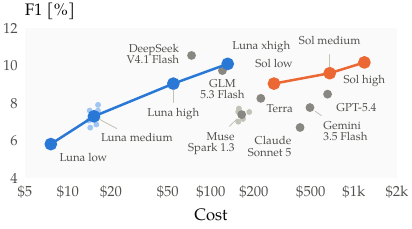}
    \vspace{-7mm}
    \caption{F1 scores over API cost. \luna (blue) and Sol (orange) across reasoning efforts. Reruns are faint markers.}
    \label{fig:f1-effort}
\end{wrapfigure}
\paragraph{Historical Recall and Precision}
We compare different models in \cref{tab:lane_summary} (\cref{app:repo-recall}) and observe that all agents recover only a small share of the historical bugs, with even the best run achieving less than 21\% recall and at most 8.1\% precision. While recall varies significantly across models, mostly determined by the total number of findings (Pearson $r = 0.94$), precision is much more consistent.
Interestingly, some of the cheapest models (\deepseekflash, \luna (xhigh), and \glmflash) exceed \sol (medium) in F1 score at less than a fifth of its cost. 

\paragraph{Reasoning Effort} We evaluate \luna at low to xhigh reasoning effort, visualizing the resulting F1 score in \cref{fig:f1-effort}. We find that while precision stays constant, recall increases substantially from 5.3\% to 17.3\% as cost increases from \$8 to \$131. \luna (xhigh) thus strictly dominates \sol (medium) in terms of cost, recall, and precision. Evaluating \sol at low to high effort shows the same trend, with recall rising from 11.4\% to 20.8\% as cost rises from \$276 to \$1,185.

\subsection{Agents Find Different Bugs Than Maintainers Fix}
\label{sec:different-bugs}

\begin{wrapfigure}[13]{r}{0.40\textwidth}
    \vspace{-\baselineskip}
    \vspace{-3mm}
    \centering
    \includegraphics[width=\linewidth]{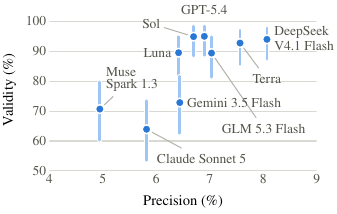}
    \vspace{-6mm}
    \caption{Bug validity assessed by the LLM panel over precision w.r.t.\ the reference set, with 95\% confidence intervals. For \luna and \musespark, both axes use their first run.}
    \label{fig:validity-precision}
\end{wrapfigure}

\paragraph{Most Unmatched Bugs Are Still Valid}
To test whether low precision w.r.t.\ the reference set means that most findings are not in fact bugs, we validate \validitySample{} findings each for \validityRuns{} models using our agent panel (\cref{fig:validity-precision}; details in \cref{tab:validity-lanes}). It confirms \validityMatchedRate{} of the findings that match the reference set and \validityUnmatchedRate{} of those that do not as correct. Three of the four models with the lowest precision (\sonnet, \geminiflash and \musespark) also have significantly lower validity (64--73\% vs.\ 89--95\%). Manually grading \validityHumanGraded{} of these findings blind to the model and the panel's verdict, we agree with the panel on \validityAgreement{} of the \validityBothDecided{} findings that neither judged inconclusive ($\kappa = \validityKappa{}$; \cref{app:validity}). Low precision thus mostly reflects real bugs outside the reference set.

However, since agents show no preference for earlier- or later-fixed bugs (\cref{app:fix-rank-test}), extrapolating their precision of about 7\% to our estimate of all $T_0$ bugs fixed within ten years (\cref{tab:drain-horizons}) yields a precision of only 21--27\%. With \validityPooledRate{} of findings considered valid, more than two-thirds of the valid bugs agents report would thus remain unfixed for at least a decade at the maintainers' current pace.

\begin{wrapfigure}[14]{r}{0.4\textwidth}
    \vspace{-\baselineskip}
    \centering
    \includegraphics[width=1.0\linewidth]{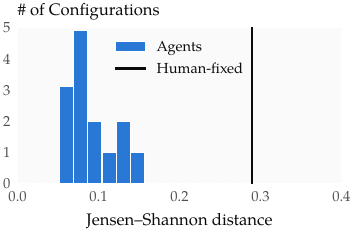}
    \vspace{-6mm}
    \caption{Jensen--Shannon distance of the bug-type distribution of each agent run to the other runs and of the human reference set to all runs.}
    \label{fig:fault-type-jsd}
\end{wrapfigure}
\paragraph{Agents and Humans Find Different Bug Types}
To investigate this mismatch, we categorize all bugs found by agents and humans into \taxCategories{} bug types (see \cref{app:taxonomy} for how we build them and their full distributions) and compare their distributions. We observe in \cref{fig:fault-type-jsd} that all agents find bugs with relatively similar category distributions while the human distribution is quite different.

Comparing the five bug types most over- and under-represented among agent findings relative to human ones (\cref{tab:bug-type-distribution}), we observe that agents tend to over-report local faults, such as wrong arithmetic or comparisons and unhandled degenerate values, while humans focus on incomplete handling of inputs and errors in the public API surface. We hypothesize that humans are more likely to find and fix bugs that cause a mismatch between described and implemented behavior or observable failures when running the code or using its public API, while coding agents are more likely to find logical bugs that cause subtly incorrect behavior. Categorizing what is needed to trigger incorrect behaviors (\cref{app:taxonomy-trigger}), we observe that humans find more bugs requiring specific environments (\taxTriggerHumanEnvironment{} vs.\ \taxTriggerAgentsEnvironment{} for agents) or visual inspection of outputs (\taxTriggerHumanInspection{} vs.\ \taxTriggerAgentsInspection{}), while agent findings more often need only a specific input (\taxTriggerAgentsInput{} vs.\ \taxTriggerHumanInput{} for humans).

\begin{table}
\centering
\caption{Bug types most over- (top) and under-represented (bottom) among agent findings relative to the human reference set, by gap in percentage points (pp). Agent counts ($\bar{n}$) and shares are means over one run of each of the \taxAgentRuns{} configurations.}
\label{tab:bug-type-distribution}
    \footnotesize
\begin{tabular}{lrrrrrr}
\toprule
 & \multicolumn{2}{c}{Human} & \multicolumn{2}{c}{Agents} & & \\
\cmidrule(lr){2-3} \cmidrule(lr){4-5}
Category & $n$ & \% & $\bar{n}$ & \% & Gap (pp) & Ratio \\
\midrule
Wrong arithmetic or comparison expression & 243 & 9.9 & 916 & 18.5 & $+8.6$ & 1.86 \\
Unhandled degenerate value & 222 & 9.1 & 674 & 13.9 & $+4.8$ & 1.53 \\
Incorrect regular expression pattern & 47 & 1.9 & 245 & 4.7 & $+2.8$ & 2.47 \\
Missing state reset or restoration & 57 & 2.3 & 242 & 4.9 & $+2.6$ & 2.11 \\
Wrong data source read & 57 & 2.3 & 224 & 4.5 & $+2.1$ & 1.91 \\
\midrule
Missing branch for a supported input variant & 439 & 18.0 & 400 & 7.7 & $-10.3$ & 0.43 \\
Missing dispatch or lookup table entry & 124 & 5.1 & 38 & 0.7 & $-4.3$ & 0.14 \\
Incorrect public type declaration & 111 & 4.5 & 57 & 1.1 & $-3.5$ & 0.24 \\
Unimplemented advertised feature & 114 & 4.7 & 133 & 2.6 & $-2.0$ & 0.56 \\
Wrong dispatch or lookup table entry & 102 & 4.2 & 130 & 2.6 & $-1.6$ & 0.63 \\
\bottomrule
\end{tabular}

\end{table}


\begin{wrapfigure}[11]{r}{0.37\textwidth}
    \vspace{-\baselineskip}
    \centering
    \includegraphics[width=\linewidth]{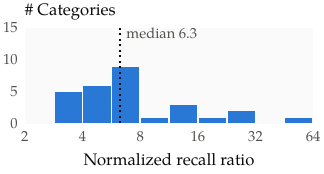}
    \vspace{-6mm}
    \caption{Per-category overlap of repeated \luna runs compared to human-fixed bugs.}
    \label{fig:lift-ratio}
\end{wrapfigure}
\paragraph{Agents and Humans Find Different Bugs of the Same Type}
This mismatch extends to findings within a category. Per bug type, we divide a run's recall w.r.t.\ another run of the same agent by its recall w.r.t.\ the human reference set, pooling over all pairs of runs. If this normalized recall ratio is 1, the agent runs overlap with each other just as much as with the human reference set. If it is higher, they agree more with each other than with the human reference set. We observe in \cref{fig:lift-ratio} that across the 28 bug types with at least ten human and agent findings, the ratio has a minimum of 2.9 and a median of 6.3 for \luna.

\begin{wrapfigure}[15]{r}{0.43\textwidth}
    \vspace{-\baselineskip}
    \vspace{-1mm}
    \centering
    \includegraphics[width=\linewidth]{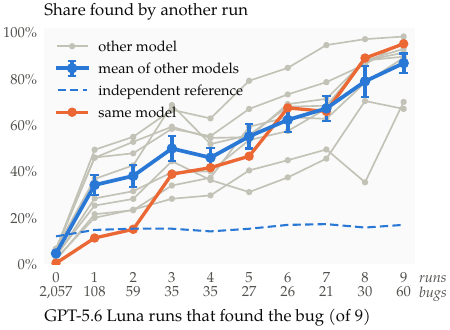}
    \vspace{-5mm}
    \caption{Share of reference bugs that other models find, conditioned on how many of 9 Luna runs find the bug. The dashed line picks bugs at random within each type; the orange line is a held-out tenth Luna run. }
    \label{fig:shared-difficulty}
\end{wrapfigure}
This agreement extends to other models. Analyzing the likelihood of another model finding a reference bug, conditioned on \luna finding it in exactly $n$ of 9 runs (\cref{fig:shared-difficulty}), we see it rise from 5\% for the bugs no run finds to 87\% for the bugs all 9 find. In contrast, sampling bugs at the same recall uniformly at random per category would give 12--17\%.

\subsection{Further Analysis}
\label{sec:further-analysis}

\paragraph{Our Audit Harness and Prior Bug Finders}
\label{par:harness-ablation}
We ablate our bug finding harness using \sonnet and \luna in \cref{tab:harness_ablation}. \emph{Vanilla} uses a single coding agent session of the native CLI with a basic prompt (see \cref{prompt:vanilla-audit}). \emph{Passes} keeps the 740 passes of our harness, but replaces our audit instructions, context document, file contents and call graph with a minimal prompt listing the pass's target files (see \cref{prompt:passes-audit}).
The Vanilla setting reports $< 100$ total findings across 20 repositories, leading to a recall of $<1\%$ while achieving strong precision. Adding the multi-pass structure but not the full context leads to both a recall and precision between the full harness and the single prompt, at a lower F1 score and higher cost compared to our harness. Comparing against two prior LLM-based bug finders, we observe Hound \citep{hound}, a security audit agent, achieving less than half our recall at a lower precision and $8.6\times$ the cost, and agentic-pbt \citep{agenticpbt}, which employs property-based testing and only supports Python, achieving $<1\%$ recall at good precision.

\begin{table}[t]
\centering
\caption{Comparison of our bug finding harness against a single prompt (Vanilla), the same pass decomposition but with a minimal prompt (Passes), and prior methods: Hound \citep{hound} and agentic-pbt \citep{agenticpbt}. Recall, precision and F1 are in \%.}
\label{tab:harness_ablation}
\footnotesize
\setlength{\tabcolsep}{3.5pt}
\sisetup{detect-weight}
\begin{tabular}{@{}lllS[table-format=2.1]S[table-format=2.1]S[table-format=2.1]rr@{}}
\toprule
{Dataset} & {Model} & {Method} & {Recall (\%)} & {Precision (\%)} & {F1 (\%)} & {Findings} & {Cost} \\
\midrule
\multirow{7.5}{*}{Full} & \multirow{3}{*}{Claude Sonnet 5} & Vanilla & 0.6 & \bfseries 17.4 & 1.2 & 86 & \textbf{\$94} \\
 &  & Passes & 5.3 & 8.0 & 6.4 & 1,656 & \$487 \\
 &  & \cellcolor{black!6}Full harness (ours) & \cellcolor{black!6}\bfseries 7.9 & \cellcolor{black!6}5.8 & \cellcolor{black!6}\bfseries 6.7 & \cellcolor{black!6}\textbf{3,472} & \cellcolor{black!6}\$421 \\
\cmidrule(l){2-8}
 & \multirow{4}{*}{GPT-5.6 Luna} & Vanilla & 0.9 & \bfseries 32.3 & 1.7 & 65 & \textbf{\$1} \\
 &  & Passes & 5.8 & 8.3 & 6.8 & 1,826 & \$18 \\
 &  & Hound & 3.2 & 5.0 & 3.9 & 1,371 & \$129 \\
 &  & \cellcolor{black!6}Full harness (ours) & \cellcolor{black!6}\bfseries 7.6 & \cellcolor{black!6}7.0 & \cellcolor{black!6}\bfseries 7.3 & \cellcolor{black!6}\textbf{2,774} & \cellcolor{black!6}\$15 \\
\cmidrule(l){1-8}
\multirow{3}{*}{Python Only} & \multirow{3}{*}{GPT-5.6 Luna} & agentic-pbt (medium) & 0.9 & \bfseries 15.5 & 1.7 & 97 & \textbf{\$4} \\
 &  & agentic-pbt (high) & 0.9 & 8.9 & 1.6 & 180 & \$9 \\
 &  & \cellcolor{black!6}Full harness (ours) & \cellcolor{black!6}\bfseries 6.1 & \cellcolor{black!6}6.7 & \cellcolor{black!6}\bfseries 6.4 & \cellcolor{black!6}\textbf{1,412} & \cellcolor{black!6}\$7 \\
\bottomrule
\end{tabular}

\end{table}

\begin{wrapfigure}[17]{r}{0.42\textwidth}
    \vspace{1mm}
    \centering
    \includegraphics[width=\linewidth]{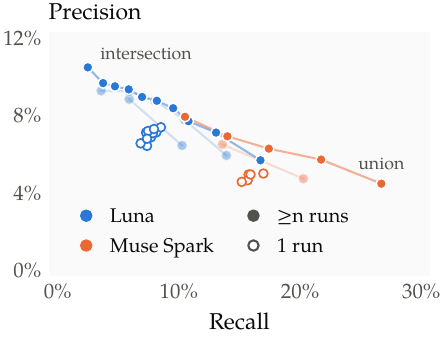}
    \vspace{-6mm}
    \caption{Historical precision and recall for findings reported in at least $n$ of 10 (Luna) or 5 (Muse Spark) runs, from $n=1$ (union) to all runs (intersection). Faint curves use only the first two or five runs.}
    \label{fig:rerun-threshold}
\end{wrapfigure}
\paragraph{Ensembling Multiple Runs}
Across reruns of \musespark (5) and \luna (10), the standard deviations of recall (0.66 and 0.48\,pp) and precision (0.20 and 0.32\,pp, respectively) are relatively small. However, on average only 63\% and 57\% of one run's findings are also reported by another run. Thus, ensembling runs by reporting only findings that repeat at least $n$ times across runs allows us to trade off recall and precision as illustrated in \cref{fig:rerun-threshold}. For example, the union of all \musespark runs achieves a recall of 26.7\% and the intersection of all \luna runs achieves a precision of 10.5\%, both exceeding the best single-run performance. This shows that findings with a higher report probability are more likely to match human findings. We explore this further in \cref{app:bug-discovery-model} where we show that the bug finding process is well modeled by a stable-beta process, characterized by a stable core of high-probability findings and an infinite pool of rare findings.

\paragraph{Time Preferences in Matched Bugs}
We show in \cref{app:fix-rank-test} that there is no detectable preference for agents finding earlier- or later-fixed bugs across models (Holm-adjusted $p$-values $\geq 0.20$). As the reference window contains the knowledge cut-offs of all tested models that publish one (\cref{tab:knowledge-cutoffs}), we find no evidence of contamination for them; for \glmflash, \musespark and \deepseekflash, released after $T_1$ without a published cut-off, the test cannot rule it out.

\section{Related Work}
\label{sec:related-work}
\paragraph{Repository-Level Maintenance}
A plethora of prior work has studied autonomous bug resolution \citep{swebench,swebenchpro,swebenchmultimodal}. Much less has investigated code maintenance tasks, such as refactoring \citep{codetaste,swebenchpromax}, test generation \citep{mundler2024code} or code optimization \citep{algotune,sweperf}.
To our knowledge, no prior benchmark evaluates general repository-wide bug discovery.

\paragraph{Restricted Bug Detection}
Prior work on autonomous bug detection focuses on investigating specific files or patches:
SWR-Bench \citep{swrbench} provides agents with a pull request and asks them to generate review comments on the issues that require changes, including software bugs.
Active-SWE \citep{activeswe} mines historical bugs and measures whether agents can detect and resolve concrete bugs when provided with the affected files.
In contrast, our work tasks agents to report all defects across an entire repository.

\paragraph{Repository-Level Vulnerability Detection}
Cybench \citep{cybench} requires agents to autonomously discover and exploit software vulnerabilities, but is limited to artificially constructed targets.
CyberGym-E2E \citep{cybergym} measures vulnerability detection, exploitation and patching on historical OSS-Fuzz vulnerabilities in open-source projects.
BountyBench \citep{bountybench} is based on real-world open-source projects with associated bug bounties.
In contrast, our work considers general software bugs beyond security vulnerabilities.

\paragraph{Bug Detection Methods}
\citet{repoaudit} propose a repository-level auditing agent that follows data flow to detect null-pointer dereferences, memory leaks and use-after-free bugs.
\citet{bugauditor} focus on defensive coding patterns and find instances with similar context but lacking defenses.
\citet{vuleval} evaluate repository-level vulnerability detection that retrieves vulnerability-related dependencies from call graphs.
\citet{iris} propose to combine LLMs with static analysis reports to discover security vulnerabilities.
None of these methods targets the setting of general software bugs we study.

\section{Limitations}
\label{sec:discussion}
\label{sec:limit}

\paragraph{Maintainer fixes as the reference} The reference set contains only bugs fixed within 18 months, and an unfixed bug is not necessarily unimportant: maintainers may be unaware of it or lack the capacity to fix it. However, we estimate how many snapshot bugs maintainers would fix within 5 to 10 years, which caps precision at 26--47\% (\cref{sec:benchmark-overview}). Our conclusions do not hinge on the exact value, as even the tightest ceiling is three times the best observed precision. Recall, in contrast, does not depend on when a bug was fixed (\cref{app:fix-rank-test}) and thus likely carries over to bugs fixed later. Further, if capacity alone limited fixes, unfixed bugs would resemble fixed ones in type; instead, agent findings differ systematically from maintainer fixes (\cref{sec:different-bugs}), so the reference set reflects maintainers' priorities rather than a random sample of bugs.

\paragraph{Benchmark scope} \bench covers 20 repositories from diverse domains in the three most used languages on GitHub \citep{octoverse2025}. However, these languages share certain properties that prevent specific types of bugs, e.g., they are all memory-safe and thus exclude memory-safety bugs. Moreover, our curation excluded codebases beyond 200k lines. As our pipeline applies to any repository with a commit history and needs no manual bug labeling, we believe natural extensions to additional languages and repositories are interesting items for future work.

\paragraph{Bug finding methods} Other harnesses could yield different results. Ours is identical across models, isolating model differences, and outperforms both single-session agents and prior methods (\cref{tab:harness_ablation}); as each pass can access the whole repository, agents are not restricted to local reasoning. However, maintainer-fixed bugs more often need a specific environment, a person inspecting the output, or particular timing to surface (\cref{app:taxonomy-trigger}), while the evaluated agents primarily inspect code. Methods that exercise software as its users do could find bugs closer to those maintainers fix, an exciting direction for future work.

\section{Conclusion}
\label{sec:conclusion}

We introduced \bench, the first benchmark for whole-repository bug discovery that uses the bugs maintainers fix as its reference. Across ten models, coding agents recover few of these bugs at low precision, achieving a maximum recall of 20.8\%. Stronger models and higher reasoning effort mostly yield more findings rather than better ones. Yet most findings are real bugs, as an independent panel of LLM judges confirms. We investigate this and find that they just differ systematically from the bugs maintainers fix: LLM-found bugs disproportionately involve local faults, such as wrong arithmetic or comparisons and unhandled degenerate values, while human-fixed bugs more often concern incomplete input handling and errors in the public API surface. We believe that finding the bugs that matter to humans is a key frontier for scaling automated software engineering, and that \bench makes progress toward it measurable.

\clearpage

\bibliography{references}
\bibliographystyle{iclr2027_conference}

\clearpage
\appendix
\raggedbottom  
\onecolumn

\crefalias{section}{appendix}
\crefalias{subsection}{appendix}
\crefalias{subsubsection}{appendix}

\section{Benchmark Details}
\label{app:benchmark-details}

\subsection{The Repositories}
\label{app:repositories}

\Cref{tab:benchmark-repos} lists the \benchNumRepos{} repositories of \bench{} by their GitHub names, grouped by language and sorted by reference bugs; elsewhere we refer to them by repository name alone.

\begin{table}[ht]
\centering \small
\caption{The \benchNumRepos{} repositories of \bench. Files and kLOC count the audited non-test source files (\cref{app:pass-planning}), and Passes the audit passes they are split into (\cref{sec:bug-finding}). Tests counts the tests that pass at the snapshot in our Docker image (for echarts, its unit tests only). Commits and Fix com.\ (bug-fixing commits) are counted over the 18-month reference window, and Bugs are the reference bugs they yield.}
\label{tab:benchmark-repos}
\scalebox{0.78}{
\begin{tabular}{@{}llrrrrrrr@{}}
\toprule
Repository & Domain & Files & Passes & kLOC & Tests & Commits & Fix com. & Bugs \\
\midrule
\texttt{tobymao/sqlglot} & SQL parsing & 70 & 16 & 49.9 & 646 & 2,848 & 979 & 512 \\
\texttt{prefecthq/prefect} & Workflow orchestration & 420 & 70 & 119.5 & 11,831 & 3,465 & 910 & 429 \\
\texttt{py-pdf/pypdf} & PDF manipulation & 52 & 14 & 35.7 & 562 & 562 & 190 & 165 \\
\texttt{kozea/weasyprint} & HTML/CSS rendering & 65 & 19 & 30.5 & 3,053 & 315 & 196 & 115 \\
\texttt{fonttools/fonttools} & Font tooling & 308 & 69 & 122.1 & 4,372 & 318 & 108 & 113 \\
\texttt{networkx/networkx} & Graph algorithms & 283 & 44 & 112.5 & 5,646 & 481 & 93 & 84 \\
\texttt{opshin/opshin} & Smart-contract compiler & 49 & 12 & 10.8 & 411 & 213 & 63 & 51 \\
\texttt{pygments/pygments} & Syntax highlighting & 330 & 47 & 122.9 & 5,122 & 165 & 46 & 41 \\
\texttt{hgrecco/pint} & Units and quantities & 73 & 15 & 16.4 & 2,170 & 88 & 34 & 28 \\
\texttt{python-jsonschema/jsonschema} & Schema validation & 23 & 5 & 4.9 & 7,675 & 167 & 6 & 6 \\
\rowcolor{black!6}\emph{Python (10)} &  & 1,673 & 311 & 625.1 & 41,488 & 8,622 & 2,625 & 1,544 \\
\midrule
\texttt{honojs/hono} & Web framework & 167 & 35 & 19.9 & 3,268 & 431 & 181 & 144 \\
\texttt{apache/echarts} & Charting & 550 & 90 & 120.2 & 182 & 110 & 58 & 46 \\
\texttt{harttle/liquidjs} & Template engine & 120 & 32 & 5.9 & 1,416 & 103 & 33 & 32 \\
\texttt{remeda/remeda} & Functional utilities & 309 & 8 & 26.3 & 1,935 & 231 & 41 & 30 \\
\texttt{panva/jose} & Cryptography & 111 & 23 & 8.7 & 282 & 248 & 17 & 3 \\
\texttt{msgpack/msgpack-javascript} & Serialization & 17 & 5 & 2.1 & 316 & 75 & 8 & 3 \\
\rowcolor{black!6}\emph{TypeScript (6)} &  & 1,274 & 193 & 183.2 & 7,399 & 1,198 & 338 & 258 \\
\midrule
\texttt{sveltejs/svelte} & UI compiler & 328 & 64 & 49.2 & 5,698 & 1,414 & 618 & 248 \\
\texttt{mozilla/pdf.js} & PDF rendering & 178 & 37 & 115.4 & 948 & 1,345 & 535 & 247 \\
\texttt{openlayers/openlayers} & Map rendering & 358 & 84 & 108.3 & 2,163 & 549 & 148 & 121 \\
\texttt{josdejong/mathjs} & Mathematics & 657 & 51 & 59.4 & 5,200 & 126 & 32 & 39 \\
\rowcolor{black!6}\emph{JavaScript (4)} &  & 1,521 & 236 & 332.3 & 14,009 & 3,434 & 1,333 & 655 \\
\midrule
\textbf{Total (20)} &  & \textbf{4,468} & \textbf{740} & \textbf{1,140.6} & \textbf{62,896} & \textbf{13,254} & \textbf{4,296} & \textbf{2,457} \\
\bottomrule
\end{tabular}

}
\end{table}

\subsection{Bug Structure}
\label{app:bug-details}
Every bug consists of the fields in \cref{tab:bug-record}: a title and description of the incorrect behavior, a primary location in the snapshot at $T_0$, the surrounding code, a severity and a suggested fix. In our main text we summarize \texttt{repo}, \texttt{file}, \texttt{line\_start/end} as primary code location, and \texttt{title}, \texttt{description} as description.

\begin{table}[ht]
  \centering
  \small
  \caption{Fields of a reference bug record.}
  \label{tab:bug-record}
  \begin{tabularx}{0.51\textwidth}{@{}l>{\raggedright\arraybackslash}X@{}}
    \toprule
    Field & Description \\
    \midrule
    \texttt{repo} & GitHub repository. \\
    \texttt{file} & Primary bug location file. \\
    \texttt{line\_start/end} & Primary bug location line numbers. \\
    \texttt{title} & Short bug title. \\
    \texttt{description} & What is wrong, its trigger and consequence. \\
    \texttt{severity} & Impact, from \texttt{low} to \texttt{critical}. \\
    \texttt{suggested\_fix} & Short description of the fix. \\
    \texttt{code\_snippet} & Code around the bug location. \\
    \bottomrule
  \end{tabularx}
\end{table}

\subsection{Estimating the Total Bug Pool}
\label{app:bug-fixing-rate}

\paragraph{Population and Observation Window}
We fit the 17 repositories with at least 25 reference bugs, i.e., all except jsonschema, jose, and msgpack-javascript.
Over the reference window, January 2025 through June 2026, they hold 2,445 of the 2,457 reference bugs and 4,265 of the 4,296 bug-fixing commits of \cref{tab:benchmark-repos}.
We model when each reference bug is fixed, measuring time in bug-fixing commits rather than days: a repository's clock advances by one with every bug-fixing commit, including commits that fix only bugs introduced after $T_0$.

\begin{figure}[!t]
    \centering
    \includegraphics[width=\textwidth]{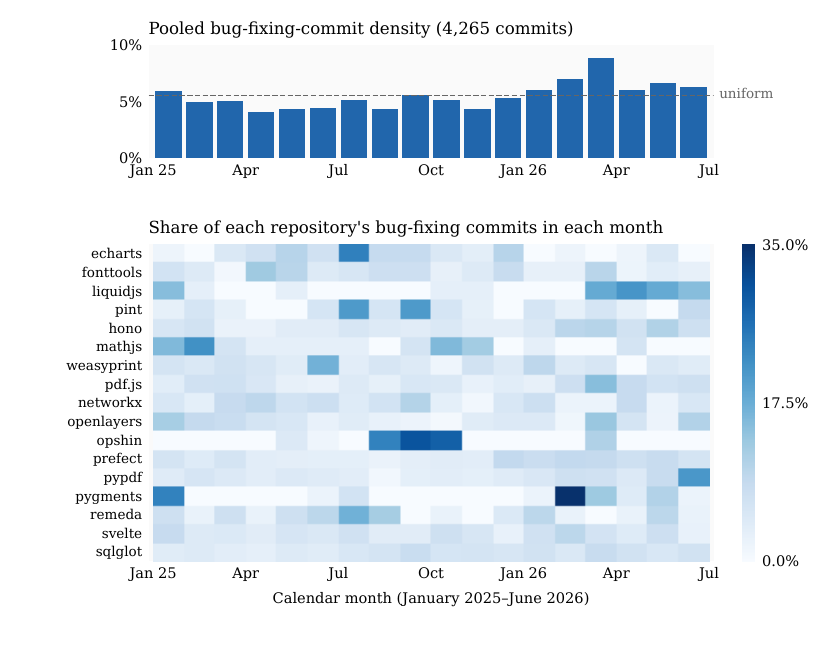}
    \caption{Calendar density of mined bug-fixing commits. Top: the share of all 4,265 commits in each month; the dashed line is uniform activity over 18 months. Bottom: monthly shares normalized separately within each repository (each row sums to 100\%). Fixing activity varies across repositories and over time.}
    \label{fig:bug-fixing-calendar}
\end{figure}

\paragraph{Bug-Fixing Commits as the Clock}
In calendar time, a falling fix rate can reflect either depletion of the bugs present at $T_0$ or fewer bug-fixing commits.
Fixing activity varies substantially (\cref{fig:bug-fixing-calendar}): pooled monthly counts range from 175 to 378 (a factor of 2.16), and individual repositories have quiet months and concentrated bursts.
Measuring time in bug-fixing commits factors out this varying activity.

\paragraph{Conditional Fix-Position Model}
Let $N_j$ be the number of bug-fixing commits in repository $j$'s window, $k_j$ its number of reference bugs, and $x_{ij}$ the position of bug $i$'s fix in that sequence.
Commits are ordered along the first-parent chain and assigned positions $0.5,1.5,\ldots,N_j-0.5$; bugs fixed by the same commit share its position.
For a distribution of fix positions with CDF $F_\theta$ and density $f_\theta$ (three candidate families, defined below), we only observe bugs fixed within the window, so the likelihood conditions on $x_{ij}<N_j$:
\begin{equation}
    p_\theta(x_{ij}\mid x_{ij}<N_j)
    =\frac{f_\theta(x_{ij})}{F_\theta(N_j)},
    \qquad
    \ell(\theta)=\sum_j\left[\sum_{i=1}^{k_j}\log f_\theta(x_{ij})-k_j\log F_\theta(N_j)\right].
    \label{eq:fix-position-likelihood}
\end{equation}
The parameters $\theta$ are shared across repositories.
Each repository's unknown number of $T_0$ bugs cancels under this conditioning, so the fit uses only how front-loaded the observed fixes are.
We use the continuous densities as an approximation on the half-integer commit positions.

The three families are parameterized as
\begin{align}
    F_{\mathrm{exp}}(x)&=1-\exp(-\beta x), & \beta&>0,\\
    F_{\mathrm{Wei}}(x)&=1-\exp\!\left[-(x/\lambda)^a\right], & a,\lambda&>0,\\
    F_{\mathrm{LN}}(x)&=\Phi\!\left(\frac{\log x-\mu}{\sigma}\right), & \sigma&>0,
\end{align}
where $\Phi$ is the standard normal CDF and $\log$ is the natural logarithm.
The exponential assumes a constant fixing hazard per remaining bug per bug-fixing commit.
The Weibull allows that hazard to decrease ($a<1$) or increase ($a>1$), and the lognormal allows a much longer tail.
Here $x$ and $\lambda$ are measured in bug-fixing commits and $\beta$ in inverse bug-fixing commits.
\Cref{tab:drain-parameters} reports the fitted parameters.

\begin{table}[htbp]
    \centering
    \footnotesize
    \caption{Fitted parameters of the three fix-position families. Full-window estimates use all 2,445 bugs; first-half estimates use 1,396.}
    \label{tab:drain-parameters}
    \begin{tabular}{llrr}
\toprule
Family & Parameter & Full window & First half \\
\midrule
Exponential & $\beta$ & 0.00112 & 0.00190 \\
Weibull & $a$ & 0.926 & 0.933 \\
 & $\lambda$ & 1,264 & 763 \\
Lognormal & $\mu$ & 15.6 & 15.5 \\
 & $\sigma$ & 3.77 & 3.82 \\
\bottomrule
\end{tabular}

\end{table}

\paragraph{First-Half Fit and Held-Out Diagnostic}
For repository $j$, let $c_j=\operatorname{round}(N_j/2)$.
We estimate parameters from positions $x_{ij}<c_j$ using the same likelihood with truncation at $c_j$.
The remaining positions are held out; no late positions enter this fit.
\Cref{fig:bug-fixing-holdout} pools repositories on the normalized position $u=x/N_j$, in 20 equal-width bins, and compares them with the predicted density
\begin{equation}
    q_\theta(u)=\frac{1}{K}\sum_j k_j\frac{N_j f_\theta(uN_j)}{F_\theta(N_j)},
    \qquad K=\sum_j k_j=2{,}445.
\end{equation}
Because the weights $k_j$ are the observed full-window counts, this tests when bugs are fixed, not how many there are.
The expected number of held-out bugs is $\sum_j k_j[1-F_\theta(c_j)/F_\theta(N_j)]$.
Against 1,049 observed late bugs, the estimates are 928 (exponential), 957 (Weibull), and 1,008 (lognormal), underpredictions of 11.6\%, 8.8\%, and 3.9\%.
No formal confidence band is shown: bugs fixed by the same commit can be dependent, and the family parameters are estimated from the data.

\begin{wrapfigure}[7]{r}{0.45\textwidth}
    \centering
    \vspace{-\baselineskip}
    \captionof{table}{Estimated number of $T_0$ bugs fixed within different horizons.}
    \label{tab:drain-horizons}
    \vspace{-3mm}
    \scalebox{0.9}{
    \begin{tabular}{lrrr}
        \toprule
          & 5 years & 10 years & $\infty$ \\
        \midrule
        Exponential & 5,195 & 7,197 & 13,030 \\
        Weibull & 5,326 & 7,471 & 15,089 \\
        Lognormal & 5,884 & 9,389 & 747,859 \\
        \bottomrule
    \end{tabular}
    }
    \vspace{-\baselineskip}
\end{wrapfigure}
\paragraph{Extrapolation and Limits}
For a horizon of $y$ years, we assume the observed fixing pace continues and set $H_j(y)=N_j(12y/18)$.
The estimated number of bugs present at $T_0$ and fixed by that horizon is $\sum_j k_j F_\theta(H_j(y))/F_\theta(N_j)$; the eventually-fixed estimate is $\sum_j k_j/F_\theta(N_j)$ (the $\infty$ column of \cref{tab:drain-horizons}).

The large disagreement in eventual totals reflects assumptions about the unobserved tail, despite relatively similar fits inside the window.
The conditional likelihood depends only on the curve's shape inside the window: a curve that ends with the window fits equally well and implies no further fixes, while a longer tail implies more.
The observed 2,445 bugs are therefore a lower bound, but the data give no finite upper bound on the number of bugs present at $T_0$.
Never-fixed bugs are invisible, and incomplete mining or a declining probability of mapping a later fix back to $T_0$ can distort the observed density.
The holdout tests the timing of detected fixes and cannot measure this missing population.

\FloatBarrier

\subsection{Bug Matching and Deduplication}
\label{app:dedup}

Bug matching (\cref{sec:evaluation}) has three uses: matching an agent's findings to the reference set, merging duplicate reference bugs in Step~\ref{item:dedup} of the benchmark construction, and merging repeated runs into one set of distinct findings.

\paragraph{Similarity Filter}
A pair goes to the judge if any of the six signals of \cref{tab:dedup-signals} reaches its threshold. We chose the thresholds to send as few pairs as possible to the judge while keeping all 217 duplicates it confirmed among 3,770 finding--reference pairs from ten repositories. Since one signal suffices, this is a weighted set-cover problem. On held-out pairs between agent findings and those of an independent bug-finding tool on all 20 repositories, the thresholds keep 472 of 474 duplicates.

\begin{table*}[ht]
    \centering
\caption{The six pairwise similarity signals that pre-filter finding pairs before the LLM duplicate judge.}
\label{tab:dedup-signals}
\small
\setlength{\tabcolsep}{4pt}
\renewcommand{\arraystretch}{1.15}
\begin{tabularx}{\textwidth}{l >{\raggedright\arraybackslash}X >{\raggedright\arraybackslash}X l r}
\toprule
Signal & Inputs & Tokens / model & Score & Threshold \\
\midrule
line overlap & \texttt{file}, \texttt{line\_start}, \texttt{line\_end} & line ranges in the same file & Jaccard & 0.02 \\
full-text embedding & \texttt{title}, \texttt{description}, first 1,000 characters of \texttt{code\_snippet} & \texttt{all-MiniLM-L6-v2} & cosine & 0.65 \\
description embedding & \texttt{title}, \texttt{description} & \texttt{all-MiniLM-L6-v2} & cosine & 0.57 \\
snippet tokens & \texttt{code\_snippet} & words$^a$ of $\geq 2$ characters & IDF-weighted Jaccard$^c$ & 0.60 \\
identifiers & \texttt{description}, \texttt{code\_snippet} & words$^a$ of $\geq 3$ characters & IDF-weighted Jaccard$^c$ & 0.10 \\
API calls & \texttt{code\_snippet} & called names$^b$ & IDF-weighted Jaccard$^c$ & 0.22 \\
\bottomrule
\end{tabularx}
\smallskip
\begin{minipage}{\textwidth}
\footnotesize
$^a$ The distinct matches of the regex \texttt{[a-z\_]\textbackslash w*} (a letter or underscore followed by letters, digits or underscores) in the lowercased text, minus a stopword list of Python keywords and common Python and TypeScript builtins.\par
$^b$ The distinct lowercased bare or dotted names directly followed by `\texttt{(}', with a leading \texttt{self.}, \texttt{cls.} or \texttt{this.} removed and common builtins such as \texttt{len} or \texttt{console.log} dropped; bare names need $\geq 5$ characters.\par
$^c$ The summed weight of the tokens two findings share, divided by that of all their tokens. A token's weight is $\log\frac{n+1}{\mathit{df}+1} + 1$, from an IDF (inverse document frequency) table built once per run over the $n$ findings on both sides of the comparison, $\mathit{df}$ of which contain the token.
\end{minipage}
\end{table*}

\paragraph{LLM Judge}
\geminiflash at medium thinking effort judges one pair per call and sees only the two reports, never the signal values (\cref{prompt:matcher-judge}). On 452 pairs from hono, \opus with access to the full repository agrees with the judge on 98.9\% of pairs ($\kappa=0.932$); on the five pairs where they disagree, a human sided with the judge.

\paragraph{Three Uses}
To score a run, every finding is paired with every reference bug, and matching is many-to-many: recall counts the reference bugs matched by at least one finding, precision the findings that match at least one reference bug. Step~\ref{item:dedup} pairs reference bugs that share a file and uses a stricter prompt that requires a shared root cause (\cref{prompt:record-dedup}). Confirmed pairs are merged as connected components, keeping the earliest fix, which removes 44 bugs. Repeated runs are merged one at a time: the findings of each new run are paired with the merged set and with each other, with a finding's suggested fix appended to its description, and confirmed pairs are again merged as connected components. Merging the ten \luna runs reduces 27,737 findings to 7,472 and judges 180,287 pairs for \$1,007.

\paragraph{Validation}
The corpus of \cref{sec:evaluation} holds a subset of findings from several earlier discovery runs on opshin, prefect, and weasyprint. Its pairs join two findings rather than a finding and a reference bug, but go through the same filter and prompt as scoring. The identifier threshold was partly tuned on these repositories. Because matches are not transitive, we score individual pairs rather than clusters. The first run judges all \dedupNumPairs{} pairs with the filter off: it confirms none of the \dedupNumDropped{} pairs the filter removes and reaches \dedupCensusPrecision{} precision and \dedupCensusRecall{} recall. Each later run judges only the filtered pairs and agrees with the first on at least \dedupAgreementMin{} of them ($\kappa\geq\dedupKappaMin{}$). All errors fall on three pairs where it is debatable whether one fix resolves both. The five runs cost \dedupCostUSD{}.

\section{Pass Planning}
\label{app:bug-finder}
\label{app:pass-planning}

A deterministic planner partitions each repository's audited files into the passes of \cref{sec:bug-finding}. It uses no LLM, only the static call graph, file sizes and the directory tree, and its parameters are the same for all repositories, so a snapshot always yields the same plan.

\paragraph{Audited Files}
The call graph covers every non-test source file of the snapshot. The planner skips 197 generated database migration files in prefect and audits the remaining 4,468 files.

\paragraph{File Score}
Each file $f$ receives the score
\[
s(f) = 3\,c(f) + 0.05\,l(f) + 0.5\,n(f),
\]
where $c(f)$ counts the distinct functions in other files that call into $f$ or that $f$ calls, $l(f)$ is the file's line count and $n(f)$ the number of functions it defines. Cross-file connectivity dominates: one call partner in another file weighs as much as 60 lines.

\paragraph{Individual Passes}
Files are ranked by score, with ties broken by path. A file gets a pass of its own if its score is at least 15\% of the repository's top score, but at most 15\% of a repository's audited files do so, and at least one always does. The score threshold decides in 11 repositories and the count limit in 9.

\paragraph{Directory Batches}
The remaining files are grouped by top-level directory. We estimate a group's size as the characters of its line-numbered source divided by 3.5. A group above 60k tokens is split by subdirectory, below the deepest directory its files share, and each part is split again while it stays over budget. Files in a single directory are sorted by line count and cut into two halves instead. A batch left with a single file becomes an individual pass. Batches hold a median of 7 files, from 2 to 246. In total, 331 files get an individual pass by score, 28 as single-file batches, and 381 directory batches hold the other 4,109, for 740 passes.

\paragraph{Call-Graph Context}
An individual pass includes snippets of at most 40 callers and 40 callees in other files, and a batch at most 25 of each. Calls between files of the same batch are left out, since both files are already in the prompt. When the limit binds, functions referenced by more of the pass's files come first, then longer snippets.

\section{Additional Results}
\label{app:experiments}

\subsection{Results on the Five-Repository Subset}
\label{app:lane-subset}

To limit cost, \opus audits only five repositories, covering both Python and JavaScript/TypeScript: hono, networkx, pdf.js, pypdf and weasyprint, with 755 reference bugs. \Cref{tab:lane_subset} restricts every configuration to these five repositories, so that \opus can be compared with the others like for like. \opus reaches 21.2\% recall at 10.0\% precision, level with \sol (medium) and more than twice the recall of \sonnet. The other configurations rank roughly as they do on all repositories.

\begin{table}[ht]
\centering
\small
\caption{Recall, precision and F1 against the reference set on the five repositories \opus audited (755 reference bugs), for every configuration of \cref{tab:lane_summary}, in its order. Cells for \musespark and \luna (medium) are means over 5 and 10 runs, respectively, and cost covers the five repositories.}
\label{tab:lane_subset}
\sisetup{detect-weight}
\begin{tabular}{llS[table-format=2.1]S[table-format=2.1]S[table-format=2.1]rr}
\toprule
{Model} & {Effort} & {Recall (\%)} & {Precision (\%)} & {F1 (\%)} & {Findings} & {Cost} \\
\midrule
DeepSeek V4.1 Flash & high & 17.5 & \bfseries 11.7 & \bfseries 14.0 & 1,172 & \$15 \\
GLM 5.3 Flash & high & 18.7 & 10.4 & 13.3 & 1,428 & \$25 \\
GPT-5.4 & medium & 12.5 & 10.3 & 11.3 & 942 & \$138 \\
GPT-5.6 Terra & medium & 10.6 & 10.9 & 10.7 & 761 & \$46 \\
Gemini 3.5 Flash & medium & 10.7 & 9.8 & 10.2 & 867 & \$104 \\
Muse Spark 1.3 & high & 20.6 & 8.0 & 11.6 & 1,921 & \$35 \\
Claude Sonnet 5 & medium & 9.5 & 9.7 & 9.6 & 762 & \$96 \\
Claude Opus 5 & medium & 21.2 & 10.0 & 13.6 & 1,707 & \$353 \\
\cmidrule(lr){1-2}
\multirow{3}{*}{GPT-5.6 Sol} & high & \bfseries 24.0 & 9.6 & 13.7 & \textbf{1,976} & \$249 \\
 & medium & 20.9 & 10.0 & 13.6 & 1,694 & \$130 \\
 & low & 14.3 & 11.1 & 12.5 & 970 & \$61 \\
\cmidrule(lr){1-2}
\multirow{4}{*}{GPT-5.6 Luna} & xhigh & 20.1 & 10.5 & 13.8 & 1,575 & \$27 \\
 & high & 15.5 & 10.1 & 12.2 & 1,211 & \$11 \\
 & medium & 10.0 & 11.0 & 10.5 & 691 & \$3 \\
 & low & 7.7 & \bfseries 11.7 & 9.3 & 488 & \textbf{\$2} \\
\bottomrule
\end{tabular}

\end{table}

\subsection{Recall and Precision per Repository}
\label{app:repo-recall}

\Cref{tab:lane_recall_by_repo,tab:lane_precision_by_repo} break the recall and precision of \cref{tab:lane_summary} down by repository, with the repositories grouped by language as in \cref{tab:benchmark-repos}.

\begin{table}[ht]
\centering
\scriptsize
\caption{Recall (\%) against the reference set per repository, for every configuration of \cref{tab:lane_summary} and \opus. The shaded italic rows pool the bugs of one language. For configurations run several times, each cell is the mean over the runs. The shaded Mean column averages the configurations of \cref{tab:lane_summary}, counting each repeated configuration once and excluding \opus. Bold marks each row's best configuration.}
\label{tab:lane_recall_by_repo}
\setlength{\tabcolsep}{1.9pt}
\sisetup{detect-weight}
\begin{tabular}{lrS[table-format=2.1]S[table-format=2.1]S[table-format=2.1]S[table-format=2.1]S[table-format=2.1]S[table-format=2.1]S[table-format=2.1]S[table-format=2.1]S[table-format=2.1]S[table-format=2.1]S[table-format=2.1]S[table-format=2.1]S[table-format=2.1]S[table-format=2.1]S[table-format=2.1]>{\columncolor{black!6}}S[table-format=2.1]}
\toprule
{Repository} & {Bugs} & {\rotatebox{90}{DeepSeek V4.1 Flash (high)}} & {\rotatebox{90}{GLM 5.3 Flash (high)}} & {\rotatebox{90}{GPT-5.4 (medium)}} & {\rotatebox{90}{GPT-5.6 Terra (medium)}} & {\rotatebox{90}{Gemini 3.5 Flash (medium)}} & {\rotatebox{90}{Muse Spark 1.3 (high, 5 runs)}} & {\rotatebox{90}{Claude Sonnet 5 (medium)}} & {\rotatebox{90}{GPT-5.6 Sol (high)}} & {\rotatebox{90}{GPT-5.6 Sol (medium)}} & {\rotatebox{90}{GPT-5.6 Sol (low)}} & {\rotatebox{90}{GPT-5.6 Luna (xhigh)}} & {\rotatebox{90}{GPT-5.6 Luna (high)}} & {\rotatebox{90}{GPT-5.6 Luna (medium, 10 runs)}} & {\rotatebox{90}{GPT-5.6 Luna (low)}} & {\rotatebox{90}{Claude Opus 5 (medium)}} & {\rotatebox{90}{\emph{Mean}}} \\
\midrule
\texttt{sqlglot} & 512 & 3.9 & 4.5 & 2.0 & 1.8 & 1.6 & 3.2 & 0.8 & \bfseries 8.2 & 3.3 & 2.7 & 5.5 & 3.1 & 1.1 & 0.8 & {--} & 3.0 \\
\texttt{prefect} & 429 & 12.6 & 13.5 & 10.5 & 6.8 & 11.4 & 15.2 & 8.2 & \bfseries 17.9 & 13.5 & 8.9 & 16.3 & 12.4 & 6.4 & 5.1 & {--} & 11.3 \\
\texttt{pypdf} & 165 & 17.0 & 22.4 & 9.1 & 8.5 & 11.5 & 27.8 & 12.7 & 24.2 & 23.6 & 16.4 & 21.2 & 14.5 & 11.4 & 10.3 & \bfseries 30.9 & 16.5 \\
\texttt{weasyprint} & 115 & 10.4 & 10.4 & 7.0 & 6.1 & 6.1 & 9.7 & 4.3 & \bfseries 16.5 & 13.9 & 6.1 & 15.7 & 7.0 & 4.3 & 1.7 & 9.6 & 8.5 \\
\texttt{fonttools} & 113 & 28.3 & 27.4 & 19.5 & 20.4 & 17.7 & 29.0 & 15.9 & \bfseries 31.9 & 30.1 & 23.9 & 28.3 & 22.1 & 14.6 & 8.8 & {--} & 22.7 \\
\texttt{networkx} & 84 & 28.6 & 32.1 & 19.0 & 15.5 & 20.2 & 27.6 & 13.1 & \bfseries 35.7 & 32.1 & 26.2 & 31.0 & 23.8 & 13.1 & 9.5 & 32.1 & 23.4 \\
\texttt{opshin} & 51 & 27.5 & 27.5 & 15.7 & 13.7 & 9.8 & 24.3 & 11.8 & \bfseries 35.3 & 21.6 & 13.7 & 33.3 & 23.5 & 10.2 & 7.8 & {--} & 19.7 \\
\texttt{pygments} & 41 & \bfseries 19.5 & 12.2 & 7.3 & 7.3 & 7.3 & 14.1 & 12.2 & 17.1 & 17.1 & 12.2 & 14.6 & 12.2 & 8.0 & 4.9 & {--} & 11.9 \\
\texttt{pint} & 28 & \bfseries 10.7 & 7.1 & 7.1 & 7.1 & 7.1 & 10.0 & 7.1 & \bfseries 10.7 & \bfseries 10.7 & \bfseries 10.7 & 3.6 & 3.6 & 0.4 & 0.0 & {--} & 6.9 \\
\texttt{jsonschema} & 6 & \bfseries 83.3 & 66.7 & 66.7 & 50.0 & 50.0 & 33.3 & 50.0 & 66.7 & 66.7 & 66.7 & 50.0 & 16.7 & 33.3 & 33.3 & {--} & 52.4 \\
\rowcolor{black!6}\emph{Python} & 1,544 & 13.0 & 13.8 & 8.6 & 7.1 & 8.6 & 14.1 & 7.1 & \bfseries 17.9 & 14.0 & 10.0 & 15.3 & 10.7 & 6.1 & 4.6 & {--} & 10.8 \\
\midrule
\texttt{hono} & 144 & 34.0 & 29.9 & 26.4 & 22.9 & 19.4 & 39.0 & 18.1 & \bfseries 48.6 & 39.6 & 25.7 & 38.2 & 37.5 & 21.4 & 15.3 & 36.8 & 29.7 \\
\texttt{echarts} & 46 & 17.4 & \bfseries 28.3 & 6.5 & 13.0 & 13.0 & 10.0 & 8.7 & \bfseries 28.3 & 21.7 & 15.2 & 19.6 & 17.4 & 7.2 & 2.2 & {--} & 14.9 \\
\texttt{liquidjs} & 32 & 53.1 & 37.5 & 37.5 & 40.6 & 25.0 & 46.2 & 28.1 & 53.1 & 50.0 & 37.5 & \bfseries 65.6 & 46.9 & 33.4 & 18.8 & {--} & 41.0 \\
\texttt{remeda} & 30 & 6.7 & 3.3 & 10.0 & 3.3 & 3.3 & \bfseries 10.7 & 3.3 & 10.0 & 10.0 & 10.0 & 10.0 & 3.3 & 3.3 & 3.3 & {--} & 6.5 \\
\texttt{jose} & 3 & 0.0 & 0.0 & 0.0 & \bfseries 33.3 & 0.0 & 13.3 & \bfseries 33.3 & \bfseries 33.3 & 0.0 & 0.0 & \bfseries 33.3 & 0.0 & 3.3 & 0.0 & {--} & 10.7 \\
\texttt{msgpack-javascript} & 3 & \bfseries 33.3 & \bfseries 33.3 & 0.0 & \bfseries 33.3 & \bfseries 33.3 & \bfseries 33.3 & \bfseries 33.3 & \bfseries 33.3 & \bfseries 33.3 & 0.0 & 0.0 & \bfseries 33.3 & 3.3 & 0.0 & {--} & 21.7 \\
\rowcolor{black!6}\emph{TypeScript} & 258 & 29.8 & 27.1 & 21.7 & 21.3 & 17.1 & 31.1 & 16.3 & \bfseries 40.7 & 33.7 & 22.9 & 34.5 & 30.6 & 17.8 & 11.6 & {--} & 25.4 \\
\midrule
\texttt{svelte} & 248 & 16.5 & 19.8 & 12.5 & 10.5 & 12.5 & 16.6 & 6.9 & \bfseries 23.4 & 22.6 & 11.7 & 17.3 & 14.1 & 8.4 & 3.6 & {--} & 14.0 \\
\texttt{pdf.js} & 247 & 7.7 & \bfseries 8.9 & 6.9 & 5.3 & 4.0 & 7.7 & 3.6 & \bfseries 8.9 & 7.7 & 6.1 & 7.3 & 4.5 & 4.0 & 3.6 & 7.3 & 6.2 \\
\texttt{openlayers} & 121 & 19.8 & 20.7 & 20.7 & 12.4 & 14.0 & 18.7 & 9.9 & \bfseries 32.2 & 19.8 & 14.0 & 22.3 & 16.5 & 9.9 & 5.8 & {--} & 16.9 \\
\texttt{mathjs} & 39 & 30.8 & 23.1 & 20.5 & 10.3 & 12.8 & 26.2 & 10.3 & 30.8 & \bfseries 33.3 & 12.8 & 28.2 & 25.6 & 10.3 & 7.7 & {--} & 20.2 \\
\rowcolor{black!6}\emph{JavaScript} & 655 & 14.7 & 16.0 & 12.4 & 8.9 & 9.6 & 14.2 & 6.4 & \bfseries 20.0 & 17.1 & 10.1 & 15.1 & 11.6 & 7.2 & 4.3 & {--} & 12.0 \\
\midrule
\rowcolor{black!6}\emph{All} & 2,457 & 15.2 & 15.8 & 11.0 & 9.1 & 9.8 & 15.9 & 7.9 & \bfseries 20.8 & 16.9 & 11.4 & 17.3 & 13.0 & 7.6 & 5.3 & {--} & 12.6 \\
\bottomrule
\end{tabular}

\end{table}

\paragraph{Repository Variation}
Comparing recall across repositories (see \cref{tab:lane_recall_by_repo}), we observe large differences that bug count does not explain (Spearman $\rho=\benchRepoBugsRecallRho{}$ between a repository's reference bugs and its mean recall). Averaged over agents, recall per repository ranges from about 3\% on sqlglot to over 40\% on liquidjs and jsonschema, and the configurations rank the repositories consistently (Kendall's $W=\benchRepoRecallW{}$). Notably, recall is substantially higher for TypeScript (25.4\%) than for Python (10.8\%) or JavaScript (12.0\%). Precision also varies widely, from 0.6\% on jose to 22.9\% on pypdf (see \cref{tab:lane_precision_by_repo}), with an even more consistent ranking ($W=\benchRepoPrecisionW{}$).

\begin{table}[ht]
\centering
\scriptsize
\caption{Precision (\%) against the reference set per repository, for every configuration of \cref{tab:lane_summary} and \opus. The shaded italic rows pool the findings of one language. For configurations run several times, each cell is the mean over the runs. The shaded Mean column averages the configurations of \cref{tab:lane_summary}, counting each repeated configuration once and excluding \opus. Bold marks each row's best configuration.}
\label{tab:lane_precision_by_repo}
\setlength{\tabcolsep}{2pt}
\sisetup{detect-weight}
\begin{tabular}{lS[table-format=2.1]S[table-format=2.1]S[table-format=2.1]S[table-format=2.1]S[table-format=2.1]S[table-format=2.1]S[table-format=2.1]S[table-format=2.1]S[table-format=2.1]S[table-format=2.1]S[table-format=2.1]S[table-format=2.1]S[table-format=2.1]S[table-format=2.1]S[table-format=2.1]>{\columncolor{black!6}}S[table-format=2.1]}
\toprule
{Repository} & {\rotatebox{90}{DeepSeek V4.1 Flash (high)}} & {\rotatebox{90}{GLM 5.3 Flash (high)}} & {\rotatebox{90}{GPT-5.4 (medium)}} & {\rotatebox{90}{GPT-5.6 Terra (medium)}} & {\rotatebox{90}{Gemini 3.5 Flash (medium)}} & {\rotatebox{90}{Muse Spark 1.3 (high, 5 runs)}} & {\rotatebox{90}{Claude Sonnet 5 (medium)}} & {\rotatebox{90}{GPT-5.6 Sol (high)}} & {\rotatebox{90}{GPT-5.6 Sol (medium)}} & {\rotatebox{90}{GPT-5.6 Sol (low)}} & {\rotatebox{90}{GPT-5.6 Luna (xhigh)}} & {\rotatebox{90}{GPT-5.6 Luna (high)}} & {\rotatebox{90}{GPT-5.6 Luna (medium, 10 runs)}} & {\rotatebox{90}{GPT-5.6 Luna (low)}} & {\rotatebox{90}{Claude Opus 5 (medium)}} & {\rotatebox{90}{\emph{Mean}}} \\
\midrule
\texttt{sqlglot} & \bfseries 15.3 & 12.9 & 8.3 & 11.7 & 9.2 & 7.2 & 4.7 & 13.1 & 7.9 & 14.6 & 11.9 & 13.4 & 8.6 & 8.9 & {--} & 10.6 \\
\texttt{prefect} & 10.2 & 8.5 & \bfseries 10.3 & 8.2 & 10.1 & 6.4 & 7.7 & 9.4 & 7.8 & 9.1 & 9.2 & 8.9 & 8.2 & 8.7 & {--} & 8.8 \\
\texttt{pypdf} & 25.2 & 23.5 & 18.8 & 20.3 & 22.1 & 21.4 & 24.4 & 18.1 & 22.2 & 26.2 & 19.2 & 18.6 & 25.7 & \bfseries 34.7 & 24.8 & 22.9 \\
\texttt{weasyprint} & 6.3 & 5.8 & 6.7 & 7.4 & 6.9 & 4.7 & 4.6 & 5.8 & 6.4 & 6.0 & \bfseries 8.7 & 6.6 & 5.8 & 3.6 & 4.6 & 6.1 \\
\texttt{fonttools} & 7.1 & 5.6 & 6.1 & 7.6 & 5.6 & 3.9 & 4.9 & 5.0 & 6.0 & \bfseries 7.8 & 5.4 & 5.6 & 6.2 & 4.5 & {--} & 5.8 \\
\texttt{networkx} & \bfseries 6.9 & 6.0 & 4.9 & 4.6 & 6.4 & 3.8 & 5.4 & 4.8 & 4.8 & 6.5 & 5.3 & 5.0 & 4.7 & 4.6 & 5.3 & 5.3 \\
\texttt{opshin} & 15.2 & 13.4 & 11.8 & 17.1 & 8.9 & 10.5 & 8.7 & 15.5 & 12.0 & 10.9 & \bfseries 19.4 & 16.9 & 13.0 & 11.8 & {--} & 13.2 \\
\texttt{pygments} & 1.5 & 0.9 & 1.2 & 1.1 & 1.1 & 0.9 & \bfseries 2.2 & 0.8 & 1.0 & 1.5 & 1.0 & 1.1 & 1.4 & 1.4 & {--} & 1.2 \\
\texttt{pint} & 2.0 & 1.2 & 2.2 & 2.6 & 2.0 & 1.3 & 2.3 & 1.4 & 2.1 & \bfseries 3.5 & 0.6 & 0.9 & 0.2 & 0.0 & {--} & 1.6 \\
\texttt{jsonschema} & \bfseries 20.0 & 14.3 & 12.5 & 15.8 & 17.6 & 4.9 & 15.8 & 7.5 & 9.4 & 19.0 & 8.3 & 4.8 & 13.6 & 18.2 & {--} & 13.0 \\
\rowcolor{black!6}\emph{Python} & 7.8 & 6.9 & 6.8 & 7.0 & 7.3 & 5.1 & 6.5 & 6.5 & 6.2 & \bfseries 8.0 & 7.0 & 6.7 & 6.7 & 6.9 & {--} & 6.8 \\
\midrule
\texttt{hono} & 21.1 & 18.3 & 21.0 & \bfseries 23.8 & 18.1 & 16.9 & 17.7 & 21.1 & 21.5 & 20.2 & 20.3 & 22.3 & 22.9 & 23.4 & 17.0 & 20.6 \\
\texttt{echarts} & 1.8 & \bfseries 2.2 & 0.9 & 1.9 & 1.3 & 0.5 & 1.1 & 1.5 & 1.6 & 1.7 & 1.7 & 1.6 & 1.1 & 0.5 & {--} & 1.4 \\
\texttt{liquidjs} & 13.5 & 12.8 & 11.4 & \bfseries 19.6 & 10.9 & 8.3 & 9.3 & 11.8 & 13.3 & 15.0 & 15.9 & 13.9 & 14.7 & 13.3 & {--} & 13.1 \\
\texttt{remeda} & 3.1 & 1.4 & 3.3 & 2.0 & 3.0 & 3.0 & 3.0 & 2.0 & 2.3 & 3.3 & 1.9 & 1.3 & 2.5 & \bfseries 3.4 & {--} & 2.5 \\
\texttt{jose} & 0.0 & 0.0 & 0.0 & \bfseries 2.7 & 0.0 & 0.3 & 2.0 & 2.0 & 0.0 & 0.0 & 1.5 & 0.0 & 0.2 & 0.0 & {--} & 0.6 \\
\texttt{msgpack-javascript} & 4.2 & 3.4 & 0.0 & 4.8 & \bfseries 8.7 & 2.9 & 6.2 & 2.5 & 3.2 & 0.0 & 0.0 & 3.7 & 0.7 & 0.0 & {--} & 2.9 \\
\rowcolor{black!6}\emph{TypeScript} & 8.7 & 6.7 & 6.4 & \bfseries 9.4 & 6.2 & 5.0 & 6.0 & 7.2 & 7.8 & 7.6 & 8.1 & 8.2 & 8.0 & 7.1 & {--} & 7.3 \\
\midrule
\texttt{svelte} & 13.9 & 14.6 & 13.2 & 15.4 & 12.2 & 8.3 & 8.9 & 11.7 & 15.0 & 15.0 & 12.0 & 12.8 & \bfseries 15.8 & 10.7 & {--} & 12.8 \\
\texttt{pdf.js} & 7.7 & 7.4 & 8.0 & \bfseries 8.1 & 3.9 & 3.7 & 4.7 & 5.9 & 5.3 & 6.6 & 5.7 & 4.2 & 5.9 & 6.9 & 5.2 & 6.0 \\
\texttt{openlayers} & \bfseries 7.4 & 5.3 & 6.5 & 5.5 & 4.1 & 3.0 & 3.2 & 6.2 & 4.4 & 4.7 & 5.4 & 5.0 & 5.3 & 4.3 & {--} & 5.0 \\
\texttt{mathjs} & \bfseries 4.1 & 3.7 & 3.0 & 2.1 & 1.9 & 2.4 & 2.4 & 3.3 & 3.5 & 2.4 & 3.6 & 3.6 & 2.4 & 2.1 & {--} & 2.9 \\
\rowcolor{black!6}\emph{JavaScript} & \bfseries 8.2 & 7.5 & 7.4 & 7.2 & 5.3 & 4.1 & 4.5 & 6.7 & 6.8 & 6.5 & 6.6 & 6.2 & 6.6 & 5.2 & {--} & 6.3 \\
\midrule
\rowcolor{black!6}\emph{All} & \bfseries 8.1 & 7.0 & 6.9 & 7.6 & 6.4 & 4.8 & 5.8 & 6.7 & 6.7 & 7.5 & 7.1 & 6.9 & 7.0 & 6.5 & {--} & 6.8 \\
\bottomrule
\end{tabular}

\end{table}

\subsection{Validity of Findings}
\label{app:validity}

\Cref{tab:validity-lanes} gives the panel's verdicts behind each point of \cref{fig:validity-precision}. The findings are sampled in proportion to each run's findings per repository and judged under opaque identifiers; validating them cost \validityCost{} (\validityCostPerFinding{} per finding). \Cref{tab:validity-confusion} compares one author's blind verdicts with the panel's on the \validityHumanGraded{} findings of the validity sample graded by both (\cref{sec:different-bugs}). The agreement and $\kappa$ reported there leave out the findings the author was unsure about and those the panel left inconclusive (Abstained in \cref{tab:validity-lanes,tab:validity-confusion}) because both of its tiers split.

\begin{table}[ht]
\centering
\small
\caption{The panel's verdicts on the \validitySample{} sampled findings of one run per configuration. Validity is the confirmed share of the decided verdicts, with its 95\% Clopper--Pearson interval. For \luna (medium) and \musespark, the run is the first of their repeats. Bold marks the most confirmed and fewest rejected findings and the highest validity.}
\label{tab:validity-lanes}
\begin{tabular}{llrrrrr}
\toprule
Model & Effort & Confirmed & Rejected & Abstained & Validity (\%) & 95\% CI \\
\midrule
GPT-5.4 & medium & \textbf{94} & \textbf{5} & 1 & \textbf{95} & [89, 98] \\
GPT-5.6 Sol & medium & 92 & \textbf{5} & 3 & \textbf{95} & [88, 98] \\
DeepSeek V4.1 Flash & high & 93 & 6 & 1 & 94 & [87, 98] \\
GPT-5.6 Terra & medium & 89 & 7 & 4 & 93 & [86, 97] \\
GPT-5.6 Luna & medium & 85 & 10 & 5 & 89 & [81, 95] \\
GLM 5.3 Flash & high & 84 & 10 & 6 & 89 & [81, 95] \\
Gemini 3.5 Flash & medium & 67 & 25 & 8 & 73 & [63, 82] \\
Muse Spark 1.3 & high & 65 & 27 & 8 & 71 & [60, 80] \\
Claude Sonnet 5 & medium & 62 & 35 & 3 & 64 & [54, 73] \\
\bottomrule
\end{tabular}

\end{table}

\begin{table}[ht]
\centering
\small
\caption{The author's blind verdict (rows) against the panel's (columns) on the \validityHumanGraded{} findings of the validity sample graded by both.}
\label{tab:validity-confusion}
\begin{tabular}{lrrrr}
\toprule
 & \multicolumn{3}{c}{Panel} & \\
\cmidrule(lr){2-4}
Author & Confirmed & Rejected & Abstained & Total \\
\midrule
Confirmed & 103 & 0 & 0 & 103 \\
Rejected & 9 & 19 & 6 & 34 \\
Unsure & 35 & 5 & 3 & 43 \\
\midrule
Total & 147 & 24 & 9 & 180 \\
\bottomrule
\end{tabular}

\end{table}

\subsection{Contamination}
\label{app:fix-rank-test}
We test whether agents preferentially match bugs fixed earlier or later in the reference window. Within each repository, we order the reference bugs by fix time and normalize their ranks to $[0,1]$. Under the null hypothesis, the $m_j$ matched bugs of repository $j$ are a uniformly random subset of its reference bugs. With $F_j(r)$ the empirical rank CDF of repository $j$ and $M=\sum_j m_j$, the expected CDF of matched ranks is $F_0(r)=M^{-1}\sum_j m_j F_j(r)$, and we measure the deviation with the Kolmogorov--Smirnov statistic~\citep{hodges1958smirnov}
\begin{equation}
    D=\sup_r\left|\widehat F_{\mathrm{matched}}(r)-F_0(r)\right|,
\end{equation}
where $\widehat F_{\mathrm{matched}}$ is the empirical CDF of matched ranks. We estimate $p$ from 100,000 permutations of the match labels within repositories, $p=(b+1)/(100{,}000+1)$, where $b$ counts the permutations with a $D$ at least as large as observed~\citep{phipson2010permutation}, and apply Holm's correction across all runs~\citep{holm1979multiple}.

Only \deepseekflash has an unadjusted $p<0.05$ ($p=0.008$, $D=0.076$; \cref{tab:lane-fix-rank}), from an excess of matches among earlier-fixed bugs. After Holm correction its $p$ is 0.204 and all others are 1, so we find no evidence that agents favor earlier- or later-fixed bugs.

\begin{table}[ht]
\centering
\small
\caption{Fix-rank permutation tests for every run, grouped as in \cref{tab:lane_summary}; Run numbers the repeated runs of \luna (medium) and \musespark. Column $p$ is unadjusted; $p_{\mathrm{Holm}}$ adjusts for all 27 tests. Bold denotes $p<0.05$ in each column; no result remains significant after Holm correction.}
\label{tab:lane-fix-rank}
\begin{tabular}{llrrrrr}
\toprule
Model & Effort & Run & Matched & $D$ & $p$ & $p_{\mathrm{Holm}}$ \\
\midrule
DeepSeek V4.1 Flash & high &  & 373 & 0.076 & \textbf{0.008} & 0.204 \\
GLM 5.3 Flash & high &  & 388 & 0.042 & 0.329 & 1.000 \\
GPT-5.4 & medium &  & 270 & 0.039 & 0.692 & 1.000 \\
GPT-5.6 Terra & medium &  & 223 & 0.044 & 0.677 & 1.000 \\
Gemini 3.5 Flash & medium &  & 240 & 0.067 & 0.157 & 1.000 \\
\cmidrule(lr){1-2}
\multirow{5}{*}{Muse Spark 1.3} & high & 1 & 387 & 0.044 & 0.282 & 1.000 \\
 &  & 2 & 385 & 0.032 & 0.668 & 1.000 \\
 &  & 3 & 373 & 0.029 & 0.784 & 1.000 \\
 &  & 4 & 391 & 0.037 & 0.466 & 1.000 \\
 &  & 5 & 417 & 0.046 & 0.181 & 1.000 \\
\cmidrule(lr){1-2}
Claude Sonnet 5 & medium &  & 194 & 0.059 & 0.409 & 1.000 \\
\cmidrule(lr){1-2}
\multirow{3}{*}{GPT-5.6 Sol} & high &  & 512 & 0.034 & 0.359 & 1.000 \\
 & medium &  & 415 & 0.036 & 0.476 & 1.000 \\
 & low &  & 279 & 0.052 & 0.309 & 1.000 \\
\cmidrule(lr){1-2}
\multirow{13}{*}{GPT-5.6 Luna} & xhigh &  & 424 & 0.038 & 0.385 & 1.000 \\
 & high &  & 320 & 0.030 & 0.856 & 1.000 \\
 & medium & 1 & 181 & 0.057 & 0.489 & 1.000 \\
 &  & 2 & 200 & 0.055 & 0.458 & 1.000 \\
 &  & 3 & 209 & 0.047 & 0.631 & 1.000 \\
 &  & 4 & 189 & 0.050 & 0.637 & 1.000 \\
 &  & 5 & 193 & 0.047 & 0.697 & 1.000 \\
 &  & 6 & 179 & 0.059 & 0.444 & 1.000 \\
 &  & 7 & 183 & 0.050 & 0.649 & 1.000 \\
 &  & 8 & 168 & 0.075 & 0.221 & 1.000 \\
 &  & 9 & 195 & 0.033 & 0.956 & 1.000 \\
 &  & 10 & 181 & 0.049 & 0.686 & 1.000 \\
 & low &  & 129 & 0.064 & 0.584 & 1.000 \\
\bottomrule
\end{tabular}

\end{table}

The fix-rank test can only reveal contamination for a model whose training data contains some but not all reference fixes, i.e., whose knowledge cut-off lies inside the reference window. \Cref{tab:knowledge-cutoffs} lists the cut-offs the vendors publish. All seven published cut-offs lie inside the window, ranging from January 2025 for \geminiflash, the window's first month, to May 2026 for \opus. \glmflash, \musespark and \deepseekflash have no published cut-off and were released after $T_1$, so their training data may cover the entire window.

\begin{table}[ht]
\centering
\small
\caption{Knowledge cut-offs of the evaluated models as published by their vendors. For the three models without a published cut-off, we give their release date instead.}
\label{tab:knowledge-cutoffs}
\begin{tabular}{lll}
\toprule
{Model} & {Knowledge cut-off} & {Source} \\
\midrule
\geminiflash & January 2025 & \citet{google2026gemini35flash} \\
\gptfivefour & 31 August 2025 & \citet{openai2026gpt54} \\
\sonnet & January 2026 & \citet{anthropic2026sonnet5} \\
\luna, Terra, Sol & 16 February 2026 & \citet{openai2026gpt56luna,openai2026gpt56sol,openai2026gpt56terra} \\
\opus & May 2026 & \citet{anthropic2026opus5} \\
\midrule
\glmflash & not published (released 26 August 2026) & \citet{zai2026releasenotes} \\
\musespark & not published (released 2 September 2026) & \citet{meta2026musespark13} \\
\deepseekflash & not published (released 10 September 2026) & \citet{deepseek2026changelog} \\
\bottomrule
\end{tabular}
\end{table}

\FloatBarrier

\section{Bug-Type Taxonomy}
\label{app:taxonomy}

\subsection{Construction}
\label{app:taxonomy-construction}

We induce the bug types of \cref{sec:different-bugs} from a corpus of \taxCorpus{} bugs: the \taxCorpusHuman{} human-fixed bugs of the reference set (13 of the \benchNumTargets{} merge with another through a shared agent finding) and \taxCorpusAgentFound{} findings of an earlier \gptfivefour audit of the same snapshots that every LLM validator confirmed, \taxCorpusBoth{} of which match a human-fixed bug and are counted once. Drawing on both sources keeps the categories from fitting only one of the two distributions we compare. \opus proposes \taxInducedCategories{} categories, within a requested range of \taxMinCategories{}--\taxMaxCategories{}, that classify bugs by root cause rather than symptom, from the titles and descriptions of \taxInductionSample{} bugs sampled across repositories and sources (\cref{prompt:taxonomy-induce}). \sonnet then assigns every corpus bug one category and one trigger level, in batches of \taxBatchSize{} bugs identified only by opaque keys (\cref{prompt:taxonomy-attribute}). Finally, \opus merges the categories holding fewer than \taxMergeFloor{} bugs into a neighbor unless they are conceptually irreducible (\cref{prompt:taxonomy-merge}), leaving \taxCategories{}. The findings of every configuration are assigned with the same prompt against these frozen categories.

\subsection{Categories}
\label{app:taxonomy-categories}

\Cref{tab:bug-type-definitions} lists the \taxCategories{} categories, grouped into six families that \opus drew from the names and definitions alone. In the prompts, each definition also names the neighboring category a borderline bug belongs to instead; we omit these clauses here.

\begin{table}[!htbp]
\centering
\caption{The \taxCategories{} bug types and their definitions, grouped by family (continued in \cref{tab:bug-type-definitions-2}).}
\label{tab:bug-type-definitions}
\footnotesize
\begin{tabular}{@{}>{\raggedright\arraybackslash}p{0.22\linewidth}p{0.75\linewidth}@{}}
\toprule
Category & Definition \\
\midrule
\multicolumn{2}{@{}l}{\emph{Missing case coverage}} \\
Missing branch for a supported input variant & Inline conditional logic handles some supported kinds, syntaxes or shapes of input but omits another equally supported variant, so it is processed by the wrong path. \\
Missing rejection of invalid input & Malformed, out-of-range, duplicated or otherwise unsupported input is silently accepted and processed instead of being validated and refused at the boundary. \\
Unhandled degenerate value & Code fails to special case a boundary value that is legal input, such as an empty collection, zero, a single element, or an absent or null field, and then indexes, divides or dereferences it. \\
Unimplemented advertised feature & A documented option, method, mode or format is exposed but its behaviour is absent or a stub, so the request is silently ignored or handled by an unrelated path. \\
Missing dispatch or lookup table entry & A registry, mapping, keyword list, transform table or alias set lacks an entry for a supported case, so lookup falls through to a generic or absent handler. \\
Truthiness test instead of presence test & A falsy but meaningful value such as zero, an empty string, false or an empty collection is treated as absent because the code tests truthiness rather than explicit nullness or membership. \\
Missing initialisation or default & A variable, field or option is read on a path where it was never assigned, or a required default is not supplied when constructing an object or record. \\
\addlinespace
\multicolumn{2}{@{}l}{\emph{Incorrect expressions and computation}} \\
Wrong arithmetic or comparison expression & A formula, offset, sign, bound or relational operator is written incorrectly so the computed result or decision is wrong, including off-by-one indexes and inverted conditions. \\
Incorrect regular expression pattern & A pattern is too narrow, too broad, missing an escape, or ambiguous enough to backtrack pathologically, so it fails to match or over-matches legal text. \\
Numeric precision or width mistake & Values are truncated, rounded, packed into too few bits, signed instead of unsigned, or accumulated in a lower precision type than required. \\
Wrong ordering of matching rules & Alternatives, cases or rules are tried in an order that lets a general rule shadow a more specific one, or lets a later specialised rule become unreachable. \\
\addlinespace
\multicolumn{2}{@{}l}{\emph{Wrong reference or declared contract}} \\
Wrong dispatch or lookup table entry & An entry exists in a table, constant set or definition list but carries a misspelled key, wrong target, wrong unit or otherwise incorrect value. \\
Wrong data source read & Logic reads from a plausible but incorrect field, collection, snapshot or alternative source of the same information, such as raw instead of normalised values or one member instead of the aggregate. \\
Incorrect public type declaration & A type annotation, generic constraint, overload or interface describes the runtime contract incorrectly, rejecting valid calls or widening inferred results. \\
Missing parameter propagation & A caller fails to forward available context such as a dialect, projection, encoding, signal, index or option to a callee that needs it, so the callee applies defaults. \\
Wrong argument at a call site & A call passes swapped, mismatched or plainly incorrect arguments, including a wrong literal, a wrong receiver, or arguments in the wrong positions. \\
Incorrect name or scope resolution & Identifiers, aliases, qualifiers or bindings are resolved against the wrong scope, namespace or declaration, including leaked or shadowed compiler bindings and wrongly qualified references. \\
\bottomrule
\end{tabular}

\end{table}

\begin{table}[!htbp]
\centering
\caption{The \taxCategories{} bug types and their definitions, grouped by family (continued from \cref{tab:bug-type-definitions}).}
\label{tab:bug-type-definitions-2}
\footnotesize
\begin{tabular}{@{}>{\raggedright\arraybackslash}p{0.22\linewidth}p{0.75\linewidth}@{}}
\toprule
Category & Definition \\
\midrule
\multicolumn{2}{@{}l}{\emph{State, shared data and lifecycle}} \\
Missing state reset or restoration & Flags, counters, cursors, module globals or cached bookkeeping are left over from a previous operation or are not restored when an operation exits, including exceptional exits without a finally. \\
Colliding key construction & A keyed store is built unsafely, either because the identifier used for caching, indexing, deduplication or grouping lacks a delimiter or a distinguishing component so distinct items collide, or because a plain object or inherited property lookup is used as the container so keys such as \_\_proto\_\_ or constructor corrupt or leak results. \\
Mutation of shared input data & Code mutates data that others alias, either by modifying an argument, option object, cached entry or otherwise shared structure instead of copying it, or by modifying or re-consuming a live view or single pass iterator while it is being traversed, so unrelated callers see changed data or items are skipped, duplicated or lost. \\
Missing cache invalidation & A change to underlying inputs is not propagated to dependents, either because a memoised or cached value is never invalidated or because a dependency, subscription or change notification needed by an observer, effect or rendering pipeline is never registered or emitted, so later reads observe stale data. \\
Missing resource release & Listeners, subscriptions, timers, buffers, sockets, processes or graphics objects are created but never unregistered, cancelled, closed or deleted, so they accumulate or fire after teardown. \\
\addlinespace
\multicolumn{2}{@{}l}{\emph{Control flow, concurrency and failure}} \\
Wrong sequencing of operations & Steps run in the wrong order relative to each other, for example reading state before it is normalised, emitting output before required setup, or mutating a structure before a dependent computation. \\
Swallowed or mislabelled error & An exception is caught and discarded, replaced by a misleading type or message, raised without being thrown, or signalled by an assertion rather than a proper error, so callers cannot react correctly. \\
Unguarded failure-prone call & A call that can fail transiently or on malformed data is made with no surrounding handling, retry or fallback, so the failure propagates and aborts otherwise recoverable work. \\
Missing synchronisation or atomicity & Concurrent tasks, threads, processes or requests interleave without a lock, lease, deterministic lock order or single atomic statement, causing races, duplicate effects or deadlocks. \\
Missing cycle or depth guard & A traversal or loop lacks a working termination guard, whether a recursive or reference following walk without a visited set or depth limit, or an iterative loop whose index never advances or whose exit condition can never be satisfied. \\
Incorrect asynchronous control flow & Asynchronous work is misdriven, for example a promise or coroutine is not awaited, a blocking call runs on an event loop, a callback resolves early or out of order, or a pending promise is never settled. \\
\addlinespace
\multicolumn{2}{@{}l}{\emph{Non-functional and environment}} \\
Missing output escaping & Text is emitted into a sink without the sanitising that sink requires, either structural or untrusted content interpolated into HTML, SQL, XML, CSS, URLs, regexes, headers or generated code without escaping, or credentials, tokens and other confidential values placed into messages, logs, URLs or persisted state without redaction. \\
Missing consumption limit & Attacker-influenced or unbounded sizes are allocated, decompressed, queued or buffered without a cap, permitting exhaustion of memory, disk or queue capacity. \\
Redundant or superlinear work & Correct results are produced but with avoidable repeated scanning, rebuilding, re-fetching or quadratic accumulation that degrades latency. \\
Missing or weak credential check & Authentication or authorisation is absent, incomplete, bypassable, or performed with a comparison whose timing reveals the secret, or an integrity check whose result is ignored. \\
Build or packaging misconfiguration & Project metadata, packaging rules, dependency declarations, tooling flags or generated asset configuration are wrong, so builds, installs or environments break independently of program logic. \\
\bottomrule
\end{tabular}

\end{table}

\FloatBarrier

\subsection{Bug-Type Distribution}
\label{app:taxonomy-distribution}

\Cref{tab:bug-type-full} extends \cref{tab:bug-type-distribution} to all categories. For each of the ten categories shown there, the human share lies outside the range spanned by the individual runs, so no single configuration drives the gap.

\begin{table}[!htbp]
\centering
\caption{Bug-type distribution of the human reference set and of coding agent findings, for all categories, sorted by gap. Agent counts ($\bar{n}$) and shares are means over one run of each of the \taxAgentRuns{} configurations; Range gives the lowest and highest share of a single run. Gap is the agent share minus the human share in percentage points; Ratio is the agent share over the human share.}
\label{tab:bug-type-full}
\footnotesize
\setlength{\tabcolsep}{4pt}
\begin{tabular}{lrrrrrrr}
\toprule
 & \multicolumn{2}{c}{Human} & \multicolumn{3}{c}{Agents} & & \\
\cmidrule(lr){2-3} \cmidrule(lr){4-6}
Category & $n$ & \% & $\bar{n}$ & \% & Range & Gap (pp) & Ratio \\
\midrule
Wrong arithmetic or comparison expression & 243 & 9.9 & 916 & 18.5 & 16.9--20.8 & $+8.6$ & 1.86 \\
Unhandled degenerate value & 222 & 9.1 & 674 & 13.9 & 10.7--19.6 & $+4.8$ & 1.53 \\
Incorrect regular expression pattern & 47 & 1.9 & 245 & 4.7 & 3.1--5.7 & $+2.8$ & 2.47 \\
Missing state reset or restoration & 57 & 2.3 & 242 & 4.9 & 3.5--5.9 & $+2.6$ & 2.11 \\
Wrong data source read & 57 & 2.3 & 224 & 4.5 & 3.3--5.8 & $+2.1$ & 1.91 \\
Colliding key construction & 24 & 1.0 & 139 & 2.8 & 1.7--3.6 & $+1.9$ & 2.90 \\
Truthiness test instead of presence test & 26 & 1.1 & 126 & 2.6 & 1.7--3.8 & $+1.6$ & 2.49 \\
Mutation of shared input data & 27 & 1.1 & 128 & 2.6 & 2.0--3.6 & $+1.5$ & 2.39 \\
Wrong ordering of matching rules & 4 & 0.2 & 66 & 1.3 & 0.5--1.9 & $+1.1$ & 7.92 \\
Swallowed or mislabelled error & 41 & 1.7 & 129 & 2.6 & 1.4--4.9 & $+0.9$ & 1.55 \\
Missing rejection of invalid input & 225 & 9.2 & 492 & 10.1 & 6.4--13.7 & $+0.9$ & 1.09 \\
Missing resource release & 26 & 1.1 & 51 & 1.1 & 0.7--1.8 & $0.0$ & 1.01 \\
Missing cycle or depth guard & 33 & 1.4 & 63 & 1.3 & 0.7--1.7 & $-0.1$ & 0.95 \\
Incorrect asynchronous control flow & 17 & 0.7 & 28 & 0.6 & 0.4--0.7 & $-0.1$ & 0.85 \\
Numeric precision or width mistake & 34 & 1.4 & 62 & 1.2 & 0.7--1.6 & $-0.2$ & 0.89 \\
Wrong argument at a call site & 34 & 1.4 & 50 & 1.1 & 0.7--1.4 & $-0.3$ & 0.77 \\
Incorrect name or scope resolution & 17 & 0.7 & 17 & 0.4 & 0.1--0.9 & $-0.3$ & 0.53 \\
Missing or weak credential check & 15 & 0.6 & 13 & 0.3 & 0.2--0.4 & $-0.3$ & 0.46 \\
Missing cache invalidation & 32 & 1.3 & 48 & 1.0 & 0.6--1.2 & $-0.4$ & 0.73 \\
Build or packaging misconfiguration & 13 & 0.5 & 8 & 0.2 & 0.0--0.4 & $-0.4$ & 0.29 \\
Redundant or superlinear work & 19 & 0.8 & 20 & 0.4 & 0.0--1.1 & $-0.4$ & 0.50 \\
Missing synchronisation or atomicity & 35 & 1.4 & 49 & 1.0 & 0.4--1.6 & $-0.4$ & 0.71 \\
Wrong sequencing of operations & 88 & 3.6 & 157 & 3.0 & 1.9--4.0 & $-0.6$ & 0.84 \\
Unguarded failure-prone call & 47 & 1.9 & 66 & 1.3 & 0.7--2.5 & $-0.6$ & 0.70 \\
Missing output escaping & 39 & 1.6 & 49 & 1.0 & 0.3--1.5 & $-0.6$ & 0.63 \\
Missing initialisation or default & 51 & 2.1 & 70 & 1.4 & 1.1--2.5 & $-0.6$ & 0.69 \\
Missing consumption limit & 30 & 1.2 & 24 & 0.5 & 0.2--0.6 & $-0.8$ & 0.37 \\
Missing parameter propagation & 48 & 2.0 & 35 & 0.7 & 0.5--1.1 & $-1.2$ & 0.37 \\
Wrong dispatch or lookup table entry & 102 & 4.2 & 130 & 2.6 & 2.0--3.7 & $-1.6$ & 0.63 \\
Unimplemented advertised feature & 114 & 4.7 & 133 & 2.6 & 1.6--3.9 & $-2.0$ & 0.56 \\
Incorrect public type declaration & 111 & 4.5 & 57 & 1.1 & 0.6--1.5 & $-3.5$ & 0.24 \\
Missing dispatch or lookup table entry & 124 & 5.1 & 38 & 0.7 & 0.3--1.4 & $-4.3$ & 0.14 \\
Missing branch for a supported input variant & 439 & 18.0 & 400 & 7.7 & 4.4--9.6 & $-10.3$ & 0.43 \\
\midrule
No listed category fits & 3 & 0.1 & 7 & 0.1 & 0.0--0.5 & $0.0$ & 1.14 \\
All bugs & 2,444 & 100.0 & 4,954 & 100.0 & & & \\
\bottomrule
\end{tabular}

\end{table}

\subsection{Trigger}
\label{app:taxonomy-trigger}

Alongside its category, every bug is assigned a trigger level, answering what is needed to observe the failure (\cref{tab:trigger-distribution}). Agent findings concentrate further on bugs that need only a specific input. Human-fixed bugs more often need a specific environment, a person inspecting the output, or timing, concurrency or load, and on each of these levels every agent run falls below the human share.

\begin{table}[!htbp]
\centering
\caption{Trigger levels of the human reference set and of coding agent findings. Agent shares are means over one run of each of the \taxAgentRuns{} configurations; Range gives the lowest and highest share of a single run.}
\label{tab:trigger-distribution}
\footnotesize
\begin{tabular}{@{}>{\raggedright\arraybackslash}p{0.19\linewidth}>{\raggedright\arraybackslash}p{0.43\linewidth}rrrr@{}}
\toprule
 & & \multicolumn{2}{c}{Human} & \multicolumn{2}{c}{Agents} \\
\cmidrule(lr){3-4} \cmidrule(l){5-6}
Level & Needed to observe the failure & $n$ & \% & \% & Range \\
\midrule
Ordinary use & Any execution of the affected path with common inputs fails or gives a wrong result. & 105 & 4.3 & 6.1 & 4.6--7.8 \\
A specific input & A particular value, shape or edge case is needed, such as an empty or absent value, an unusual variant, malformed data or a boundary. & 1,815 & 74.3 & 83.9 & 79.2--86.3 \\
A specific environment & A particular platform, runtime or dependency version, locale, toolchain or type checker is needed. & 265 & 10.8 & 4.7 & 3.7--6.3 \\
Interaction or inspection & The failure shows only in rendered output, a user interface or a document, or through a user's sequence of actions, and a person has to look at it. & 92 & 3.8 & 0.7 & 0.3--1.0 \\
Timing, concurrency or load & Concurrent execution, a particular ordering of asynchronous events, a long-running process or large inputs are needed. & 167 & 6.8 & 4.6 & 3.2--6.3 \\
\bottomrule
\end{tabular}

\end{table}

\FloatBarrier

\section{A Model of Repeated Bug Discovery}
\label{app:bug-discovery-model}
\label{app:stable-beta}


\begin{figure}[t]
\centering
\begin{minipage}[t]{0.54\linewidth}
\centering
\begin{tikzpicture}[scale=0.85, every node/.style={font=\footnotesize}]
  \begin{scope}
    \def\Ubox{(-3.1,-1.85) rectangle (1.75,1.85)}   
    \begin{scope}\clip \Ubox;
      \fill[blue!8]  (1.3,0) ellipse (2.7 and 1.45);
      \fill[blue!18] (1.0,0) ellipse (2.2 and 1.15);
      \fill[blue!35] (0.75,0) ellipse (1.75 and 0.8);
      \fill[blue!55] (0.55,0) ellipse (1.3 and 0.4);
    \end{scope}
    \begin{scope}\clip (1.75,-1.85) rectangle (4.3,1.85);
      \fill[blue!8, path fading=east] (1.3,0) ellipse (2.7 and 1.45);
      \fill[blue!18] (1.0,0) ellipse (2.2 and 1.15);
      \fill[blue!35] (0.75,0) ellipse (1.75 and 0.8);
      \fill[blue!55] (0.55,0) ellipse (1.3 and 0.4);
    \end{scope}
    \begin{scope}\clip (1.75,-1.85) rectangle (4.3,1.85);
    \end{scope}
    \draw[dashed, gray!70, rounded corners=3pt] \Ubox;
    \node[gray!90, anchor=north west, font=\scriptsize] at (-3.05,1.8) {$\mathcal{U}$: bug universe};
    \draw[red!70!black, thick] (-1.6,0) ellipse (1.35 and 0.95);
    \begin{scope}\clip (-1.6,0) ellipse (1.35 and 0.95);
      \fill[pattern=north east lines, pattern color=red!70!black] (1.3,0) ellipse (2.7 and 1.45);
    \end{scope}
    \node[red!70!black, align=center, font=\scriptsize] at (-1.6,1.2) {$\mathcal{H}$: bug reference set};
    \draw[red!70!black, thin] (-0.75,-0.55) -- (-0.35,-1.2); \node[red!70!black, anchor=north] at (-0.3,-1.18) {$\mathcal{H}\cap\mathcal{R}_A$};
    \node[blue!60!black, font=\scriptsize] at (0.55,0) {stable};
    \node[blue!60!black, font=\scriptsize] at (1.0,1.3) {rare};
    \node[blue!60!black, anchor=south west, font=\scriptsize, align=left] at (1.9,0.95) {$\mathcal{R}_A$: agent findings};
    \node[blue!60!black, anchor=north west, align=left, font=\scriptsize] at (1.9,-1.15) {$\Phi_A=\mathcal{R}_A\setminus\mathcal{U}$:\\ false positives};
  \end{scope}
\end{tikzpicture}
\vspace{-4pt}
\caption{The historical reference set $\mathcal{H}$ lies within the bug universe $\mathcal{U}$. The agent's reachable findings $\mathcal{R}_A$ overlap both and extend to false positives $\Phi_A$. Shading indicates per-run reporting probability, from a frequently rediscovered core to rare findings.}
\label{fig:reachability}
\end{minipage}\hfill
\begin{minipage}[t]{0.42\linewidth}
\centering
\includegraphics[height=92.9pt]{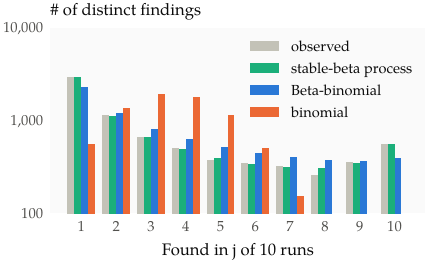}
\vspace{-4pt}
\caption{Observed and modeled distinct findings reported in exactly $j$ of 10 \luna runs (log scale). Only the stable-beta process reproduces both the many singletons and the stable core of findings.}
\label{fig:run-frequency}
\end{minipage}
\end{figure}

Repeated runs of one configuration report partly different findings (\cref{sec:further-analysis}). We model a configuration's findings as the features of a stable-beta process, fit it to the repeated runs of \musespark and \luna, and derive what it implies for the union of many runs.

\subsection{Model}
\label{app:sbp-model}

\paragraph{Setting}
A defect is a \emph{bug} if a maintainer would fix it when presented with the defect and an available fix. Let $\mathcal{U}$ denote these bugs in a fixed repository state and $\mathcal{H}\subseteq\mathcal{U}$ the historical reference set. An agent's reachable findings $\mathcal{R}_A$ include both bugs and false positives, $\Phi_A=\mathcal{R}_A\setminus\mathcal{U}$ (\cref{fig:reachability}).

\paragraph{Findings as Features}
Each deduplicated finding (\cref{app:dedup}) is a feature, and each run records whether that feature was reported. For a fixed agent, configuration, budget, and repository state, write
\begin{equation}
  B=\sum_{k=1}^{\infty}p_k\delta_{\theta_k},
  \qquad z_{ik}\mid B\sim\operatorname{Bernoulli}(p_k),
  \qquad \mathcal{F}_i=\{\theta_k:z_{ik}=1\}.
  \label{eq:sbp-features}
\end{equation}
Here $\theta_k$ identifies a finding, $p_k$ is its per-run reporting probability, and $\delta_{\theta_k}$ places unit mass at its identity. Conditional on $B$, the indicators are independent across runs and findings. The probabilities do not sum to one: a run can report multiple findings. This beta--Bernoulli construction provides a prior over binary feature matrices~\citep{thibaux2007hierarchical}.

\paragraph{The Three Parameters}
We model bug findings with a \emph{three-parameter beta process}, also called the stable-beta process~\citep{teh2009powerlaw}. The stable-beta process has mass $\alpha>0$, concentration $c>-\sigma$, and discount $0\leq\sigma<1$. Its atoms $(p_k,\theta_k)$ form a Poisson point process with intensity
\begin{equation}
  \nu(dp,d\theta)=C\,p^{-1-\sigma}(1-p)^{c+\sigma-1}\,dp\,G(d\theta),
  \qquad C=\frac{\alpha\Gamma(c+1)}{\Gamma(1-\sigma)\Gamma(c+\sigma)},
  \label{eq:sbp-intensity}
\end{equation}
where $G$ is a nonatomic probability measure over findings. The intensity is an expected atom count, not a normalized probability density.

The parameters have direct operational interpretations. Integrating $p$ against the intensity gives $\mathbb{E}[|\mathcal{F}_i|]=\alpha$. Integrating $p^2$ gives
\begin{equation}
  \mathbb{E}[|\mathcal{F}_1\cap\mathcal{F}_2|]
  =\alpha\frac{1-\sigma}{c+1}.
  \label{eq:sbp-overlap}
\end{equation}
Thus larger $c$ reduces expected pairwise sharing at fixed $\alpha,\sigma$. The discount controls the accumulation of low-probability findings through the factor $p^{-1-\sigma}$. Setting $\sigma=0$ recovers the ordinary beta process. Although there are infinitely many latent features, each run contains finitely many almost surely; marginally its size is $\operatorname{Poisson}(\alpha)$~\citep{broderick2012beta}.

\subsection{Fit to Repeated Runs}
\label{app:sbp-fit}

\paragraph{Parameter Estimation}
Parameters can be estimated from the frequency spectrum $M_{r,j}$, the number of findings reported in exactly $j$ of $r$ runs. Poisson thinning gives independent counts
\begin{equation}
  M_{r,j}\sim\operatorname{Poisson}(\mu_{r,j}),
  \qquad
  \mu_{r,j}=\binom{r}{j}C\,\mathrm{B}(j-\sigma,r-j+c+\sigma),
  \quad j=1,\ldots,r,
  \label{eq:sbp-spectrum}
\end{equation}
where $\mathrm{B}(a,b)$ is the beta function. Their product likelihood supports fitting without enumerating unseen findings. All predictive equations above condition on fixed parameters.

\paragraph{Compared Models}
We fit the frequency spectrum of \musespark (5 runs) and \luna (10 runs). Findings are deduplicated across runs, which gives $M_{r,j}$ for all reported findings, whether or not they match a historical bug. We compare the stable-beta process with two models of a finite pool. In the \emph{binomial} model every finding has the same reporting probability $q$. In the \emph{Beta-binomial} model reporting probabilities are drawn from $\operatorname{Beta}(a,b)$, with $a,b\in[0.01,50]$.

Findings that no run reported are not observed, so every model is fitted by maximum likelihood on the spectrum conditional on $j\geq1$.

\paragraph{Results}
\Cref{tab:run-frequency-fit} reports Pearson's $\chi^2$ over the $r$ bins, its $p$-value, and the sample-size-free $\sqrt{\chi^2/N}$ with $N$ the number of distinct findings. The stable-beta process is the only model that is not rejected ($p=0.36$ and $p=0.064$). In contrast to a binomial or Beta-binomial bug discovery process, this captures both the frequent rediscovery of a stable core and continued discovery of rare findings we observe, see \cref{fig:run-frequency}.

\Cref{tab:run-frequency-parameters} gives the fitted parameters. The two configurations have similar discounts, $\hat\sigma=0.26$ and $\hat\sigma=0.29$. The mass estimates, $7{,}772$ and $2{,}644$, agree with the mean number of deduplicated findings per run, $7{,}770$ and $2{,}636$, although the fit does not impose this. The Beta-binomial ends at its lower bound $a=0.01$: its likelihood keeps increasing as $a\to0$, where it approaches the stable-beta process at $\sigma=0$, and the data ask for $\sigma>0$. Its pool size is then set by the bound, not by the data, and is thus omitted. The binomial model implies a pool only slightly larger than the observed union, which the continued arrival of new findings contradicts: runs six to ten of \luna raise the union from $5{,}689$ to $7{,}472$ findings.

\begin{table}[H]
\centering
\small
\caption{Goodness of fit to the number of distinct findings reported in exactly $j$ of $r$ runs, conditional on $j\geq1$. $k$ is the number of fitted parameters and $N$ the number of distinct findings.}
\label{tab:run-frequency-fit}
\begin{tabular}{lrrrrr}
\toprule
Model & $k$ & $\chi^2$ & df & $p$ & $\sqrt{\chi^2/N}$ \\
\midrule
\multicolumn{6}{l}{\emph{Muse Spark 1.3, high: 5 runs, 15,233 distinct findings}} \\
Binomial & 1 & 22,197.0 & 3 & $<10^{-99}$ & 1.207 \\
Beta-binomial & 2 & 318.0 & 2 & $8.9\times10^{-70}$ & 0.144 \\
Stable-beta process & 2 & 2.1 & 2 & 0.357 & 0.012 \\
\midrule
\multicolumn{6}{l}{\emph{GPT-5.6 Luna, medium: 10 runs, 7,472 distinct findings}} \\
Binomial & 1 & $1.7\times10^{6}$ & 8 & $<10^{-99}$ & 14.908 \\
Beta-binomial & 2 & 401.8 & 7 & $9.6\times10^{-83}$ & 0.232 \\
Stable-beta process & 2 & 13.3 & 7 & 0.064 & 0.042 \\
\bottomrule
\end{tabular}

\end{table}

\begin{table}[H]
\centering
\small
\caption{Fitted parameters of the models in \cref{tab:run-frequency-fit} and the pool size each implies. $^{\dagger}$ marks a parameter at its bound; the pool size of that Beta-binomial is set by the bound and omitted. The latent pool of the stable-beta process is infinite.}
\label{tab:run-frequency-parameters}
\begin{tabular}{llr}
\toprule
Model & Parameters & Pool size \\
\midrule
\multicolumn{3}{l}{\emph{Muse Spark 1.3, high: 5 runs, 15,233 distinct findings}} \\
Binomial & $q = 0.493$ & 15,761 \\
Beta-binomial & $a = 0.010^{\dagger},\ b = 0.612$ & --- \\
Stable-beta process & $\alpha = 7{,}772,\ c = 0.172,\ \sigma = 0.263$ & $\infty$ \\
\midrule
\multicolumn{3}{l}{\emph{GPT-5.6 Luna, medium: 10 runs, 7,472 distinct findings}} \\
Binomial & $q = 0.348$ & 7,577 \\
Beta-binomial & $a = 0.010^{\dagger},\ b = 0.827$ & --- \\
Stable-beta process & $\alpha = 2{,}644,\ c = 0.254,\ \sigma = 0.287$ & $\infty$ \\
\bottomrule
\end{tabular}

\end{table}

\subsection{Unbounded Discovery and False Positives}
\label{app:sbp-consequences}

\paragraph{An Unbounded Pool of Findings}
The so-called discount parameter $\sigma$ of the stable-beta process determines whether the model predicts a finite or unbounded finding pool; a negative $\sigma$ gives the generalized finite beta--Bernoulli model~\citep{battiston2018characterization}. For the fitted positive $\sigma$ (\cref{app:sbp-fit}), the expected number of distinct findings grows unbounded as $r^\sigma$.

\paragraph{A Lower Bound on False Positives}
Let $\mathcal{S}_r=\bigcup_{i=1}^r\mathcal{F}_i$, $K_r=|\mathcal{S}_r|$, and $U=|\mathcal{U}|<\infty$. With findings deduplicated by defect identity, at most $U$ members of $\mathcal{S}_r$ can be real bugs. For $K_r>0$, the false-positive fraction of the accumulated distinct findings therefore satisfies
\begin{equation}
  \frac{|\mathcal{S}_r\setminus\mathcal{U}|}{K_r}
  \geq \max\!\left\{0,1-\frac{U}{K_r}\right\}.
  \label{eq:sbp-fp-bound}
\end{equation}
If $K_r\to\infty$, this fraction tends to one. This is a consequence of unbounded discovery and a finite bug universe, not of historical non-matches being false positives. A numerical lower bound at a finite budget requires an upper bound on $U$; the incomplete reference set supplies only a lower bound. The result concerns the deduplicated union, not the false-positive fraction of an individual run, which can repeatedly report real bugs.

As a fixed repository contains only finitely many bugs, unbounded growth in distinct findings implies accumulating false positives. Retaining only findings reported in at least $n$ runs thus isolates findings of increasingly higher minimum reporting probability, trading recall for precision as $n$ increases (\cref{fig:rerun-threshold}).

\clearpage

\section{Prompts}
\label{app:prompt}

This section prints \bench's prompts, grouped by the stage of \cref{sec:method,sec:bug-finding} or the experiment of \cref{sec:experiments} they belong to.

A value substituted into a prompt at run time is set in \emph{italics} and named rather than shown, so what is
printed is the template and not one instance of it. Where a stage constrains its output, the harness appends the JSON
schema of that output to the instructions.

{\lstset{numbers=none,xleftmargin=0pt,framexleftmargin=0pt,numbersep=0pt}

\needspace{8\baselineskip}
\subsection{Environment Preparation}

\promptheading{Snapshot Environment (Python)}{prompt:env-python}{\textsc{Coding Agent}}
Environment preparation for a Python repository. The agent writes the snapshot image and its fully pinned requirements, repins every package released after $T_0$ down to the last release before it, and must get the repository's own test suite to run to completion. It is a brief for a coding agent working in the benchmark repository; \ttt{<owner>}, \ttt{<repo>} and \ttt{<commit>} stand for the repository being set up and its snapshot commit.

\begin{promptbox}{Prompt}{userborder}
  \begin{lstlisting}[basicstyle=\ttfamily\fontsize{7}{7.6}\selectfont,breaklines=false,showstringspaces=false,language=,mathescape=false,escapeinside={(*@}{@*)}]
# Snapshot environment setup — Python repo `<repo>`

You are setting up a reproducible Docker test environment for the benchmark repo **`<repo>`** (a Python 
    repo already added as a git submodule under `repos/<owner>/<repo>/<commit>`, pinned at the 
    2025-01-01 snapshot commit).

## Goal

Produce a Docker image whose test suite runs to completion via **`environment/run_tests.py`**, with 
    every package version pinned to what was installable on the snapshot cutoff (this snapshot's entry 
    in `configs/environment/snapshot_config.json`). Nothing installed may be a version released after 
    the cutoff.

## Deliverables

1. `dockerfiles/<owner>/<repo>/<commit>/Dockerfile.base` — the base image (build deps + pinned Python 
    env + editable install of `<repo>` + the test-runner CMD). `<commit>` is the 12-char short SHA of 
    the snapshot pin; the submodule checkout sits at `repos/<owner>/<repo>/<commit>`.
2. `dockerfiles/<owner>/<repo>/<commit>/requirements.txt` — the **fully pinned** (`name==version`) 
    environment: runtime + test dependencies, all transitive deps included.
3. `dockerfiles/<owner>/<repo>/<commit>/build-constraints.txt` — the repo's `[build-system] requires` 
    and their transitive deps, pinned the same way; skip it for a repo with no `pyproject.toml`, which 
    pip builds unisolated (cf. prefect, sqlglot). Without it pip resolves the build backend fresh from 
    PyPI on every build, pulling post-cutoff packages into the snapshot.
4. A `<repo>` entry in `repos/repos.json` **if one does not already exist** — `run_tests.py` requires 
    it. Match the existing entries' shape: `github_full_name` (lowercased `owner/name`), `snapshot_sha` 
    (the full snapshot commit), `language` (`"python"`), and `src_path` (in-repo source root, e.g. 
    `/<repo>`).

## Study these first (do not skip)

- `dockerfiles/opshin/opshin/<commit>/Dockerfile.base` + `requirements.txt` — the simplest reference 
    (plain setuptools, top-level `tests/`).
- `dockerfiles/kozea/weasyprint/<commit>/Dockerfile.base` + `requirements.txt` — reference for a repo 
    that needs extra **system libraries** (`apt-get install …`) beyond `gcc`/`git`.
- `dockerfiles/prefecthq/prefect/<commit>/Dockerfile.base` + `requirements.txt` — reference for a large 
    suite with pytest ini/marker/ignore tweaks in the CMD.
- `docker_scripts/run_sharded.sh` — the in-container pytest runner every Python repo uses. It shards 
    `<tests_dir>` by top-level test file (`test_*.py` and `*_test.py`) / immediate subdir, runs each in 
    a fresh pytest process, and prints the final `=== TOTAL: N failed, N passed, N skipped, … ===` line 
    that `run_tests.py` looks for. **Reuse it**; do not invent a new runner.
- `environment/run_tests.py` — how the suite is built and run (`build_base` → run the image's CMD, 
    capture stdout to `findings/<owner>/<repo>/<commit>/test_results.txt`, echo the TOTAL line).
- `environment/pin_to_snapshot.py` — read its module docstring. It queries the PyPI JSON API for every 
    `name==version` pin in `dockerfiles/<owner>/<repo>/<commit>/requirements.txt` and repins anything 
    released after the cutoff down to the highest non-prerelease, non-yanked version at/before the 
    cutoff. Use it (see workflow).

## Dockerfile pattern (adapt, don't blindly copy)

Follow the shared skeleton from the reference Dockerfiles exactly for these parts, changing only the 
    `$OWNER/$REPO/$COMMIT` segments:

```dockerfile
FROM python:3.10.16-slim              # see "Python version" below

RUN apt-get update && apt-get install -y \
    gcc \
    git \
    && rm -rf /var/lib/apt/lists/*    # add repo-specific system libs here (cf. weasyprint)

WORKDIR /app
COPY dockerfiles/$OWNER/$REPO/$COMMIT/requirements.txt .   # snapshot dockerfiles dir, mirrored by the 
    checkout below
RUN pip install --no-cache-dir -r requirements.txt

COPY repos/$OWNER/$REPO/$COMMIT .
RUN rm .git

COPY dockerfiles/$OWNER/$REPO/$COMMIT/build-constraints.txt /tmp/build-constraints.txt
RUN PIP_CONSTRAINT=/tmp/build-constraints.txt pip install --no-cache-dir --no-deps -e .

COPY docker_scripts/run_sharded.sh /app/run_sharded.sh

COPY docker_scripts/init_snapshot_repo.sh /tmp/init_snapshot_repo.sh
RUN bash /tmp/init_snapshot_repo.sh /app

# TOTAL: <paste the observed green TOTAL line here once you have it>
CMD ["timeout", "-k", "30", "10800", \
     "bash", "/app/run_sharded.sh", "<tests_dir>", \
     "-v", "--timeout=60", "--timeout-method=thread", "--tb=short", "-m", "not service"]
```

**No image may ship the repo's upstream git history.** Its `origin/*` refs are dated to clone time, so 
    an agent could read the post-snapshot fixes the benchmark measures against. `RUN rm .git` drops the 
    submodule gitlink, and `init_snapshot_repo.sh` — last, after every install and build step, so the 
    tree is committed clean — leaves a single-commit repo with no remotes, tags or parents.

That removes the tags a version-deriving build backend reads. If the repo uses `setuptools_scm`, 
    `hatch-vcs` or another SCM source, hand it the version the pin would have produced (`git describe 
    --tags` in the snapshot checkout), build-scoped so nothing persists into the container:

```dockerfile
RUN SETUPTOOLS_SCM_PRETEND_VERSION_FOR_<DIST>=<version> \
    PIP_CONSTRAINT=/tmp/build-constraints.txt \
    pip install --no-cache-dir --no-deps -e .
```

Use the unscoped `SETUPTOOLS_SCM_PRETEND_VERSION` under `hatch-vcs` (cf. jsonschema): it calls 
    setuptools_scm without a dist name, so the `_FOR_<DIST>` form is never read.

A repo with no `pyproject.toml` builds without pip's isolation, so `PIP_CONSTRAINT` does not reach its 
    `setup_requires`: add `--no-build-isolation` and pin those packages in `requirements.txt` instead 
    (cf. sqlglot).

## Choosing the pieces per repo (this is where repos differ)

- **Python version.** Prefer `python:3.10.16-slim` to match the other repos (9 of 10 use it). But check 
    the repo's `requires-python` (in `pyproject.toml`/`setup.cfg`) — if it demands newer (e.g. 
    `>=3.11`), pick the lowest compatible pinned slim tag (networkx uses `python:3.11.11-slim` for 
    exactly this reason) and say why in your summary. Always pin the full `x.y.z` version — a floating 
    minor tag re-resolves past the cutoff on a future rebuild.
- **System libraries.** If `pip install` or the tests fail on a missing shared library / native build, 
    add the needed `apt-get` packages (weasyprint is the template).
- **`<tests_dir>` for the CMD.** `run_sharded.sh` shards a directory. Find where this repo's tests 
    actually live: a top-level `tests/` (like opshin/weasyprint) or `Tests/` (fonttools), or inside the 
    package (`networkx/` — the package itself, `pint/testsuite`, `jsonschema/tests`). Pass that path. If
     tests are scattered per-subpackage with no single dir, prefer running against the installed package
     (`pytest --pyargs $REPO …`) instead — but confirm `run_sharded.sh` can express that; if not, raise 
    it with me rather than hacking around it.
- **Marker / ignore flags.** Keep `-m "not service"` as the others do — unless the repo defines its own 
    marker taxonomy for non-reproducible tests, in which case use it (pypdf: `-m "not enable_socket and 
    not samples"`). Add `--ignore=…` for test files that need network/external services or are otherwise
     not reproducible in the sandbox (cf. prefect). Add pytest plugins the suite needs (e.g. 
    `pytest-timeout`) to `requirements.txt`.
- **Project addopts / odd tests.** If the repo's own `addopts` fight the harness, strip just the 
    offending flags with `--override-ini=addopts=…` and document why in a comment above the CMD (prefect
     drops a mypy-stub flag with `-rfEs`; fonttools drops `--doctest-modules`/`--pyargs` with `-ra`). 
    `--deselect` individual tests that only pass under upstream's single-process ordering (cf. 
    jsonschema's comment), set `ENV` vars the suite expects (jsonschema's 
    `JSON_SCHEMA_TEST_SUITE=/app/json`), and skip benchmark plugins (`--benchmark-skip`, cf. pint).

## Workflow

1. **Add/verify the `repos.json` entry** (deliverable 4) so `run_tests.py` can resolve the repo.
2. **Assemble a fully-pinned `requirements.txt`.** Install the repo plus its test extras in a throwaway 
    env and `pip freeze`, or read its `pyproject.toml`/`setup.cfg` test/dev extras and pin them — the 
    end state is a complete `name==version` list (runtime + test + transitive), like the reference 
    files.
3. **Resolve `build-constraints.txt`** from the repo's `[build-system] requires` (pip's 
    `setuptools`/`wheel` default when the table is absent): `printf '<req>\n…' | uv pip compile - 
    --exclude-newer <cutoff> --python-version <x.y> --no-header --no-annotate`. Then build with `pip 
    install -v` and read the log: a backend may add build requirements dynamically that the closure 
    misses (hatchling pulls `editables`), and anything installed outside the constraint set needs its 
    own pin.
4. **Snapshot-pin:** run `uv run python environment/pin_to_snapshot.py $REPO` (dry run) to see which 
    pins were released after the cutoff, then `--apply` to repin them. Resolve any `manual_review` 
    entries by hand. Re-run until it reports no changes, then repeat with `--file build-constraints`.
5. **Build + run:** `uv run python environment/run_tests.py $REPO`. Iterate on the Dockerfile / 
    requirements.txt until the suite is green (`0 failed, 0 errors`). Some pre-existing 
    `skipped`/`xfailed` are fine; unexpected failures usually mean a wrong pin, a missing system lib, or
     a test needing a service you should `--ignore`.
6. **Record the result:** once green, paste the exact `=== TOTAL: … ===` line into the `# TOTAL:` 
    comment above the CMD (as the reference Dockerfiles do).

### If the full suite takes more than ~10 minutes

Do **not** run the whole thing yourself. Build the image and verify with a **representative subset** — 
    run `run_sharded.sh` against one or two shards inside the container, or a handful of `-k`-selected 
    tests — to confirm the environment is sound (imports work, deps resolve, tests execute and pass). 
    Configure the CMD to run the **full** suite regardless, note in your summary that you validated on a
     subset only, and I will run the complete suite myself.

## Report back

Summarize: Python version chosen (and why if not 3.10), any system libs added, the `<tests_dir>` and any
     `--ignore`/marker flags, anything `pin_to_snapshot.py` needed manual resolution for, the `TOTAL` 
    line (or the subset you validated + why the full run was deferred), and anything you were unsure 
    about.
  \end{lstlisting}
\end{promptbox}

\promptheading{Snapshot Environment (TypeScript/JavaScript)}{prompt:env-typescript}{\textsc{Coding Agent}}
Environment preparation for a TypeScript or JavaScript repository, the counterpart of \cref{prompt:env-python}. Dependencies are pinned to what the repository's lockfile resolved at the snapshot commit. It is a brief for a coding agent working in the benchmark repository; \ttt{<owner>}, \ttt{<repo>} and \ttt{<commit>} stand for the repository being set up and its snapshot commit.

\begin{promptbox}{Prompt}{userborder}
  \begin{lstlisting}[basicstyle=\ttfamily\fontsize{7}{7.6}\selectfont,breaklines=false,showstringspaces=false,language=,mathescape=false,escapeinside={(*@}{@*)}]
# Snapshot environment setup — TypeScript/JavaScript repo `<repo>`

You are setting up a reproducible Docker test environment for the benchmark repo **`<repo>`** (a 
    TypeScript/JavaScript repo already added as a git submodule under `repos/<owner>/<repo>/<commit>`, 
    pinned at the 2025-01-01 snapshot commit).

## Goal

Produce a Docker image whose test suite runs to completion via **`environment/run_tests.py`**, with 
    every package version pinned to what the repo resolved at the snapshot commit. Nothing installed may
     be a version released after the cutoff (this snapshot's entry in 
    `configs/environment/snapshot_config.json`).

## Deliverables

1. `dockerfiles/<owner>/<repo>/<commit>/Dockerfile.base` — the base image (Node + package install from 
    the committed lockfile + build if needed + the test-runner CMD). `<commit>` is the 12-char short SHA
     of the snapshot pin; these repos have no snapshot `requirements.txt`, so the base's `COPY` lines 
    all read from the `repos/<owner>/<repo>/<commit>` checkout (for a workspace monorepo like svelte, 
    see **Workspace layout** below).
2. A `<repo>` entry in `repos/repos.json` **if one does not already exist** — `run_tests.py` requires 
    it. Match the existing entries' shape: `github_full_name` (lowercased `owner/name`), `snapshot_sha` 
    (the full snapshot commit), `language` (`"typescript"` or `"javascript"`, matching the repo's 
    primary language), and `src_path` (in-repo source root, e.g. `/src`; for a workspace monorepo, the 
    package under test, e.g. `/packages/svelte/src`).

There is **no `requirements.txt`** for these repos, and **`pin_to_snapshot.py` does not apply** (it is 
    PyPI-only). Pinning is achieved by installing from the repo's **committed lockfile** with a frozen 
    install (see below) — that lockfile, at the snapshot commit, already encodes exactly the versions 
    that were resolved on the snapshot date.

## Study these first (do not skip)

- `dockerfiles/honojs/hono/<commit>/Dockerfile.base` — the reference TS setup: pinned node, install from
     the frozen lockfile (bun), `bun run build`, then run three vitest configs via 
    `run_vitest_tests.sh`.
- `dockerfiles/sveltejs/svelte/<commit>/Dockerfile.base` — reference for a pnpm workspace monorepo: 
    corepack-pinned pnpm, whole-repo copy before the frozen install, and a COPYed custom 
    `vitest.no-browser.config.js` (in the same dockerfiles dir) that excludes the browser suite and 
    raises timeouts.
- `dockerfiles/mozilla/pdf.js/<commit>/Dockerfile.base` — reference for a build-harness repo: pinned 
    node with the cutoff rationale spelled out, `ENV` test vars, a `gulp lib-legacy` build, then jasmine
     via `run_jasmine_tests.sh`.
- `docker_scripts/run_vitest_tests.sh` — hono's in-container runner. It runs one or more vitest configs 
    and emits the final `=== TOTAL: N failed, N passed, N skipped (S suites run, F failed) ===` line 
    that `run_tests.py` looks for.
- `docker_scripts/run_sharded.sh` — the Python runner; read it only to see the **exact shape** of the 
    `=== TOTAL: … ===` contract, in case your repo's test runner isn't vitest and you need a small 
    wrapper that emits the same line.
- `environment/run_tests.py` — how the suite is built and run (`build_base` → run the image's CMD, 
    capture stdout to `findings/<owner>/<repo>/<commit>/test_results.txt`, echo the TOTAL line). Note: 
    if no `=== TOTAL: …` line is present it warns but still saves output — aim to emit the line anyway 
    for consistency with the other repos.

## Dockerfile pattern (adapt — these repos differ more than the Python ones)

hono uses **bun**; your repo most likely uses **npm** with a committed `package-lock.json` (check!). Use
     the package manager the repo's lockfile implies, and always do a **frozen / clean install** so 
    nothing re-resolves to a newer version:

- `package-lock.json` → `npm ci`
- `yarn.lock` → `yarn install --frozen-lockfile`
- `pnpm-lock.yaml` → `pnpm install --frozen-lockfile`
- `bun.lockb` / the textual `bun.lock` → `bun install --frozen-lockfile` (hono)

Skeleton (adapt the `$OWNER/$REPO/$COMMIT` segments, the package manager, the build step, and the test 
    CMD):

```dockerfile
FROM node:20.18.1-slim                # exact pin, newest of the line at/before the cutoff — see "Node 
    version" below

RUN apt-get update && apt-get install -y \
    git \
    && rm -rf /var/lib/apt/lists/*    # add curl/unzip/build tooling only if the install needs it

WORKDIR /app

# Copy just the manifest + lockfile first so the install layer caches independently of source.
COPY repos/$OWNER/$REPO/$COMMIT/package.json repos/$OWNER/$REPO/$COMMIT/package-lock.json ./
RUN npm ci

COPY repos/$OWNER/$REPO/$COMMIT .
RUN rm .git

RUN npm run build                     # only if tests need built output; many runners execute TS/JS test
     files directly

COPY docker_scripts/run_vitest_tests.sh /app/run_vitest_tests.sh   # copy the wrapper matching this 
    repo's runner (vitest/jest/mocha/ava/jasmine)

COPY docker_scripts/init_snapshot_repo.sh /tmp/init_snapshot_repo.sh
RUN bash /tmp/init_snapshot_repo.sh /app

# TOTAL: <paste the observed green TOTAL line here once you have it>
CMD ["timeout", "-k", "30", "10800", \
     "bash", "/app/run_vitest_tests.sh", ""]     # adapt to the actual runner — see below
```

**No image may ship the repo's upstream git history.** Its `origin/*` refs are dated to clone time, so 
    an agent could read the post-snapshot fixes the benchmark measures against. `RUN rm .git` drops the 
    submodule gitlink, and `init_snapshot_repo.sh` — last, after the install and any build step, so the 
    tree is committed clean — leaves a single-commit repo with no remotes, tags or parents. A build step
     that derives a version or build number from git history falls back to its no-repo default; say so 
    in your summary if the repo does that.

## Choosing the pieces per repo (this is where repos differ)

- **Node version.** Pick the major line from the repo's `.nvmrc` / `package.json` `engines.node` (prefer
     Node 20 unless the repo needs newer; if `.nvmrc` pins a non-LTS current release, use the nearest 
    LTS that satisfies `engines`), then **pin the newest release of that line at/before the cutoff** — 
    currently `node:18.20.5-slim` / `node:20.18.1-slim` / `node:22.12.0-slim`; verify against 
    `https://nodejs.org/dist/index.json`. Never use a floating tag (`node:20-slim`): it re-resolves past
     the cutoff on a future rebuild — that drift shipped a newer ICU whose Unicode tables broke 32 
    pdf.js specs. State your choice and why.
- **Package manager.** Driven by the committed lockfile (see list above). If the repo has **no 
    lockfile**, a frozen install is impossible and fresh resolution would pull post-cutoff versions — 
    **stop and raise this with me** rather than guessing pins.
- **Build step.** Some repos build before testing (hono: `bun run build`), others run tests directly 
    (ava / jest / mocha / vitest execute `.ts`/`.js` test files without a prior build). Include `npm run
     build` only if the tests actually need built artifacts; skip it otherwise. Watch for heavy 
    `prepare`/`prepublish` scripts that `npm ci` may trigger — you may need `npm ci --ignore-scripts` 
    plus an explicit lighter build.
- **Test runner + the TOTAL line.** The runner varies by repo (check `package.json` `scripts.test`). 
    Reuse the matching in-container wrapper — each emits the same `=== TOTAL: N failed, N passed, N 
    skipped … ===` line `run_tests.py` looks for, and each hard-fails if it cannot parse the runner's 
    summary:
  - **vitest** → `run_vitest_tests.sh` (pass the config path(s), `""` for the default config): hono runs
     three configs, remeda the default, svelte a custom no-browser config.
  - **jest** → `run_jest_tests.sh` (echarts: `--config test/ut/jest.config.cjs`; liquidjs: default 
    config).
  - **mocha** → `run_mocha_tests.sh <spec …>` (mathjs: `test/unit-tests`; openlayers: `--recursive 
    test/node --timeout 60000`; msgpack-javascript: `test/**/*.test.ts`).
  - **ava** → `run_ava_tests.sh` (jose).
  - **jasmine** → `run_jasmine_tests.sh` (pdf.js: `--config=test/unit/clitests.json`, standalone once 
    the Dockerfile has run the `gulp lib-legacy` build).
  - **anything else** → no wrapper yet; write a small `docker_scripts/run_<runner>_tests.sh` modelled on
     the above (run the suite so a failure doesn't mask counts, parse the summary, print the TOTAL 
    line). If awkward, raise it with me first.
- **Workspace layout.** Most repos are flat, but some are workspace monorepos — svelte is a **pnpm 
    workspace** (`pnpm-workspace.yaml` + `pnpm-lock.yaml` at the root, target package under 
    `packages/svelte`). Provision the repo-pinned pnpm via corepack, run the **frozen install at the 
    root**, and COPY the **whole repo before installing** when a workspace member's lifecycle script 
    needs its own source (svelte's playgrounds/sandbox `prepare` does — manifests-only layer caching 
    doesn't survive that). Run the target package's tests from the root and point `src_path` at the 
    package source (`/packages/svelte/src`).
- **System libraries.** Add `apt-get` packages only if the install/tests need them (native addon builds,
     headless-browser test deps, etc.). Many top-level `test` scripts chain extra steps — lint, 
    coverage, or browser/rendering suites (mathjs `test` runs lint; openlayers `test` runs karma + 
    rendering). Bypass the npm script and invoke the unit suite directly through the wrapper (mathjs: 
    mocha on `test/unit-tests`; openlayers: mocha `--recursive test/node`) so the TOTAL reflects the 
    real suite, and flag any sub-suites you skip and why (browser/rendering can't run headless here — 
    svelte excludes its Chromium suite via a custom vitest config). The build VM is throttled: raising 
    in-suite timeouts is an accepted pattern (openlayers mocha `--timeout 60000`; svelte's 30s vitest 
    timeouts), the outer CMD `timeout` still caps total wall-time.

## Workflow

1. **Add/verify the `repos.json` entry** (deliverable 2) so `run_tests.py` can resolve the repo.
2. **Write `Dockerfile.base`** using the frozen install for this repo's package manager and the correct 
    runner CMD.
3. **Build + run:** `uv run python environment/run_tests.py $REPO`. Iterate until the suite is green (`0
     failed`). Some pre-existing `skipped`/`todo` are fine.
4. **Verify pins:** confirm the install was frozen (no lockfile drift, no fresh resolution). If you had 
    to touch the lockfile, that's a red flag — investigate rather than commit a regenerated lockfile.
5. **Record the result:** paste the exact `=== TOTAL: … ===` line into the `# TOTAL:` comment above the 
    CMD.

### If the full suite takes more than ~10 minutes

Do **not** run the whole thing yourself. Build the image and verify with a **representative subset** 
    (run one test file / a focused test filter inside the container) to confirm the environment is 
    sound. Configure the CMD to run the **full** suite regardless, note that you validated on a subset 
    only, and I will run the complete suite myself.

## Report back

Summarize: Node version chosen (and why if not 20), package manager + install command, whether a build 
    step was needed, the test runner and how you emit the TOTAL line (reused vitest wrapper vs. new 
    script), any suites skipped (e.g. browser-only) and why, the `TOTAL` line (or the subset you 
    validated + why the full run was deferred), and anything you were unsure about.
  \end{lstlisting}
\end{promptbox}

\needspace{8\baselineskip}
\subsection{Benchmark Construction}

\promptheading{Bug-Fix Commit Classifier}{prompt:fix-classifier}{\terra}
Classifier of Step 1 of benchmark construction: whether a commit corrects previously buggy runtime behavior.

\begin{promptbox}{System}{systemborder}
  \begin{lstlisting}[basicstyle=\ttfamily\fontsize{7}{7.6}\selectfont,breaklines=false,showstringspaces=false,language=,mathescape=false,escapeinside={(*@}{@*)}]
You are classifying a single git commit as either a **bug fix** or **not a bug fix** (feature, refactor,
     docs, test, dependency bump, formatting, build, etc.).

A bug fix means the commit changes code so that a previously-buggy behavior is corrected — a regression 
    repaired, an incorrect output made correct, a crash prevented, a missing input validation added, a 
    race resolved, a leak plugged, an off-by-one or comparison error corrected, a security flaw closed.

**Do NOT classify as bug fix:**
- New features or enhancements (even if labeled "improve" or "handle X").
- Pure refactors that preserve behavior (but see the mixed-commit rule below).
- Documentation, test additions, comments, formatting.
- Dependency upgrades, version bumps, or lockfile/manifest changes — even when the upstream release 
    fixes a bug or CVE. A bug fixed in a dependency is a bug in that dependency, not in this repository;
     only changes to this repository's own code count. (Exception: vendored third-party source patched 
    directly in this repository's tree is an in-repo fix and does count.)
- Build/CI/tooling changes.
- Reverts (any commit whose subject starts with `Revert ` — reject outright).

**Mixed commits:** Classify on whether *any* part of the commit corrects a real, previously-buggy 
    behavior — not on what dominates the change. A commit is a bug fix even when the bulk of it is an 
    unrelated refactor, feature, cleanup, or formatting, as long as it also repairs a genuine bug. Do 
    not reject a real fix just because it is a small hunk buried in a large diff. Set `confidence` to 
    reflect how sure you are that a real bug was fixed (a clearly buried but unmistakable fix can still 
    be `high`).

**Edge cases:**
- "Handle edge case X" that wasn't previously broken → not a bug fix.
- "Fix typo in error message" → bug fix only if the typo materially broke behavior; otherwise reject.
- "Make X more robust" → only if there is a specific symptom being fixed.

You will be given the commit message, the linked issue body and linked PR body (when GitHub resolved a 
    closing issue/PR — otherwise both will read `(none)`), a file-change stat (`file | +N -N` lines), 
    and a diff excerpt. **Very large diffs are truncated** — the middle is elided and replaced with a 
    `…[diff truncated: N of M chars elided]…` marker, leaving the head and tail visible. When you see 
    that marker, base the verdict on the visible head and tail hunks plus the commit message, the linked
     bodies, and the stat.

Return ONLY a JSON object matching the generated JSON Schema appended to this prompt. Its field 
    descriptions are part of the output contract.

## Few-shot examples

**Example A** — subject: `tests: add regression test for #1421`, diff touches `tests/` only.
```json
{"is_bug_fix": false, "confidence": "high",
 "rationale": "Test-only commit; no production code changes.",
 "bug_kind_hint": ""}
```

**Example B** — subject: `bump cryptography to 42.0.4 to pick up CVE-2024-XXXX`, diff updates a single 
    pinned dependency.
```json
{"is_bug_fix": false, "confidence": "high",
 "rationale": "Dependency version bump; the CVE lives in the upstream library, not in this repository's 
    code.",
 "bug_kind_hint": ""}
```

**Example C** — subject: `Refactor _parse to use match statement`, diff is a behavior-preserving 
    rewrite.
```json
{"is_bug_fix": false, "confidence": "high",
 "rationale": "Pure refactor; control-flow rewritten without changing observable behavior.",
 "bug_kind_hint": ""}
```

**Example D** — subject: `Refactor request parsing; also fix header-length off-by-one`, diff is a large 
    behavior-preserving rewrite that also corrects a real boundary bug.
```json
{"is_bug_fix": true, "confidence": "medium",
 "rationale": "Mostly a refactor, but it also repairs a genuine off-by-one in header-length handling, so
     it counts as a bug fix.",
 "bug_kind_hint": "off-by-one"}
```
  \end{lstlisting}
\end{promptbox}

\begin{promptbox}{User}{userborder}
  \begin{lstlisting}[basicstyle=\ttfamily\fontsize{7}{7.6}\selectfont,breaklines=false,showstringspaces=false,language=,mathescape=false,escapeinside={(*@}{@*)}]
Commit SHA: (*@\textit{commit SHA}@*)
Subject: (*@\textit{commit subject}@*)

Body:
(*@\textit{commit message body}@*)

Linked issue body:
(*@\textit{linked issue body, or (none)}@*)

Linked PR body:
(*@\textit{linked pull request body, or (none)}@*)

File-change stat:
(*@\textit{file-change stat}@*)

Diff:
(*@\textit{the commit diff, head-and-tail truncated to the configured budget}@*)
  \end{lstlisting}
\end{promptbox}

\promptheading{Bug Extractor}{prompt:splitter}{\terra}
Bug extractor of Step 2: it turns one bug-fixing commit into zero or more candidate bugs, one per distinct root cause.

\begin{promptbox}{System}{systemborder}
  \begin{lstlisting}[basicstyle=\ttfamily\fontsize{7}{7.6}\selectfont,breaklines=false,showstringspaces=false,language=,mathescape=false,escapeinside={(*@}{@*)}]
You are splitting a bug-fix commit into one or more **logical bugs**.

A logical bug means a distinct underlying defect in the pre-fix code. Do not split by symptom, affected 
    entry point, failing test, or user-visible behavior. Multiple symptoms caused by the same root cause
     are one bug.

The commit is purported to fix at least one bug. Some commits bundle multiple unrelated fixes, and some 
    get mislabeled by the upstream classifier. Your job is to:

1. Decide whether this commit really fixes a bug at all.
2. If yes, identify whether it fixes one distinct root-cause bug or several distinct root-cause bugs.
3. Emit one record per logical bug.
4. For each bug, pinpoint the **parent-side** line range that contains the buggy code: the lines the fix
     removes or modifies. For a hunk that only **adds** lines (a missing check or guard the fix 
    inserts), the parent has no buggy line to point at — set `parent_lines` to the insertion point `[A, 
    A]`, where `A` is the parent line the new code follows (the last context line shown immediately 
    before the `+` run; use `[1, 1]` when the insertion precedes the first line). Reserve `[0, 0]` for a
     bug you genuinely cannot localize to any hunk.

**Be CONSERVATIVE about splitting.** Default to one bug per commit.

Only split when it is obvious that the commit fixes two or more distinct bugs: separate root causes, 
    separate broken invariants, unrelated correctness issues, or independent code paths that do not 
    share the same fix rationale.

Do not split merely because one bug has multiple symptoms.

Examples:
- A URL parsing bug causes both the CLI and the backend server to crash on certain URLs. This is one 
    bug: the root cause is the URL parsing defect. Do not create separate records for the CLI crash and 
    server crash.
- A commit fixes URL parsing and also fixes input sanitization. These are two bugs if the fixes address 
    independent defects.

**One logical bug, even if it spans files**:
- Caller/callee mismatch: the caller passes the wrong shape and the callee misinterprets it. List both 
    files, with the file where the wrong behavior originates first.
- Shared helper repaired in one place that fixes several call sites.
- One root cause that manifests through several entry points, commands, APIs, or user-visible failures.
- A primary source file plus its test file (`tests/...` or `*_test.py`). **Omit the test file from 
    `files`** — the test only verifies the fix, it is not part of the buggy code.

**Split into multiple bugs only** when:
- The commit message itself lists distinct bugs, and the diff supports that they are separate root 
    causes.
- Different files address visibly unrelated defects, such as a parser bug and an unrelated auth bug 
    bundled in one commit.
- Hunks touch unrelated functions whose only commonality is being in the same module.
- The fixes correct independent wrong assumptions, checks, invariants, or data-handling paths.

**Reject the commit entirely (emit `"bugs": []`)** when the diff is not actually fixing a bug:
- Pure refactor, rename, formatting, or style change.
- Documentation, comments, or type hints with no behavioral change.
- Test-only edits where the fix is not in this commit.
- Performance improvement with no correctness impact.
- Dependency or configuration bump with no code-behavior change.
- Adding new functionality without correcting prior behavior.

Inputs you will see: the commit message, including subject and body; the linked issue body, if any; the 
    linked PR body, if any; and the unified diff. Return ONLY a JSON object:

Follow the generated JSON Schema appended to this prompt. Its field descriptions are part of the output 
    contract.

```json
{
  "bugs": [
    {
      "summary": "One-sentence description of what is wrong (snapshot-fresh, present tense — not 'was 
    wrong').",
      "files": ["path/to/primary_file.py", "path/to/secondary_file.py"],
      "hunks": [
        {"file": "path/to/primary_file.py", "parent_lines": [142, 156]}
      ],
      "bug_kind_hint": "off-by-one in loop bound"
    }
  ]
}
```

If the bug is truly cross-cutting (no single file dominates and multiple files share the wrong 
    behavior), use a single record with `files` listing every affected path and `hunks` containing one 
    entry per file. Do not invent splits.
  \end{lstlisting}
\end{promptbox}

\begin{promptbox}{User}{userborder}
  \begin{lstlisting}[basicstyle=\ttfamily\fontsize{7}{7.6}\selectfont,breaklines=false,showstringspaces=false,language=,mathescape=false,escapeinside={(*@}{@*)}]
Commit subject: (*@\textit{commit subject}@*)

Commit body:
(*@\textit{commit message body}@*)

Linked issue body:
(*@\textit{linked issue body, or (none)}@*)

Linked PR body:
(*@\textit{linked pull request body, or (none)}@*)

Diff:
(*@\textit{the fix diff, head-and-tail truncated to the configured budget}@*)
  \end{lstlisting}
\end{promptbox}

\promptheading{Bug Presence Check}{prompt:agent-gate}{\terra}
Presence check of Step 4, run at high reasoning effort. The agent works in a read-only checkout at the snapshot with the later history available, and drops a candidate bug only when it can establish that the snapshot was not already wrong.

\begin{promptbox}{System}{systemborder}
  \begin{lstlisting}[basicstyle=\ttfamily\fontsize{7}{7.6}\selectfont,breaklines=false,showstringspaces=false,language=,mathescape=false,escapeinside={(*@}{@*)}]
**Output format — read this first.** Your response MUST be exactly one JSON object matching the 
    generated JSON Schema appended to this prompt — its field descriptions are part of the output 
    contract. No prose, no preamble, no markdown fences, no explanation outside the JSON. The very first
     character of your response must be `{`. The very last character must be `}`. Put all reasoning 
    inside the `reasoning` field. Any text outside the JSON object is a protocol violation.

---

You are deciding one candidate in a benchmark of historical bugs. A commit fixed something; the 
    benchmark pins the code at an earlier **snapshot** revision and claims the bug was already there. 
    Your job is to rule on that claim: **was this a real bug at the snapshot revision (`keep_A` or 
    `keep_B`), or was it not (`drop_C` or `drop_not_present`)?**

Nobody has judged this candidate before you. There is no earlier verdict to agree with or overturn, and 
    no queue clause telling you what to look at. You have the repository, you run the checks below on 
    every candidate, and your citations decide.

## Your working directory

You are running **read-only** in a checkout of this repository, pinned at the snapshot revision: `HEAD` 
    is the snapshot. Do not attempt to write anything — a command that wants to write the index will 
    fail.

Read-only bounds what you may change, not what you may run. Besides the git commands below, **you may 
    run the snapshot's own code**: import a module and call it, enumerate a registry, or run the fix's 
    own reproducer against the snapshot. A reproducer that passes at the snapshot is direct evidence the
     defect is not there, and a wrong result you produce yourself is the strongest realizability answer 
    available. Cite what you ran as `execution`, with the command as the locator and its output as the 
    quote.

The object store also holds history **after** the snapshot, including the fix commit itself and the 
    branch it landed on. That is the point of running you here: questions of the form "did this exist 
    yet at the snapshot?" are decidable, and you must decide them by running git, not by inference.

The commands that answer most candidates:

- `git show <fix_sha>` — the fix diff. `git show <fix_sha> --stat` first if it is large.
- `git show HEAD:<path>` — the file as it stood at the snapshot. Use `git grep -n <name> HEAD -- <path>`
     to find a line.
- `git log -S '<token>' --oneline HEAD..<fix_sha> -- <path>` — commits **after** the snapshot that added
     or removed that token. A hit here is a candidate for the commit that introduced the repaired 
    precondition.
- `git log -S '<token>' --oneline HEAD -- <path>` — the same search **before** the snapshot. A hit here 
    means the token was already in play.
- `git merge-base --is-ancestor <sha> HEAD && echo ANCESTOR || echo NOT-ANCESTOR` — the decisive 
    temporal test. Run it on the commit you identified and report both the command and its answer.
- `git grep -n <name> HEAD` — every site of an identifier at the snapshot, across the repo. This is how 
    you refute a claim that something is absent.

The prompt below gives you the candidate's pinned line ranges, the splitter's summary, the fix commit 
    message, the linked issue and PR bodies, machine-generated mapping notes and deterministic 
    provenance signals. It does **not** give you a symbol scan or an embedded diff: run `git grep` and 
    `git show` yourself, which is better evidence than either — line-level, uncapped, and taken from the
     snapshot rather than from a file state months later.

## The two errors this stage exists to catch

The dominant failure is **temporal**: the fix repairs code, a parameter, an option, a compatibility 
    target or a call path that arrived **after** the snapshot. The fix's diff looks like a real bug fix 
    — because it is one, against the *later* code. The candidate is still wrong, because the defect 
    could not happen at the snapshot.

The second failure runs the other way: a real, triggerable defect is **killed** by an absence claim 
    built from a partial window. The code called missing is in the file, a few hundred lines outside the
     pinned range. One `git grep` settles that too.

Both are decidable here. Neither is decidable by reading the excerpt.

## Required checks

Run every check below and add at least one `citations` entry naming it in that entry's `check` field. A 
    required check with no citation of its own, or one whose citation is a paraphrase instead of a 
    command's output, invalidates the verdict. Name the check that settled it in `decisive_check`; it 
    must be one you cited.

- **`fix_diff`** — read `git show <fix_sha>`. Cite the hunk that carries the repair.
- **`fix_shape`** — read `git show <fix_sha> --stat`, then answer four questions about the change 
    itself. Does it add a test, or did it have to **rewrite an existing one's assertions**? Does it 
    touch documentation or a declared support surface? Does the linked issue carry a label affirming a 
    defect? Does it leave equivalent call sites untouched? Cite the stat line, the test hunk, or the 
    absence of one. This check asks what the author treated as required; it never says what is at the 
    snapshot.
- **`snapshot_file`** — read the candidate's file at the snapshot around the pinned line ranges, and 
    wider if the ranges look like a hull. Cite a numbered line that bears on the claim. If the file does
     not exist at the snapshot, that itself is the citation.
- **`temporal_ancestry`** — identify the commit that introduced the precondition the fix repairs, then 
    run `git merge-base --is-ancestor` on it against the snapshot. Cite the commit (as 
    `introducing_commit`) and the ancestry answer (as `ancestry`), and repeat both in the top-level 
    `introducing_commit` and `ancestry_answer` fields whenever you run this check, whether or not it 
    decides the resolution. If no such commit exists — the precondition was always there — say so and 
    cite the pre-snapshot `git log -S` that shows it.
- **`symbol_sites`** — `git grep -n` at the snapshot every identifier that is load-bearing for the 
    defective path. Cite `path:line` and say whether the hit is a definition, a use, or a comment. A 
    name that genuinely returns nothing is the strongest evidence of absence there is; report that as 
    the citation. Imports, typing aliases and helpers the fix itself introduces are expected to be 
    absent and prove nothing.
- **`upstream_text`** — read the commit message and the linked issue and PR bodies given below, and 
    quote the sentence that bears on the change. Read them for **leads to chase in the checkout**, above
     all a claim that some later PR or commit caused the problem. The sentence itself settles nothing: 
    `upstream_text` may corroborate a resolution and may never be its `decisive_check`.

## The four resolutions

- **`keep_A`** — the defect exists at the snapshot, is triggerable by an input or call path that exists 
    at the snapshot, and a competent reviewer reading the snapshot code could identify it as wrong.
- **`keep_B`** — the same, except recognizing it needs outside context: a runtime's quirk, a documented 
    support commitment, a spec corner, a real nonconforming sample. B is a real bug; C is not.
- **`drop_C`** — the snapshot behavior was not a bug.
- **`drop_not_present`** — the defective code, control flow or data path the fix rewrites did not exist 
    at the snapshot at all.

## What decides it: keep freely, drop only on a cited ground

There is no prior verdict here, so nothing decides a candidate by default except the evidence you 
    gather. The two directions do not carry the same burden.

**A keep needs the realizability answer every verdict carries.** Answer it in `snapshot_misbehavior` 
    before you choose a resolution: would the pre-fix code crash or produce a wrong result on a 
    realizable input at the snapshot? Realizable means the triggering input or call path exists *at the 
    snapshot* — place it there with `git grep` or `git show HEAD:<path>`, never from the fix diff. Name 
    the concrete crash or wrong result and the path that reaches it. A `keep_A` or `keep_B` that names 
    one is a confident keep; one that cannot is a keep you should mark low-confidence and explain, never
     a drop you cannot ground.

**A drop must name one of exactly four grounds in `drop_ground`, and cite evidence that establishes 
    it.** Each ground is a fact about the snapshot code, and you must state it as one. How the commit or
     its PR describes itself is never a ground.

- **`precondition_postdates`** → `drop_C`. The trigger the fix repairs arrived after the snapshot. 
    Establish it with an `ancestry` citation answering `NOT-ANCESTOR` on the commit that introduced the 
    precondition, or with a `symbol_site` citation showing the load-bearing name absent repo-wide at 
    `HEAD`. An `ancestry_answer` of `ANCESTOR` removes this ground outright — the precondition was 
    already there, and you must then judge the defect on its merits.
- **`deliberate_rejection`** → `drop_C`. The snapshot rejects the input on purpose, through validation 
    you can point at, and the fix relaxes it for broken producers. Establish it with a `snapshot_line` 
    citation of that validation. A crash the snapshot performs *unintentionally* on malformed input is a
     robustness defect, not a deliberate rejection.
- **`capability_absent`** → `drop_C`. The fix adds a capability, a compatibility target or a support 
    surface the snapshot never had, and its absence broke nothing the snapshot itself promised. 
    Establish it with a `snapshot_line` or `symbol_site` citation — of the project's own declared 
    surface (its build target, its runtime or platform configuration, its documented commitments, **its 
    own committed tests**), or of equivalent call sites the fix left untouched. Code that is correct 
    against everything the project declares it supports is not defective because some environment 
    outside that surface cannot run it.
- **`code_absent`** → `drop_not_present`. The code, control flow or data path the fix rewrites is not at
     the snapshot: the snapshot has a different implementation, already contains the check the fix adds,
     or lacks the code the fix modifies. Establish it with a `symbol_site` or `snapshot_line` citation. 
    Prefer this over `precondition_postdates` when the code itself is missing, not merely the 
    requirement.

**If you cannot name and cite a ground, keep.** That is not deference to anyone — it is the burden of 
    proof. A keep that turns out wrong leaves one bad record in a benchmark that reports its 
    contamination; an unproven drop removes a real bug for good.

Every ground above is settled by locating something, or by failing to locate it: a commit's ancestry, a 
    validation site, a declared support surface, an absent symbol. **There is deliberately no ground for
     "this changes no observable behavior".** That claim rests on judging two code paths equivalent, 
    which no command can settle, and a reader who believes it is often reading only part of what the 
    change repairs. So if the change looks like pure hygiene to you and none of the four grounds fits, 
    the answer is `keep_A` or `keep_B`, not a drop. Say in `snapshot_misbehavior` that you could not 
    establish a concrete misbehavior, answer with low confidence, and name the drop you considered in 
    `runner_up_resolution`. Reporting an uncertain keep costs the benchmark one flagged record; the drop
     you cannot establish costs it a real bug.

## Five standing rules

**The bug summary is a claim, not a finding.** An earlier automated step wrote it by reading the *fix 
    commit*, so it describes this candidate as a defect before anyone judged whether it was one, and its
     wording carries the fixer's framing. Verify that the defect it describes is the defect at the 
    pinned ranges: a commit often repairs several things, and the summary can name a different one than 
    the ranges hold. And when the summary names more than one defect, account for every one of them 
    before you drop — ruling out one and ignoring the rest is not a ground.

**The pinned ranges are a pointer, and sometimes a wrong one.** When they hold code the summary 
    describes, judge that code. When they hold nothing it describes, that is not absence: a defect a few
     hundred lines outside them is the commonest way a real bug is killed here. `git grep` the defect 
    the summary names across the whole snapshot, and if it is present and triggerable there, that is a 
    keep — say in `reasoning` that the ranges pointed elsewhere.

**The fix diff is not evidence about what is at the snapshot.** It is taken against a parent revision 
    months later. Its context lines routinely contain parameters, options, branches and helpers the 
    snapshot lacks, and a realizability argument built out of them is the single most common way this 
    benchmark leaks a non-bug. Code you have seen only in the fix diff is post-snapshot code until you 
    place it at the snapshot with `git grep` or `git show HEAD:<path>`. Its tests and reproducers run 
    against the later code and place no input at the snapshot.

**The fix diff is evidence about what the project treated as required.** What the author changed, what 
    the author left alone, and which tests the author had to touch are facts about the fix, and they 
    bear on whether the snapshot behavior was *wrong* rather than on whether it was *there*. That is the
     `fix_shape` check. Use the diff for that question and never for the other one.

**A real crash or wrong result stays a keep however the change was packaged.** A `STY:`, `MAINT:`, 
    `ROB:` or "gracefully handle" framing, a maintenance label, a typing sweep, a release merge — none 
    of these demote a defect that produces a crash or a wrong answer on input the snapshot accepts. 
    Neither does the opposite packaging: a `feat:` or `ENH:` prefix, an "adds support for" body, a 
    version-milestone note or a release-notes line. A commit can add a capability and repair a defect at
     once, and which one this candidate names is a question about the snapshot code, not about the 
    commit's own words.

**Absence must be proven, never inferred from the pinned ranges.** The ranges are a partial view; most 
    of the file is never named. Shifted line numbers, renamed identifiers and reorganised surroundings 
    are not absence — a variant spelling of the same defect is the same defect. Only a repo-wide `git 
    grep` that returns nothing for a load-bearing name supports `code_absent`.

## Reading the deterministic provenance signals

Each is a prior; none decides on its own, and none is a `drop_ground`.

- `fix arrived via a merge commit`: about a quarter of true records arrive via merges — a merge alone 
    means nothing. It becomes evidence in combination: a release or `next`-branch merge subject plus a 
    feature-shaped diff and no bug-affirming issue is the release-merge pattern, and it points you at 
    `temporal_ancestry`.
- `commit-subject type prefix`: upstream's own label (`BUG:`, `ROB:`, `ENH:`, `STY:`, `DEV:`, `MAINT:`).
     Most `ROB:` commits repair genuine crash-on-malformed-input defects; some relax deliberate 
    validation. The prefix picks which check to run, never the answer.
- `SZZ introducer status`: `ancestral` — the blamed lines existed at the snapshot. `none_ancestral` — 
    the blamed commits are not ancestors of the snapshot, which *suggests* post-snapshot origin but does
     not prove it, since a pre-snapshot commit merged late is non-ancestral without postdating. Either 
    way, run `temporal_ancestry` and let git answer. `no_introducers` — no signal: added-only and 
    formatting-only fixes blame nothing.
- `linked issue labels` / `linked PR labels`: upstream's own framing, verbatim. "Not bug", wontfix, 
    enhancement and feature framing all point at a check; none of them is a ground. Labels can be sparse
     or stale. Read the label text rather than any bug-affirming flag derived from it: that flag is a 
    substring match, so a label reading "not bug" sets it.

## How to weigh what you find

- An ancestry answer is the strongest evidence available to you. `NOT-ANCESTOR` on the commit that 
    introduced the repaired precondition establishes `precondition_postdates` unless something else 
    places the precondition at the snapshot; `ANCESTOR` removes the temporal explanation entirely.
- A `git grep` hit at the snapshot refutes an absence claim outright, and the kind of hit matters: a 
    definition or a use places the code there, a match inside a comment or a string does not.
- Upstream text may corroborate a code-level finding; it never decides one. If you cannot state the 
    resolution as a fact about the snapshot code, the upstream wording has not given you one.
- What the fix left alone is evidence. A commit that repairs one site while leaving equivalent sites 
    untouched is usually accommodating one caller rather than repairing a defect the project recognises;
     `git grep` the same construct across the snapshot and say what you found. The mirror holds too: a 
    fix that changes every site of a construct is repairing the construct.
- **A test in the snapshot tree is part of the snapshot's declared surface.** A snapshot test that 
    asserts the behavior the fix changes is a `snapshot_line` citation, and it establishes 
    `capability_absent` or `deliberate_rejection`: the project had pinned that behavior, so changing it 
    is a decision rather than a repair. It runs the other way too — a fixture committed at the snapshot 
    that the snapshot code mishandles places a realizable input there without touching the fix diff, and
     that is a keep.
- A fix that had to **rewrite an existing test's assertions** is the sharpest form of the same evidence:
     the behavior it changed was codified. A fix that only **adds** a test pinning corrected behavior of
     a function that already exists shows the opposite. A fix that keeps the old default and adds an 
    opt-in beside it, or adds a new entry point beside an untouched old one, is adding a capability. A 
    fix carrying its own backward-compatibility guard is telling you the patch is a no-op on the 
    snapshot-era target.
- Where the checks genuinely do not settle it, say so in `reasoning`, answer with low confidence, and 
    name the close rival in `runner_up_resolution`. Do not manufacture a reading.

## Examples (synthetic — the shas, paths and names below belong to no real repository)

**Example 1** — a fix corrects a bounds comparison inside a per-chunk quota accounting path.
```json
{"citations": [
   {"check": "fix_diff", "kind": "fix_diff", "locator": "git show 7c3f19ab -- pkg/archive/chunker.py", 
    "quote": "-    if used > chunk_quota:\n+    if used >= chunk_quota:"},
   {"check": "fix_shape", "kind": "fix_diff", "locator": "git show 7c3f19ab --stat", "quote": " 
    pkg/archive/chunker.py    | 2 +-\n tests/test_chunker.py     | 41 +++++++"},
   {"check": "snapshot_file", "kind": "snapshot_line", "locator": "pkg/archive/chunker.py:88", "quote": 
    "    while remaining:"},
   {"check": "symbol_sites", "kind": "symbol_site", "locator": "git grep -n chunk_quota HEAD", "quote": 
    "(no output)"},
   {"check": "temporal_ancestry", "kind": "introducing_commit", "locator": "git log -S 'chunk_quota' 
    --oneline HEAD..7c3f19ab -- pkg/archive/chunker.py", "quote": "e28d4f60 add per-chunk quota 
    accounting"},
   {"check": "temporal_ancestry", "kind": "ancestry", "locator": "git merge-base --is-ancestor e28d4f60 
    HEAD", "quote": "NOT-ANCESTOR"},
   {"check": "upstream_text", "kind": "upstream_text", "locator": "PR body", "quote": "follow-up to the 
    quota work landed last month"}],
 "resolution": "drop_C",
 "drop_ground": "precondition_postdates",
 "decisive_check": "temporal_ancestry",
 "introducing_commit": "e28d4f60",
 "ancestry_answer": "NOT-ANCESTOR",
 "snapshot_misbehavior": "None. The quota the off-by-one governs does not exist at the snapshot — the 
    name has no occurrence anywhere in the tree — so no input reaches the comparison.",
 "decisive_evidence": "e28d4f60, which introduced the quota accounting the fix repairs, is not an 
    ancestor of the snapshot",
 "runner_up_resolution": "drop_not_present",
 "confidence": "high",
 "reasoning": "The repaired branch belongs to quota accounting added after the snapshot, and the 
    snapshot tree has no occurrence of the name at all, so the defect had no way to trigger there."}
```

**Example 2** — a fix flips `return LoadFailure(...)` to `raise LoadFailure(...)` deep inside a large 
    annotation sweep; the pinned range covers only the file's imports.
```json
{"citations": [
   {"check": "fix_diff", "kind": "fix_diff", "locator": "git show 93b1c07e -- lib/manifest/loader.py", 
    "quote": "-        return ManifestError(path)\n+        raise ManifestError(path)"},
   {"check": "symbol_sites", "kind": "symbol_site", "locator": "git grep -n load_manifest HEAD -- 
    lib/manifest/loader.py", "quote": "742:def load_manifest(self, path):"},
   {"check": "fix_shape", "kind": "fix_diff", "locator": "git show 93b1c07e --stat", "quote": "34 files 
    changed, 902 insertions(+), 871 deletions(-); no test file touched"},
   {"check": "snapshot_file", "kind": "snapshot_line", "locator": "lib/manifest/loader.py:751", "quote":
     "            return ManifestError(path)"},
   {"check": "snapshot_file", "kind": "execution", "locator": "python -c \"import lib.manifest.loader as
     m; print(type(m.load_manifest(None, 'truncated.mf')))\"", "quote": "<class 
    'lib.manifest.loader.ManifestError'>"},
   {"check": "temporal_ancestry", "kind": "ancestry", "locator": "git merge-base --is-ancestor a4f28d15 
    HEAD", "quote": "ANCESTOR"},
   {"check": "upstream_text", "kind": "upstream_text", "locator": "commit message", "quote": "STY: 
    annotate the loader module"}],
 "resolution": "keep_A",
 "drop_ground": null,
 "decisive_check": "snapshot_file",
 "introducing_commit": "a4f28d15",
 "ancestry_answer": "ANCESTOR",
 "snapshot_misbehavior": "Loading a truncated manifest through the public path returns the error object 
    instead of raising it, so the caller treats the failure as a value and carries on with a half-loaded
     manifest.",
 "decisive_evidence": "lib/manifest/loader.py:751 at the snapshot returns the exception object instead 
    of raising it, inside the function defined at 742",
 "runner_up_resolution": null,
 "confidence": "high",
 "reasoning": "The function the fix rewrites exists at the snapshot at line 742, far outside the pinned 
    range, and at 751 it returns the error rather than raising it, so the caller's failure is silently 
    swallowed. The sweep packaging does not change that."}
```

**Example 3** — a fix makes a reader tolerate a repeated header field the format forbids.
```json
{"citations": [
   {"check": "fix_diff", "kind": "fix_diff", "locator": "git show 05e9b3ca -- src/headers.py", "quote": 
    "-        raise HeaderError(...)\n+        continue"},
   {"check": "snapshot_file", "kind": "snapshot_line", "locator": "src/headers.py:4417", "quote": "
            raise HeaderError(f\"repeated field at {pos}\")"},
   {"check": "fix_shape", "kind": "fix_diff", "locator": "git show 05e9b3ca -- src/headers_test.py", 
    "quote": "-    assert_raises(HeaderError, parse, doubled)\n+    assert parse(doubled).fields == 1"},
   {"check": "symbol_sites", "kind": "symbol_site", "locator": "git grep -n HeaderError HEAD -- 
    src/headers.py", "quote": "4417:        raise HeaderError(f\"repeated field at {pos}\")"},
   {"check": "temporal_ancestry", "kind": "ancestry", "locator": "git merge-base --is-ancestor d71a6e84 
    HEAD", "quote": "ANCESTOR"},
   {"check": "upstream_text", "kind": "upstream_text", "locator": "issue body", "quote": "our writer 
    emits the field twice; please accept these files"}],
 "resolution": "drop_C",
 "drop_ground": "deliberate_rejection",
 "decisive_check": "snapshot_file",
 "introducing_commit": "d71a6e84",
 "ancestry_answer": "ANCESTOR",
 "snapshot_misbehavior": "None. src/headers.py:4417 raises deliberately on a repeated field, which the 
    format forbids; the snapshot rejects the input on purpose rather than mishandling it.",
 "decisive_evidence": "src/headers.py:4417 at the snapshot raises deliberately, through validation of a 
    rule the format states",
 "runner_up_resolution": "keep_B",
 "confidence": "high",
 "reasoning": "The snapshot rejects the input through explicit validation of a stated rule, and the fix 
    relaxes it for a nonconforming writer, which is a leniency change rather than a defect repair. The 
    ancestry answer rules out the temporal reading."}
```

**Example 4** — a fix rewrites one call to avoid a language builtin so an embedded runtime can load the 
    module.
```json
{"citations": [
   {"check": "fix_diff", "kind": "fix_diff", "locator": "git show 1a9f47c2 -- src/shaping/shaper.ts", 
    "quote": "-    const tail = runs.findLast(isVisible)\n+    const tail = 
    runs.slice().reverse().find(isVisible)"},
   {"check": "snapshot_file", "kind": "snapshot_line", "locator": "src/shaping/shaper.ts:214", "quote": 
    "    const tail = runs.findLast(isVisible)"},
   {"check": "snapshot_file", "kind": "snapshot_line", "locator": "tsconfig.json:4", "quote": "    
    \"target\": \"ES2023\","},
   {"check": "fix_shape", "kind": "fix_diff", "locator": "git show 1a9f47c2 --stat", "quote": " 
    src/shaping/shaper.ts | 2 +-\n 1 file changed, 1 insertion(+), 1 deletion(-)"},
   {"check": "symbol_sites", "kind": "symbol_site", "locator": "git grep -n findLast HEAD -- src", 
    "quote": "src/shaping/shaper.ts:214\nsrc/layout/lines.ts:88\nsrc/layout/lines.ts:132"},
   {"check": "temporal_ancestry", "kind": "ancestry", "locator": "git merge-base --is-ancestor 6b8e05d3 
    HEAD", "quote": "ANCESTOR"},
   {"check": "upstream_text", "kind": "upstream_text", "locator": "PR body", "quote": "our embedded 
    runtime has no findLast, so I patched it locally"}],
 "resolution": "drop_C",
 "drop_ground": "capability_absent",
 "decisive_check": "snapshot_file",
 "introducing_commit": "6b8e05d3",
 "ancestry_answer": "ANCESTOR",
 "snapshot_misbehavior": "None on any runtime the project declares. The call is correct against the 
    declared target, and only an embedded runtime outside that surface lacks the builtin.",
 "decisive_evidence": "tsconfig.json:4 at the snapshot declares the target the call is valid against, 
    and the fix left two equivalent call sites in src/layout/lines.ts untouched",
 "runner_up_resolution": "keep_B",
 "confidence": "high",
 "reasoning": "The snapshot call is valid against the project's own declared target, and the fix changed
     one site while leaving two identical uses in place, so it accommodates one runtime outside the 
    declared surface rather than repairing a defect the project recognises."}
```

---

Now run the required checks, then emit the JSON object. Do not write any text before the opening `{`. Do
     not write any text after the closing `}`. Do not wrap the JSON in a code fence.
  \end{lstlisting}
\end{promptbox}

\begin{promptbox}{User}{userborder}
  \begin{lstlisting}[basicstyle=\ttfamily\fontsize{7}{7.6}\selectfont,breaklines=false,showstringspaces=false,language=,mathescape=false,escapeinside={(*@}{@*)}]
Bug summary (from splitter): (*@\textit{the splitter's one-line summary of the candidate bug}@*)
Snapshot file: (*@\textit{snapshot file}@*)

Working directory: a checkout of (*@\textit{repository}@*) pinned at the snapshot; HEAD is (*@\textit{snapshot commit}@*).
Fix commit: (*@\textit{fix commit}@*)
Fix commit's first parent (the diff base): (*@\textit{first parent of the fix commit}@*)

Pinned candidate ranges (snapshot coordinates):
(*@\textit{one line per pinned candidate range, in snapshot coordinates}@*)

Candidate locations (numbered snapshot source, one block per candidate):
```
(*@\textit{numbered snapshot source, one block per candidate location}@*)
```

Fix commit message:
(*@\textit{fix commit message}@*)

Linked issue body:
(*@\textit{linked issue body, or (none)}@*)

Linked PR body:
(*@\textit{linked pull request body, or (none)}@*)

Mapping notes:
(*@\textit{one line per mapping note, or (none)}@*)

Deterministic provenance signals (machine-derived priors, not verdicts):
- fix arrived via a merge commit: (*@\textit{yes or no}@*)
- commit-subject type prefix: (*@\textit{commit-subject type prefix, or (none)}@*)
- SZZ introducer status: (*@\textit{the SZZ introducer status line}@*)
- linked issue labels: (*@\textit{linked issue labels, or (none)}@*)
- linked PR labels: (*@\textit{linked pull request labels, or (none)}@*)
  \end{lstlisting}
\end{promptbox}

\promptheading{Bug Report Synthesizer}{prompt:synthesiser}{\terra}
Report synthesizer of Step 5. It writes the snapshot-time bug report of a surviving candidate bug against the snapshot code, without citing the fix.

\begin{promptbox}{System}{systemborder}
  \begin{lstlisting}[basicstyle=\ttfamily\fontsize{7}{7.6}\selectfont,breaklines=false,showstringspaces=false,language=,mathescape=false,escapeinside={(*@}{@*)}]
You are writing a **bug finding record** from a fix commit. The record must describe the bug **as it 
    exists at a specific code snapshot**, using present-tense language — not as a historical fix.

You will be given:

- The bug summary from the splitter step.
- A numbered list of **candidate locations** — grounded line ranges the fix touched — each shown as 
    `[index] path:start-end` followed by the snapshot source with absolute line numbers.
- The unified diff of the fix (showing what changed).
- The commit message, linked issue body, and linked PR body, if available.

Write the record as if you discovered the bug fresh by reading the snapshot code. The audience is a 
    security/code reviewer auditing the snapshot.

First decide **where** the bug is: set `primary_candidate_index` to the index of the single candidate 
    whose lines a reviewer would point to as the defect site. Ignore candidates that are unrelated 
    refactors, renames, or reformatting bundled into the same commit. Return `0` when only one candidate
     is listed. Then write the `title` and `description` about that candidate's lines.

**Omission bugs.** When the fix diff is purely additive (it only inserts lines — a guard, check, or 
    branch — with no `-` deletions in the relevant hunk), the defect is the *absence* of that code at 
    the snapshot. The candidate lines show the region where the missing check belongs, not a defective 
    visible statement. Write the `title` and `description` as an omission — "missing X check in Y before
     Z" — describing what is not there and the consequence, rather than pointing at a correct-looking 
    line as if it were wrong.

**Hard rules**:

- Title and description must be present tense and describe the bug as currently present in the snippet.
- Do not reference the fix at all in `title` or `description`. Phrases like "was fixed by", "the patch",
     "the commit", "previously", "before this change" are forbidden in those two fields.
- Cite identifiers and line numbers from the snapshot snippet, not from the diff.
- The only field where you may reference the fix is `suggested_fix` — there it is acceptable (and 
    useful) to say "replace `xs[i]` with `xs[i % len(xs)]`" etc.

### Examples

**Title — bad (past tense, references fix)**:
> "Fixed off-by-one when iterating range bounds in `compile_loop`"

**Title — good (present tense, names bug + locus)**:
> "Off-by-one in `compile_loop` skips the final element of `bounds`"

**Description — bad (narrates the patch)**:
> "The original code subtracted 1 from the length, which was wrong. The commit changes it to use the 
    full length, fixing the iteration."

**Description — good (snapshot-fresh)**:
> "`compile_loop` iterates `range(len(bounds) - 1)` at line 142, which omits the final element of 
    `bounds`. When a caller passes a non-empty list, the last item is silently dropped. Triggered on 
    every call with `len(bounds) >= 1`."

Return ONLY a JSON object matching the generated JSON Schema appended to this prompt. Its field 
    descriptions are part of the output contract.
  \end{lstlisting}
\end{promptbox}

\begin{promptbox}{User}{userborder}
  \begin{lstlisting}[basicstyle=\ttfamily\fontsize{7}{7.6}\selectfont,breaklines=false,showstringspaces=false,language=,mathescape=false,escapeinside={(*@}{@*)}]
Bug summary (from splitter): (*@\textit{the splitter's one-line summary of the candidate bug}@*)
Snapshot file: (*@\textit{snapshot file}@*)

Candidate locations (numbered snapshot source, one block per candidate):
```
(*@\textit{numbered snapshot source, one block per candidate location}@*)
```

Fix commit message:
(*@\textit{fix commit message}@*)

Linked issue body:
(*@\textit{linked issue body, or (none)}@*)

Linked PR body:
(*@\textit{linked pull request body, or (none)}@*)

Fix diff:
(*@\textit{the fix diff}@*)
  \end{lstlisting}
\end{promptbox}

\promptheading{Deduplication Judge}{prompt:record-dedup}{\geminiflash}
Deduplication judge of Step 6. It carries the same record block as the matcher judge of \cref{prompt:matcher-judge}, under a stricter rule: the two records must describe one root-cause defect rather than merely admit one fix, and identical snippets and coincident line ranges count as weak evidence because a snippet is grown to its enclosing function.

\begin{promptbox}{Prompt}{userborder}
  \begin{lstlisting}[basicstyle=\ttfamily\fontsize{7}{7.6}\selectfont,breaklines=false,showstringspaces=false,language=,mathescape=false,escapeinside={(*@}{@*)}]
You are reviewing two historical bug records mined from the commit history of the same repository 
    snapshot. Each was reconstructed from a nearby post-snapshot fix commit. Decide whether the two 
    records describe the **same underlying root-cause defect**.

Two records are duplicates only if they are the *same flaw* — a single fix resolves both because they 
    are one and the same defect, not merely because the two are related, adjacent, or one region 
    contains the other. Two genuinely different defects that happen to live in the same function are NOT
     duplicates.

Important cautions when weighing the evidence:
- The code snippets are grown to the enclosing function or class, so two distinct defects in the same 
    function can carry byte-identical snippets. Do not treat snippet overlap as evidence of duplication.
- Line ranges are approximate and can coincide for the same reason. Weigh the titles and descriptions — 
    the described mechanism, trigger, and faulty behaviour — over snippet or line-range overlap.
- When the two descriptions point at different mechanisms or different lines of the same function, treat
     them as distinct defects.

=== BUG A ===
ID: (*@\textit{id of A}@*)
Title: (*@\textit{title of A}@*)
File: (*@\textit{file of A}@*):(*@\textit{first line of A}@*)-(*@\textit{last line of A}@*)
Description:
(*@\textit{description of A}@*)

Code snippet:
```
(*@\textit{code snippet of A}@*)
```

=== BUG B ===
ID: (*@\textit{id of B}@*)
Title: (*@\textit{title of B}@*)
File: (*@\textit{file of B}@*):(*@\textit{first line of B}@*)-(*@\textit{last line of B}@*)
Description:
(*@\textit{description of B}@*)

Code snippet:
```
(*@\textit{code snippet of B}@*)
```

First, briefly reason through the evidence: compare the described defects, their triggers, and the 
    faulty behaviour, treating identical snippets and coincident line ranges as weak signals. Then 
    output your conclusion as a single fenced JSON block matching the JSON Schema appended below; its 
    field descriptions are part of the output contract.
  \end{lstlisting}
\end{promptbox}

\needspace{8\baselineskip}
\subsection{Benchmark Validation}

\promptheading{False-Positive Auditor}{prompt:fp-auditor}{\terra}
False-positive auditor of the benchmark validation, run at high reasoning effort over every shipped reference bug. It works read-only in a checkout pinned at the snapshot, with the later history reachable, and may drop a record only on a ground it establishes with a citation of an accepted kind; the fix's commit message, issue and pull-request text are inadmissible as citations.

\begin{promptbox}{System}{systemborder}
  \begin{lstlisting}[basicstyle=\ttfamily\fontsize{7}{7.6}\selectfont,breaklines=false,showstringspaces=false,language=,mathescape=false,escapeinside={(*@}{@*)}]
**Output format — read this first.** Your response MUST be exactly one JSON object matching the 
    generated JSON Schema appended to this prompt — its field descriptions are part of the output 
    contract. No prose, no preamble, no markdown fences, no explanation outside the JSON. The very first
     character of your response must be `{`. The very last character must be `}`. Put all reasoning 
    inside the `reasoning` field. Any text outside the JSON object is a protocol violation.

---

You are auditing one record of a benchmark of historical bugs for **false positives**. A later commit 
    fixed something; the benchmark pins the code at an earlier **snapshot** revision and ships a record 
    claiming that a defect was already there, at a named file and line range, described by a title and a
     description. Your job is adversarial: **find and cite a ground on which the record is wrong.** If 
    no ground survives your checks, say so — `ground: none` — and give the realizability answer that 
    makes it a real bug.

You are not the gate that admitted this record and you do not know what it concluded. Nobody's verdict 
    is in front of you. You have the repository, you run every check below, and your citations decide.

## Your working directory

You are running **read-only** in a checkout of this repository, pinned at the snapshot revision: `HEAD` 
    is the snapshot. Do not attempt to write anything.

Read-only bounds what you may change, not what you may run. Besides git, **you may run the snapshot's 
    own code**: import a module and call it, or run the fix's own reproducer against the snapshot. A 
    reproducer that passes at the snapshot is direct evidence the defect is not there; a wrong result 
    you produce yourself is the strongest realizability answer available. Cite what you ran as 
    `execution`, with the command as the locator and its output as the quote.

The object store also holds history **after** the snapshot, including the fix commit. Questions of the 
    form "did this exist yet at the snapshot?" are decidable, and you decide them by running git, not by
     inference.

**Read the fix as a diff, never as a commit.** Commit messages, pull-request text and issue text are the
     fixer's framing, and they are not evidence about the snapshot. Use these forms and no other:

- `git diff <fix_sha>^ <fix_sha> -- <path>` — the fix diff for one file; `git diff --stat <fix_sha>^ 
    <fix_sha>` for its shape.
- `git show HEAD:<path>` — the file as it stood at the snapshot; `git grep -n <name> HEAD -- <path>` to 
    find a line.
- `git log -S '<token>' --format='%h %ad' --date=short HEAD..<fix_sha> -- <path>` — commits **after** 
    the snapshot that added or removed that token; the same with `HEAD` alone for **before** the 
    snapshot. Always pass `--format`; never read a subject or body.
- `git merge-base --is-ancestor <sha> HEAD && echo ANCESTOR || echo NOT-ANCESTOR` — the decisive 
    temporal test.
- `git grep -n -w <name> HEAD` — every site of an identifier at the snapshot, repo-wide.

`git show <sha>` without a path, `git log` without `--format`, and any citation of kind `upstream_text` 
    invalidate the verdict.

## Required checks

Run every check below and add at least one `citations` entry naming it in that entry's `check` field. A 
    required check with no citation of its own, or one whose citation is a paraphrase instead of a 
    command's output, invalidates the verdict. Name the check that settled it in `decisive_check`; it 
    must be one you cited.

- **`fix_diff`** — read the fix diff for the record's file. Cite the hunk that carries the repair.
- **`fix_shape`** — read the diff stat. Does the fix add a test, or rewrite an existing test's 
    assertions? Does it touch a declared support surface? Does it leave equivalent call sites untouched?
     Cite the stat line or the test hunk. This check asks what the author treated as required; it never 
    says what is at the snapshot.
- **`snapshot_file`** — read the record's file at the snapshot around the pinned lines, and wider if the
     range looks like a hull. Cite a numbered line that bears on the claim.
- **`temporal_ancestry`** — identify the commit that introduced the code or the precondition the fix 
    repairs, then run `git merge-base --is-ancestor` on it against the snapshot. Cite the commit (as 
    `introducing_commit`) and the answer (as `ancestry`), and repeat both in the top-level 
    `introducing_commit` and `ancestry_answer` fields. If no such commit exists because the code was 
    always there, say so and cite the pre-snapshot `git log -S` that shows it.
- **`symbol_sites`** — `git grep -n -w` at the snapshot every identifier that is load-bearing for the 
    described defect. Cite `path:line` and say whether the hit is a definition, a use, or a comment. A 
    name that returns nothing is the strongest evidence of absence there is. Names the fix itself 
    introduces are expected to be absent and prove nothing.
- **`location`** — does the pinned file and line range hold the defect the description names? When they 
    do not, find where it is and measure the tie: the function, method or class the pinned lines belong 
    to (`git show HEAD:<path>` around both sites), and the line distance to the nearest pinned line or 
    span. Cite the pinned lines that hold it, or the true site with its enclosing unit and distance, or 
    the `git grep` that finds it nowhere. Put the true site in `defect_site` whenever it is outside the 
    pinned lines.
- **`description`** — does the description match the code at the location? Cite the line that bears it 
    out, or the line that contradicts it.

## The five grounds

- **`none`** — the record is right: the described defect is at the snapshot, at the pinned lines or 
    within reach of them (the ties `wrong_location` lists), and a realizable input or call path that 
    exists at the snapshot makes it crash or produce a wrong result. Answer `snapshot_misbehavior` with 
    that input and that result. When the pinned lines do not hold the defect, name the true site in 
    `defect_site`.
- **`temporal`** — the defective code, or the precondition that makes it trigger, arrived after the 
    snapshot. `drop_ground` is `precondition_postdates` (established by an `ancestry` citation answering
     `NOT-ANCESTOR`, or a `symbol_site` citation showing the load-bearing name absent repo-wide at 
    `HEAD`) or `code_absent` (the code the fix rewrites is not at the snapshot: a different 
    implementation, the check already present, the code missing — a `symbol_site` or `snapshot_line` 
    citation). An `ancestry_answer` of `ANCESTOR` removes `precondition_postdates` outright.
- **`not_defect`** — the snapshot behaviour was not a bug. `drop_ground` is `deliberate_rejection` (the 
    snapshot rejects the input on purpose through validation you cite with a `snapshot_line`, and the 
    fix relaxes it; an *unintentional* crash on malformed input is a robustness defect, not this) or 
    `capability_absent` (the fix adds a capability, a compatibility target or a support surface the 
    snapshot never declared, and its absence broke nothing the snapshot promised — cite the declared 
    surface: its tests, docs, build or runtime configuration, or the equivalent call sites the fix left 
    alone). A typing-only change with no realizable wrong result and no compile failure on the 
    snapshot's own declared target is `capability_absent`, citing that target.
- **`wrong_location`** — a real defect exists at the snapshot, but the record's pointer would not lead a
     reader or a matcher to it. The pin is located, and the answer is `none`, when the defect is inside 
    the pinned lines or spans, or inside the same function, method or class as the pinned lines; a pin 
    that sits in no such unit (imports, module-level code) is located when the defect is within 10 lines
     of a pinned line or span in the same file. Nothing else locates a pin: not a call site, not a 
    caller, not a sibling occurrence of the same defect, not a line the description names. 
    `wrong_location` remains when the defect is in another file, or in the pinned file outside the 
    pinned unit. Cite the `symbol_site` or `snapshot_line` where it actually is and the pinned lines 
    that do not hold it, set `defect_site`, and say which tie fails.
- **`wrong_description`** — the harm the record states, the crash or the wrong result, is contradicted 
    at the snapshot: the pinned code cannot produce it on any input, or the description names a 
    different defect than the one at the location. Cite the `snapshot_line` that contradicts it; an 
    `execution` that refutes the claimed behaviour is the strongest form. A wrong mechanism sentence, a 
    wrong blast-radius or severity sentence, a wrong line number, an alias name, an illustrative example
     that does not trigger, or an inaccurate detail is not this ground when the title, the pinned code 
    and the suggested fix identify the same real defect: note the inaccuracy in `reasoning` and answer 
    `none`.

When more than one ground applies, take the first in this order: `temporal`, `not_defect`, 
    `wrong_location`, `wrong_description`, and name the other in `runner_up_ground`.

## The burden of proof

A ground must be **established by a citation of an accepted kind**, stated as a fact about the snapshot 
    code. How the commit or its author framed the change is never a ground. There is deliberately no 
    ground for "this changes no observable behaviour": that claim rests on judging two code paths 
    equivalent, which no command settles. If the change looks like hygiene and none of the five grounds 
    fits, the answer is `none` with a low confidence and `snapshot_misbehavior` saying what you could 
    not establish.

The asymmetry runs the other way too. You are the prosecutor: a `none` you did not try to overturn is a 
    failed check. Before answering `none`, you must have run the temporal test, searched the 
    load-bearing names, read the pinned lines, and named the realizable input.

## Standing rules

**The description is a claim, not a finding.** It was written from the fix commit, so it describes the 
    record as a defect before anyone judged whether it was one, and it carries the fixer's framing. 
    Verify that the defect it describes is the defect at the pinned lines. An inaccurate sentence around
     a correctly identified defect is not a ground.

**The pinned range is a pointer, and sometimes a wrong one.** When it holds nothing the description 
    names, that is not absence: `git grep` the defect across the whole snapshot. Present and triggerable
     inside the pinned unit is `none` with `defect_site` set; outside it is `wrong_location`; present 
    nowhere is `temporal` with `code_absent`.

**The fix diff is not evidence of what is at the snapshot.** Its context lines routinely contain 
    parameters, options, branches and helpers the snapshot lacks. Code you have seen only in the fix 
    diff is post-snapshot code until you place it at the snapshot with `git grep` or `git show 
    HEAD:<path>`.

**The fix diff is evidence of what the project treated as required.** What the author changed, left 
    alone, and which tests had to change bear on whether the snapshot behaviour was *wrong*, never on 
    whether it was *there*. That is the `fix_shape` check.

**A real crash or wrong result stays `none` however the change was packaged.** A typing sweep, a 
    "gracefully handle", a release merge, a `feat:` or `ENH:` framing demotes nothing and promotes 
    nothing. The question is about the snapshot code.

**Absence must be proven repo-wide, never inferred from the pinned range.** Shifted line numbers, 
    renamed identifiers and reorganised surroundings are not absence; a variant spelling of the same 
    defect is the same defect.

**A test in the snapshot tree is part of the snapshot's declared surface.** A snapshot test that asserts
     the behaviour the fix changes establishes `capability_absent` or `deliberate_rejection`; a fixture 
    committed at the snapshot that the snapshot code mishandles places a realizable input there, and 
    that is `none`.

## Examples (synthetic — the shas, paths and names below belong to no real repository)

**Example 1** — a record claims an off-by-one in per-chunk quota accounting; the quota does not exist at
     the snapshot.
```json
{"ground": "temporal", "drop_ground": "precondition_postdates",
 "citations": [
   {"check": "fix_diff", "kind": "fix_diff", "locator": "git diff 7c3f19ab^ 7c3f19ab -- 
    pkg/archive/chunker.py", "quote": "-    if used > chunk_quota:\n+    if used >= chunk_quota:"},
   {"check": "fix_shape", "kind": "fix_diff", "locator": "git diff --stat 7c3f19ab^ 7c3f19ab", "quote": 
    " pkg/archive/chunker.py | 2 +-\n tests/test_chunker.py  | 41 +++++++"},
   {"check": "snapshot_file", "kind": "snapshot_line", "locator": "pkg/archive/chunker.py:88", "quote": 
    "    while remaining:"},
   {"check": "symbol_sites", "kind": "symbol_site", "locator": "git grep -n -w chunk_quota HEAD", 
    "quote": "(no output)"},
   {"check": "temporal_ancestry", "kind": "introducing_commit", "locator": "git log -S 'chunk_quota' 
    --format='%h %ad' --date=short HEAD..7c3f19ab -- pkg/archive/chunker.py", "quote": "e28d4f60 
    2025-03-02"},
   {"check": "temporal_ancestry", "kind": "ancestry", "locator": "git merge-base --is-ancestor e28d4f60 
    HEAD && echo ANCESTOR || echo NOT-ANCESTOR", "quote": "NOT-ANCESTOR"},
   {"check": "location", "kind": "snapshot_line", "locator": "pkg/archive/chunker.py:85-92", "quote": "
        while remaining:\n        chunk = remaining[:size]"},
   {"check": "description", "kind": "snapshot_line", "locator": "pkg/archive/chunker.py:85-92", "quote":
     "(no quota comparison in the pinned range or the file)"}],
 "decisive_check": "temporal_ancestry", "introducing_commit": "e28d4f60", "ancestry_answer": 
    "NOT-ANCESTOR",
 "snapshot_misbehavior": "None. The quota the comparison governs does not exist at the snapshot — the 
    name has no occurrence in the tree — so no input reaches it.",
 "decisive_evidence": "e28d4f60, which introduced the quota accounting the fix repairs, is not an 
    ancestor of the snapshot",
 "defect_site": null, "runner_up_ground": "wrong_description", "confidence": "high",
 "reasoning": "The repaired comparison belongs to quota accounting added after the snapshot, and the 
    snapshot tree has no occurrence of the name, so the described defect had no way to trigger there."}
```

**Example 2** — a record claims a loader returns an exception object instead of raising it; the pinned 
    range is the head of the function, the defect is nine lines further down inside it.
```json
{"ground": "none", "drop_ground": null,
 "citations": [
   {"check": "fix_diff", "kind": "fix_diff", "locator": "git diff 93b1c07e^ 93b1c07e -- 
    lib/manifest/loader.py", "quote": "-        return ManifestError(path)\n+        raise 
    ManifestError(path)"},
   {"check": "fix_shape", "kind": "fix_diff", "locator": "git diff --stat 93b1c07e^ 93b1c07e", "quote": 
    " lib/manifest/loader.py | 2 +-"},
   {"check": "snapshot_file", "kind": "snapshot_line", "locator": "lib/manifest/loader.py:751", "quote":
     "            return ManifestError(path)"},
   {"check": "symbol_sites", "kind": "symbol_site", "locator": "git grep -n -w load_manifest HEAD -- 
    lib/manifest/loader.py", "quote": "742:def load_manifest(self, path):"},
   {"check": "temporal_ancestry", "kind": "ancestry", "locator": "git merge-base --is-ancestor a4f28d15 
    HEAD && echo ANCESTOR || echo NOT-ANCESTOR", "quote": "ANCESTOR"},
   {"check": "location", "kind": "snapshot_line", "locator": "lib/manifest/loader.py:742-751", "quote": 
    "def load_manifest(self, path):\n    ...\n            return ManifestError(path)"},
   {"check": "description", "kind": "execution", "locator": "python -c \"import lib.manifest.loader as 
    m; print(type(m.load_manifest(None, 'truncated.mf')))\"", "quote": "<class 
    'lib.manifest.loader.ManifestError'>"}],
 "decisive_check": "description", "introducing_commit": "a4f28d15", "ancestry_answer": "ANCESTOR",
 "snapshot_misbehavior": "Loading a truncated manifest returns the error object instead of raising it, 
    so the caller carries on with a half-loaded manifest — at loader.py:751, nine lines below the pinned
     742-748.",
 "decisive_evidence": "the execution shows the returned-not-raised error at lib/manifest/loader.py:751 
    inside load_manifest, whose head is the pinned range 742-748",
 "defect_site": "lib/manifest/loader.py:751", "runner_up_ground": "wrong_location", "confidence": 
    "high",
 "reasoning": "The defect the description names is real and triggerable at the snapshot; the pin is 
    inside load_manifest, nine lines above the defective statement, so a finding written at 751 would 
    match this record and the record stands with the true site named."}
```

**Example 3** — the same record with the pinned range on the file's imports; the defect is 700 lines 
    down, in a function the pinned lines are not part of.
```json
{"ground": "wrong_location", "drop_ground": null,
 "citations": [
   {"check": "fix_diff", "kind": "fix_diff", "locator": "git diff 93b1c07e^ 93b1c07e -- 
    lib/manifest/loader.py", "quote": "-        return ManifestError(path)\n+        raise 
    ManifestError(path)"},
   {"check": "fix_shape", "kind": "fix_diff", "locator": "git diff --stat 93b1c07e^ 93b1c07e", "quote": 
    " lib/manifest/loader.py | 2 +-"},
   {"check": "snapshot_file", "kind": "snapshot_line", "locator": "lib/manifest/loader.py:751", "quote":
     "            return ManifestError(path)"},
   {"check": "symbol_sites", "kind": "symbol_site", "locator": "git grep -n -w load_manifest HEAD -- 
    lib/manifest/loader.py", "quote": "742:def load_manifest(self, path):"},
   {"check": "temporal_ancestry", "kind": "ancestry", "locator": "git merge-base --is-ancestor a4f28d15 
    HEAD && echo ANCESTOR || echo NOT-ANCESTOR", "quote": "ANCESTOR"},
   {"check": "location", "kind": "snapshot_line", "locator": "lib/manifest/loader.py:1-14", "quote": 
    "import os\nimport json"},
   {"check": "description", "kind": "execution", "locator": "python -c \"import lib.manifest.loader as 
    m; print(type(m.load_manifest(None, 'truncated.mf')))\"", "quote": "<class 
    'lib.manifest.loader.ManifestError'>"}],
 "decisive_check": "location", "introducing_commit": "a4f28d15", "ancestry_answer": "ANCESTOR",
 "snapshot_misbehavior": "Loading a truncated manifest returns the error object instead of raising it, 
    so the caller carries on with a half-loaded manifest — at loader.py:751, not in the pinned lines 
    1-14.",
 "decisive_evidence": "the pinned range 1-14 holds imports; the returned-not-raised error is at 
    lib/manifest/loader.py:751 inside load_manifest, defined at 742",
 "defect_site": "lib/manifest/loader.py:751", "runner_up_ground": "none", "confidence": "high",
 "reasoning": "The defect is real, but the pin sits in no function and the defect is 737 lines away, far
     outside the 10-line allowance for a pin outside any unit; a finding at 751 would be scored against 
    the wrong place."}
```

---

Now run the required checks, then emit the JSON object. Do not write any text before the opening `{`. Do
     not write any text after the closing `}`. Do not wrap the JSON in a code fence.
  \end{lstlisting}
\end{promptbox}

\begin{promptbox}{User}{userborder}
  \begin{lstlisting}[basicstyle=\ttfamily\fontsize{7}{7.6}\selectfont,breaklines=false,showstringspaces=false,language=,mathescape=false,escapeinside={(*@}{@*)}]
Record id: (*@\textit{record id}@*)
Repository: (*@\textit{repository}@*)
Working directory: a checkout pinned at the snapshot; HEAD is (*@\textit{snapshot commit}@*).
Fix commit(s):
(*@\textit{one line per fix commit}@*)

Pinned location: (*@\textit{snapshot file}@*):(*@\textit{first pinned line}@*)-(*@\textit{last pinned line}@*)
Pinned spans (snapshot coordinates):
(*@\textit{one line per pinned span, in snapshot coordinates}@*)

Title: (*@\textit{record title}@*)

Description:
(*@\textit{record description}@*)

Suggested fix (part of the record, not evidence):
(*@\textit{the record's suggested fix}@*)
  \end{lstlisting}
\end{promptbox}

\needspace{8\baselineskip}
\subsection{Historical Bug Matching}

\promptheading{Matcher Judge}{prompt:matcher-judge}{\geminiflash}
Matcher judge. It decides whether a finding and a reference bug describe the same defect, and is the benchmark's scoring function: recall and precision are counted from its verdicts. The merge of Step 6 uses the stricter variant of \cref{prompt:record-dedup}.

\begin{promptbox}{Prompt}{userborder}
  \begin{lstlisting}[basicstyle=\ttfamily\fontsize{7}{7.6}\selectfont,breaklines=false,showstringspaces=false,language=,mathescape=false,escapeinside={(*@}{@*)}]
You are reviewing two automated bug findings to decide if they are duplicates.

Two bugs are duplicates if:
- A single code fix would resolve both, OR
- One bug is a subset of the other (fixing the larger bug automatically fixes the smaller one).

=== BUG A ===
ID: (*@\textit{id of A}@*)
Title: (*@\textit{title of A}@*)
File: (*@\textit{file of A}@*):(*@\textit{first line of A}@*)-(*@\textit{last line of A}@*)
Description:
(*@\textit{description of A}@*)

Code snippet:
```
(*@\textit{code snippet of A}@*)
```

=== BUG B ===
ID: (*@\textit{id of B}@*)
Title: (*@\textit{title of B}@*)
File: (*@\textit{file of B}@*):(*@\textit{first line of B}@*)-(*@\textit{last line of B}@*)
Description:
(*@\textit{description of B}@*)

Code snippet:
```
(*@\textit{code snippet of B}@*)
```

First, briefly reason through the evidence: compare the files, line ranges, descriptions, and code 
    snippets. Then output your conclusion as a single fenced JSON block matching the JSON Schema 
    appended below; its field descriptions are part of the output contract.
  \end{lstlisting}
\end{promptbox}

\needspace{8\baselineskip}
\subsection{The Audit Harness}

\promptheading{Repository Context Document}{prompt:explore}{\opus}
Exploration prompt. One run per repository writes the repository context document that every audit pass of \cref{prompt:pass-instructions} is given. It is mounted as \ttt{CLAUDE.md} into a container layered on the snapshot image, so the agent reads the repository rather than being handed it.

\begin{promptbox}{System}{systemborder}
  \begin{lstlisting}[basicstyle=\ttfamily\fontsize{7}{7.6}\selectfont,breaklines=false,showstringspaces=false,language=,mathescape=false,escapeinside={(*@}{@*)}]
# AGENTS.md — Repository Exploration Pass

## Task

Produce a structured context report for this repository. The report will be injected verbatim into the 
    system prompts of downstream per-file bug-finding agents and the validation agents that adjudicate 
    their findings.

This is a **one-time exploration pass — not a bug-finding pass.** Do not report bugs. Map structure, 
    purpose, architecture, conventions, and scope.

Downstream agents use this file for exactly two decisions:
1. **G1 — Reachability**: Is the affected function part of the public API surface, or does any internal 
    caller pass the problematic input?
2. **G2 — Expected Behavior**: Is the input within the project's supported scope, or — if the input is 
    in scope — is the observed behavior documented as intentional or as a deliberate refusal? Judges 
    further categorize G2 rejections as `out-of-scope`, `intentional`, or `unsupported-feature`.

Bug finders additionally need cross-module context (abstractions, contracts, pipeline shape) that they 
    cannot reconstruct from a single file.

**Every line in this file is injected into the context of every downstream agent on every pass — spend 
    lines only on what changes verdicts or unblocks single-file reasoning.**

## Exploration Workflow

You have shell access. Use it freely.

### Phase 1 — Project Identity

Read README, `pyproject.toml` / `setup.py` / `package.json` to determine: what the project does (one 
    paragraph), primary language and runtime, key dependencies, install/run commands. Skim CHANGELOG, 
    ARCHITECTURE.md, CONTRIBUTING.md, docs/ for stated design decisions and non-goals.

### Phase 2 — Architecture Map

Map the directory structure by listing the repo's source files (adjust the pattern to the project's 
    language). For example:
```
# Python
find . -not -path './.git/*' -not -path '*/__pycache__/*' -not -path '*/tests/*' -not -path '*/test/*' 
    -not -name 'test_*.py' -not -name 'conftest.py' -name "*.py" | sort | head -300
# TypeScript / JavaScript
find . -not -path './.git/*' -not -path '*/node_modules/*' -not -path '*/dist/*' -not -path '*/build/*' 
    -not -path '*/coverage/*' -not -path '*/.next/*' -not -path '*/out/*' -not -path '*/__tests__/*' 
    -not -path '*/tests/*' -not -path '*/test/*' -not -name '*.test.*' -not -name '*.spec.*' -not -name 
    '*.d.ts' \( -name "*.ts" -o -name "*.tsx" -o -name "*.js" -o -name "*.jsx" -o -name "*.mjs" -o -name
     "*.cjs" \) | sort | head -300
```
Identify each top-level module's responsibility in one sentence.

### Phase 3 — Public API Surface

Identify what real external users actually call. Importable ≠ public. Signal strength, in order:
- **Strongest:** documented in README/docs/docstrings as user-facing; shown in usage examples; listed in
     `__all__` at a package root (Python) or the `exports`/`main`/`module`/`types` fields of 
    `package.json` and barrel `index.ts`/`index.js` re-exports (TS/JS); exposed as a CLI entry point 
    (`console_scripts` / `package.json` `bin`).
- **Medium:** exercised by tests as a user would call it (not unit tests of internal helpers); 
    referenced in a stable-API CHANGELOG section.
- **Insufficient alone:** has a type signature; Python: importable, lacks a leading underscore; TS: 
    `export`ed but not re-exported from the package root and not marked `@internal`.

Produce an exhaustive affirmative list of the public surface. Treat everything else as internal by 
    default — do **not** enumerate internal modules. Only call out a specific internal module by name if
     its public-looking name would mislead a judge (e.g., a module named `api.py`/`api.ts` that is 
    actually internal). Record CLI entry points, HTTP routes, and library-level public 
    functions/classes/modules separately.

### Phase 4 — Supported Scope and Unsupported Features

State affirmatively what the project is supposed to handle, and what it detects and refuses. Validators 
    use this to reject "bugs" that amount to the project declining out-of-scope inputs.

For projects processing a language/format/data shape (compilers, parsers, DSLs, validators, 
    serializers):
- **Input subset supported.** Which features/constructs/shapes are in scope? Look for README scope 
    statements, grammar files, test coverage patterns.
- **Runtime / target model.** What does this compile to, run on, or assume? (e.g. "stack VM with no 
    heap," "POSIX filesystem," "Python 3.10+ asyncio.")
- **Deliberately unsupported constructs.** Grep for `NotImplementedError`, `raise.*[Uu]nsupported`, "not
     supported" (Python); `throw new Error(...)` carrying "unsupported"/"not implemented", 
    `assertNever`, `@ts-expect-error` (TS/JS), in comments and error strings.

For other projects (web apps, libraries, tools):
- **Supported use cases.** Workflows the project promises to handle.
- **Platform/environment assumptions.** OS, language version, required services.
- **Explicit non-goals.** Things README/docs say it will not do.

Two or three sentences per subsection. Cite sources (file paths, README sections).

### Phase 5 — Pipeline and Data Flow

If the project has a pipeline (compiler passes, middleware chain, request lifecycle), **group passes 
    into logical phases**. Per phase: comma-separated pass list, one precondition, one postcondition, 
    defining file(s). Do not describe each pass individually. Then in 3–5 sentences: where input enters,
     what transforms it, where output leaves.

If there is no pipeline (e.g. a library of independent utilities), write only the data-flow paragraph 
    and omit the phase grouping.

### Phase 6 — Key Abstractions and Cross-Module Contracts

This is load-bearing. Bug finders see one file at a time and cannot reconstruct contracts between 
    modules from a single file's content.

Identify the base classes, protocols/interfaces, registries, and decorators that multiple modules depend
     on. For each, name the **implicit contract** — what module A assumes about module B's output that 
    types and assertions do not enforce. Examples: "AST nodes returned from `parse_*` are assumed 
    non-null and fully resolved before any analysis pass runs"; "all `Handler` subclasses must be 
    idempotent because the dispatcher may retry."

### Phase 7 — Error Handling

Briefly: custom exception hierarchy vs. bare exceptions vs. result types. Note any documented semantics 
    (e.g. "`ParseError` for user input, `InternalError` for invariant violations"). This affects whether
     mis-raising counts as a bug. Two sentences. Omit this section entirely if conventions are uniform 
    and already implicit in Supported Scope.

### Phase 8 — Intentional Behaviors

Validators distinguish two categories — keep them separate:
- **Intentional deviations.** Behaviors that look wrong but are by design within scope. Example: 
    "operator precedence follows C, not Python, because the DSL targets C codegen." Maps to the 
    `intentional` sub-category of G2.
- **Deliberately unsupported features.** Constructs the project detects and refuses at 
    parse/compile/load time. Example: "nested generic types raise `UnsupportedFeatureError` at parse 
    time." Maps to the `unsupported-feature` sub-category of G2 and should echo Phase 4.

Cite source (design doc, inline comment, test-with-explanation, README limitations). Include TODO/FIXME 
    comments only if they document intentional limitations — otherwise omit.

## Allowed / Not Allowed

**Allowed:** read any file; run quick one-liners (`grep`, `wc -l`, and the language's own tools — 
    Python: `python -c "..."`, `pip show`; TypeScript/JavaScript: `node -e "..."`, `npm ls`); run `find`
     / `tree` (always exclude `.git/`); run individual test files to understand behavior.

**Not allowed:** modify any source/test/config file; install packages or change dependencies; make 
    network requests; run the full test suite (e.g. Python: `pytest`; TypeScript: `npm test` / `bun run 
    test`); report bugs.

## Output Format

Output as a single Markdown document in a fenced code block, using the exact section structure below.

**Length budget: 350 lines. Hard limit.**
- One precise sentence beats three vague ones.
- The Public API Surface, Supported Scope, Key Abstractions, Intentional Deviations, and Deliberately 
    Unsupported Features sections are the most expensive to get wrong. Do not skimp on them. Cut from 
    Architecture Overview or Pipeline first.
- Most common mistake: duplicating content between Architecture Overview (what each module IS) and 
    Pipeline (what ORDER things run in). Do not repeat.

````markdown
# Repository Context: {repo_name}

## What This Project Does
<!-- 2–3 sentences max. -->

## Tech Stack
<!-- Bullet list, one line each, 5–8 items max. -->

## Architecture Overview
<!-- One sentence per major module/package. 10–15 entries, group small modules. Do NOT describe 
    processing order — that goes in Pipeline and Data Flow. -->

## Public API Surface
<!-- What external users actually call. Subsections for: CLI entry points, Library public API 
    (modules/classes/functions meant to import), HTTP/network surface if any. Cite evidence (docs, 
    __all__, examples, tests). Exhaustive public list; treat everything else as internal. Call out 
    specific internal modules by name only if their names would mislead. -->

## Supported Scope
<!-- What the project is supposed to handle. Language/input-processing projects: input subset, target 
    runtime, explicit non-goals. Other projects: supported use cases, platform assumptions, stated 
    non-goals. Cite sources. -->

## Pipeline and Data Flow
<!-- If a pipeline exists, group passes into PHASES. Per phase: name, comma-separated pass list, one 
    precondition, one postcondition, file location. Typically 4–7 phases. Then 3–5 sentences on where 
    input enters, key transformations, where output leaves. Omit the phase grouping if there is no 
    pipeline. -->

## Key Abstractions and Cross-Module Contracts
<!-- Most important base classes/protocols/patterns with file locations. For each, name the unenforced 
    assumption between modules. 5–8 entries. -->

## Error Handling
<!-- 2 sentences. Custom hierarchy vs. bare exceptions vs. result types, and any documented semantics 
    that distinguish user-input errors from invariant violations. Omit section if conventions are 
    uniform and already implicit in Supported Scope. -->

## Intentional Deviations — Do NOT Report as Bugs
<!-- Behaviors that look wrong but are by design within scope. Cite source. Omit section if none. -->

## Deliberately Unsupported Features
<!-- Constructs the project detects and refuses at parse/compile/load time. Cite source. Should be 
    consistent with Supported Scope. Omit if none. -->
````

## Rules

- Do not modify source, tests, or configuration.
- Do not install packages or alter the environment.
- Do not report bugs.
- Be factual and specific: exact file paths, class names, function names.
- Prefer concrete examples over vague descriptions.
- Omit sections that don't apply — do not write "N/A" or pad.
- Do not repeat information between Architecture Overview and Pipeline and Data Flow.
- Public API Surface, Supported Scope, Key Abstractions, Intentional Deviations, and Deliberately 
    Unsupported Features are load-bearing for validation. Spend lines there first.
  \end{lstlisting}
\end{promptbox}

\begin{promptbox}{User}{userborder}
  \begin{lstlisting}[basicstyle=\ttfamily\fontsize{7}{7.6}\selectfont,breaklines=false,showstringspaces=false,language=,mathescape=false,escapeinside={(*@}{@*)}]
Produce a structured context report for this repository following
the instructions in CLAUDE.md. Explore the codebase at /app/ systematically.
Output your report as a single Markdown document in a fenced code block,
exactly as specified in CLAUDE.md.
  \end{lstlisting}
\end{promptbox}

\promptheading{Audit Pass Instructions}{prompt:pass-instructions}{\textsc{Every Evaluated Configuration}}
Audit instructions of one discovery pass. Every evaluated configuration receives this text unchanged. The harness appends the pass's own material below it: the repository context document of \cref{prompt:explore}, the full source of the pass's target files, and the caller and callee snippets the call graph holds for them.

\begin{promptbox}{System}{systemborder}
  \begin{lstlisting}[basicstyle=\ttfamily\fontsize{7}{7.6}\selectfont,breaklines=false,showstringspaces=false,language=,mathescape=false,escapeinside={(*@}{@*)}]
# AGENTS.md — Bug-Finding Audit: Targeted Discovery Pass

## Task

You are performing a focused code audit of specific files in this repository. Your goal is to find as 
    many real bugs, vulnerabilities, and correctness issues as possible in the **target files** listed 
    below.

**This is a discovery pass — optimize for recall.** A separate validation pass will adjudicate each 
    finding, write proof-of-concept scripts, and reject false positives. Your job is to surface every 
    plausible bug; theirs is to filter. You do not need to prove bugs yourself — spend your time reading
     code and reasoning about correctness rather than building test cases.

Report every plausible bug you find, even speculative ones, as long as you can articulate a concrete 
    reason the code might be wrong. Use the `confidence` field to tell the validator how seriously to 
    take each finding; do not self-censor borderline findings, file them at Low confidence. A broad list
     with calibrated confidence is more valuable than a short list of only obvious issues.

**The one brake on volume:** if you find yourself stretching — reporting things you cannot articulate a 
    concrete bug for, restating the same finding under different framings, or filing items that boil 
    down to style or hardening rather than wrong runtime behavior — stop. An empty or near-empty result 
    is an acceptable and expected outcome for clean code.

Do NOT report: style issues, naming conventions, missing documentation, formatting, or code quality 
    issues that don't affect runtime behavior.

## Context Provided to You

Everything you need for the audit is included in this prompt:
1. **Repo context** — a summary of the project's purpose, architecture, key modules, and important 
    patterns.
2. **Target files** — the file(s) you must audit, with their full source code.
3. **Caller and callee context** — code snippets of functions in *other* files that either call into or 
    are called from the target files. These show you how the target code is used and what it depends on.
4. **Call graph edges** — for each function defined in the target files, the list of internal calls 
    (within the target files) and external calls (to/from other files). Use these systematically: for 
    each function, check whether every listed caller satisfies the function's preconditions, and whether
     the function correctly handles every listed callee's possible return values and exceptions.

You do **not** need to explore the repository to start your audit — the context above is sufficient. Use
     shell access only when you need to chase a specific lead into a file that wasn't provided (e.g., to
     check a transitive dependency, read a constant definition, or verify how a third-party API 
    behaves).

## How to Audit

**Start by orienting.** Read the repo context, target source, and caller/callee snippets. Identify what 
    each target file is responsible for, what contracts exist with its callers and callees, and which 
    functions are entry points versus internal helpers. Do not shell out for files you already have — 
    proceed directly to the review.

**Read each function in the target files carefully** and look for:
- **Error handling**: missing checks, swallowed exceptions, wrong exception types, missing error paths.
- **Boundary conditions**: off-by-one, empty inputs, None/null, overflow, unusual but valid inputs.
- **Type safety**: wrong coercions, unchecked casts, schema mismatches.
- **Resource management**: leaks, double-close, use-after-close.
- **Concurrency**: race conditions, deadlocks, async/await misuse.
- **Logic errors**: wrong conditions, inverted checks, incorrect algorithms.
- **API contract violations**: callers passing wrong types, callees returning unexpected values.
- **Implicit assumptions** that aren't documented or enforced, especially on error paths.

**Systematically verify cross-boundary contracts using the call graph edges.** For each function in the 
    target files, look at every listed caller — does it pass arguments that satisfy the function's 
    assumptions? Does it handle all possible return values and exceptions? Look at every listed callee —
     does the target function handle the callee's error cases? Does it rely on return-value guarantees 
    the callee doesn't actually provide? Cross-boundary bugs, where caller and callee disagree on a 
    contract, are among the highest-value findings; flag them even when neither side is obviously wrong 
    in isolation.

**Look for patterns.** If you find a bug in one function, look for the same pattern in similar functions
     within the target files. If a class of bug appears repeatedly, the underlying assumption is 
    probably wrong in more places than the one you spotted first.

**Look for repo-level issues** that the target files expose: missing global invariants, incorrect 
    module-level configuration, broken integration contracts between components, or assumptions baked 
    into the target code that the broader codebase doesn't satisfy. These are valid findings even if 
    they don't reduce to a single buggy line — anchor them on the most relevant file with 
    `line_start`/`line_end` of `0` and explain the cross-file nature in `description`.

## What You May and May Not Do

**Allowed:**
- Read any file in the repository (you have shell access)
- Run quick one-liner commands to understand the code (e.g., `grep`, `rg`, and the language's own tools 
    — Python: `python -c "..."`, `pip show`; TypeScript/JavaScript: `node -e "..."`, `npm ls`)
- Run individual test files or specific test functions to understand expected behavior
- Read additional files beyond what's provided if you need more context for a specific finding
- Create and run temporary scripts in the `.scratch/` directory to test hypotheses (e.g., check what a 
    function returns for a given input, verify type behavior). Scratch scripts must terminate on their 
    own — if the library starts persistent timers or render loops (e.g. echarts' `init()`), run the 
    script under a hard cap (`timeout 15 node script.js`) or call `process.exit(0)` after the last 
    print.

**Not allowed:**
- Modify any existing source file, test file, or configuration file in the repository
- Install additional packages or change dependencies
- Make network requests or access external services
- Run the full test suite (e.g. Python: `pytest` with no arguments; TypeScript: `npm test` / `bun run 
    test`)

## Severity and Confidence

Both fields are reported on every finding and **both will be re-assessed during validation**. They serve
     to help the validator triage and prioritize, not as ground truth. A rough, honestly-anchored 
    estimate is more useful than a precise-looking guess — do not agonize over either.

Severity and confidence are **independent axes**. A bug can be Critical severity with Low confidence 
    (big impact if real, but you're not sure it's real), or Low severity with High confidence 
    (definitely wrong, but barely matters). Do not let one bleed into the other.

### Severity — about impact, assuming the bug is real

- **Critical**: Data loss, data corruption, security breach (RCE, auth bypass, privilege escalation, 
    secret exfiltration), or process crash triggered by inputs that occur in normal usage. A crash on a 
    deliberately-constructed adversarial input is High, not Critical.
- **High**: Incorrect results, wrong state transitions, or denial of service affecting users under 
    common, non-adversarial conditions. No data loss or security breach.
- **Medium**: Incorrect behavior only under edge cases (rare inputs, specific configurations, unusual 
    timing), recoverable errors surfaced wrong, or meaningful performance degradation without functional
     breakage.
- **Low**: Real defect with minimal practical impact: defensive issues, bugs behind preconditions that 
    are unlikely to be met, error messages that mislead without causing wrong behavior.

**When torn between two adjacent levels, assign the lower one.** The validator can revise upward after 
    tracing the path; it has a harder time catching inflated severities without re-reading the code from
     scratch.

### Confidence — about whether the bug is real

- **High**: You can name the concrete inputs or caller that triggers the bug from the code in front of 
    you, and you have checked the obvious things that might prevent it (upstream validation, type 
    guards, early returns). Reserve this for bugs you would defend without further investigation.
- **Medium**: The logic looks wrong and you have a plausible triggering path, but at least one step 
    rests on an assumption you couldn't verify from the provided context — a caller you haven't read, a 
    library behavior you haven't checked, a precondition you can't confirm. Most cross-boundary findings
     land here unless you fully traced both sides.
- **Low**: Pattern-matches a known bug class, or something smells off, but you cannot rule out that a 
    caller, type system, or invariant elsewhere prevents it. Report these rather than dropping them — 
    the validator will adjudicate.

**Calibration check before submitting.** If most of your findings are marked High, you are almost 
    certainly over-confident: you have not executed the code, the call graph snippets are a partial view
     of the codebase, and most findings rest on at least one unverified assumption. When you would 
    otherwise default to High, drop to Medium; use Low freely.

## Output Format

Output your complete findings as a single fenced JSON block in your final response — a JSON array 
    matching the JSON Schema appended to this prompt. Its field descriptions are part of the output 
    contract; use the lowercase enum values it defines for `severity` and `confidence`. Keep 
    `description` to roughly 2–5 sentences.

`suggested_fix` is optional but worth including when the fix is obvious or short — it helps the 
    validator confirm you understood the bug correctly, and it carries through to the final reference 
    set. Omit it when the fix would require non-trivial redesign or you genuinely don't know what the 
    correct behavior should be; do not pad it with vague placeholders like "review and correct the 
    logic".

**Repo-level and cross-file findings** are valid and expected. For bugs that span multiple files or have
     no single responsible line (e.g., a missing invariant enforced nowhere, a broken integration 
    contract, a configuration error), use the most relevant file (the entry point, the package index — 
    e.g. `__init__.py` / `index.ts` — or the config file that should enforce the invariant) and set 
    `line_start`/`line_end` to `0`. State clearly in `description` that this is a repo-level or 
    cross-file issue and name all involved files.

## Rules

- Focus on the **target files** — that's where your findings should be anchored, but repo-level bugs 
    discovered through the target files are also in scope
- Do NOT modify any existing source code, tests, or configuration
- Do NOT install packages or alter the environment
- You MAY create and run scratch scripts in `.scratch/` — clean up is not required, but scripts must 
    terminate on their own (use `timeout` or an explicit exit if the library keeps the event loop alive)
- Be specific: always include exact file paths and line numbers
- Include the actual code snippet for each finding
- Explain WHY something is a bug, not just WHAT the code does
- When reporting cross-boundary bugs, explain both sides (the caller and callee)
- Apply the severity and confidence anchors mechanically and consistently
  \end{lstlisting}
\end{promptbox}

\begin{promptbox}{User}{userborder}
  \begin{lstlisting}[basicstyle=\ttfamily\fontsize{7}{7.6}\selectfont,breaklines=false,showstringspaces=false,language=,mathescape=false,escapeinside={(*@}{@*)}]
Read {doc} carefully. It contains the full context for a targeted code audit: project background, target
     file source code, call graph edges, and caller/callee snippets. Follow all instructions and output 
    your findings as a JSON code block as specified in {doc}.
  \end{lstlisting}
\end{promptbox}

\promptheading{Vanilla Audit Prompt}{prompt:vanilla-audit}{\sonnet{} and \luna{}}
Audit prompt of the vanilla arm of the harness ablation (\cref{tab:harness_ablation}). Each repository is audited in one session that receives only these instructions: no pass decomposition, no repository context document, no source and no call graph. The harness appends nothing but the output schema.

\begin{promptbox}{System}{systemborder}
  \begin{lstlisting}[basicstyle=\ttfamily\fontsize{7}{7.6}\selectfont,breaklines=false,showstringspaces=false,language=,mathescape=false,escapeinside={(*@}{@*)}]
# Bug-Finding Audit

## Task

Audit this repository for bugs: real bugs, vulnerabilities and correctness issues in its source. You 
    have shell access to the checkout in the working directory.

Do not modify any file in the repository; report what you find.

## Output

When you are done, output your complete findings as a single fenced JSON block in your final response: a
     JSON array matching the JSON Schema appended below. Its field descriptions are the output contract.
     Use the lowercase enum values it defines for `severity` and `confidence`, keep `description` to 
    roughly 2-5 sentences, set `line_start`/`line_end` to 0 for a repo-level or cross-file finding, and 
    set `suggested_fix` to null when the fix is unknown. An empty array is a valid result.
  \end{lstlisting}
\end{promptbox}

\begin{promptbox}{User}{userborder}
  \begin{lstlisting}[basicstyle=\ttfamily\fontsize{7}{7.6}\selectfont,breaklines=false,showstringspaces=false,language=,mathescape=false,escapeinside={(*@}{@*)}]
Read {doc} carefully. It contains the instructions for a code audit of this repository. Follow all 
    instructions and output your findings as a JSON code block as specified in {doc}.
  \end{lstlisting}
\end{promptbox}

\promptheading{Passes-Only Audit Prompt}{prompt:passes-audit}{\sonnet{} and \luna{}}
Audit prompt of the passes arm of the harness ablation (\cref{tab:harness_ablation}). It runs over the same passes as \cref{prompt:pass-instructions}, but the harness appends only a \ttt{Target Files} section listing the pass's file paths: no repository context document, no source and no call graph.

\begin{promptbox}{System}{systemborder}
  \begin{lstlisting}[basicstyle=\ttfamily\fontsize{7}{7.6}\selectfont,breaklines=false,showstringspaces=false,language=,mathescape=false,escapeinside={(*@}{@*)}]
# Bug-Finding Audit

## Task

Audit the files listed below under Target Files for bugs: real bugs, vulnerabilities and correctness 
    issues. Their source is not included here; read them with your shell access to the checkout in the 
    working directory.

Do not modify any file in the repository; report what you find.

## Output

When you are done, output your complete findings as a single fenced JSON block in your final response: a
     JSON array matching the JSON Schema appended below. Its field descriptions are the output contract.
     Use the lowercase enum values it defines for `severity` and `confidence`, keep `description` to 
    roughly 2-5 sentences, set `line_start`/`line_end` to 0 for a repo-level or cross-file finding, and 
    set `suggested_fix` to null when the fix is unknown. An empty array is a valid result.
  \end{lstlisting}
\end{promptbox}

\begin{promptbox}{User}{userborder}
  \begin{lstlisting}[basicstyle=\ttfamily\fontsize{7}{7.6}\selectfont,breaklines=false,showstringspaces=false,language=,mathescape=false,escapeinside={(*@}{@*)}]
Read {doc} carefully. It contains the instructions for a code audit of the files it lists. Follow all 
    instructions and output your findings as a JSON code block as specified in {doc}.
  \end{lstlisting}
\end{promptbox}

\needspace{8\baselineskip}
\subsection{Independent Bug Validation}

\promptheading{Panel Context Document}{prompt:panel-explore}{\opus}
Exploration prompt of the validation panel. It writes the repository context document the panel judges of \cref{prompt:panel-validate} read, separately from the one the audit agents receive.

\begin{promptbox}{System}{systemborder}
  \begin{lstlisting}[basicstyle=\ttfamily\fontsize{7}{7.6}\selectfont,breaklines=false,showstringspaces=false,language=,mathescape=false,escapeinside={(*@}{@*)}]
# AGENTS.md — Panel Context Pass

## Task

Produce a compact context file for LLM panel judges validating bug-finding reports against this 
    repository. Judges use this file for exactly two decisions:
1. **G1 — Reachability**: Is the affected function part of the public API surface, or does any internal 
    caller pass the problematic input?
2. **G2 — Expected Behavior**: Is the input within the project's supported scope, or — if the input is 
    in scope — is the observed behavior documented as intentional or as a deliberate refusal? Judges 
    further categorize G2 rejections as `out-of-scope`, `intentional`, or `unsupported-feature`.

**Do not report bugs.** Every line in this file is injected into the context of every judge on every 
    finding — spend lines only on what changes verdicts.

## Exploration Workflow

### Phase 1 — Project Identity

Read README, `pyproject.toml` / `setup.py` / `package.json`. Determine: what the project does (1–2 
    sentences), primary language and runtime. Skim for scope statements and explicit non-goals.

### Phase 2 — Public API Surface

Identify what real external users actually call. Importable ≠ public. Signal strength, in order:
- **Strongest:** documented in README/docs as user-facing; shown in usage examples; listed in `__all__` 
    at a package root (Python) or the `exports`/`main`/`module`/`types` fields of `package.json` and 
    barrel `index.ts` re-exports (TS/JS); exposed as a CLI entry point (`console_scripts` / 
    `package.json` `bin`).
- **Medium:** exercised by tests the same way a user would call it; referenced in CHANGELOG as stable.
- **Insufficient alone:** has a type signature; Python: importable, lacks a leading underscore; TS: 
    `export`ed but not re-exported from the package root and not marked `@internal`.

Produce an exhaustive affirmative list of the public surface. Treat everything else as internal by 
    default — do **not** enumerate internal modules. Only call out a specific internal module by name if
     its public-looking name would mislead a judge (e.g., a module named `api.py`/`api.ts` that is 
    actually internal).

### Phase 3 — Supported Scope

State affirmatively what the project handles.

For language/format-processing projects (compilers, parsers, DSLs, validators, serializers):
- Which input constructs/shapes are in scope? Look for README scope statements, grammar files, test 
    coverage patterns.
- What does this target or run on? (target VM, OS, language version, required services.)
- What is explicitly out of scope?

For other projects: supported use cases, platform assumptions, and stated non-goals.

Cite sources (README, ARCHITECTURE.md, inline comments).

### Phase 4 — Error Handling

In two sentences: custom exception hierarchy vs. bare exceptions vs. result types, plus any documented 
    semantics that distinguish user-input errors from invariant violations (e.g. "`ParseError` for user 
    input, `InternalError` for invariant violations"). This affects whether mis-raising counts as a bug.
     Omit the output section entirely if conventions are uniform and already implicit in Supported 
    Scope.

### Phase 5 — Intentional Deviations and Unsupported Features

Grep for `NotImplementedError`, `raise.*[Uu]nsupported`, "not supported" (Python); `throw new 
    Error(...)` carrying "unsupported"/"not implemented", `assertNever`, `@ts-expect-error` (TS/JS), in 
    comments and error strings. Then identify two categories — keep them separate:
- **Intentional deviations:** behaviors that look wrong but are documented as by-design within scope. 
    Example: "operator precedence follows C, not Python, because the DSL targets C codegen." Maps to the
     `intentional` sub-category of G2.
- **Deliberately unsupported features:** constructs the project detects and refuses at 
    parse/compile/load time. Example: "nested generic types raise `UnsupportedFeatureError` at parse 
    time." Maps to the `unsupported-feature` sub-category of G2. Should echo Supported Scope.

Cite source for each entry: design doc, inline comment, test with explanation, README section.

## Allowed / Not Allowed

**Allowed:** read any file; run quick one-liners (`grep`, `wc -l`, and the language's own tools — 
    Python: `python -c "..."`, `pip show`; TypeScript/JavaScript: `node -e "..."`, `npm ls`); run 
    `find`/`tree` (always exclude `.git/`); run individual test files to understand behavior.

**Not allowed:** modify any source/test/config file; install packages; make network requests; run the 
    full test suite (e.g. Python: `pytest`; TypeScript: `npm test` / `bun run test`); report bugs.

## Output Format

Output as a single Markdown document in a fenced code block, using the exact section structure below.

**Length budget: 150 lines. Hard limit.**
- Public API Surface, Supported Scope, Intentional Deviations, and Deliberately Unsupported Features are
     load-bearing for validation. Spend lines there.
- What This Project Does: 1–2 sentences only.
- One precise sentence beats three vague ones.

````markdown
# Repository Context: {repo_name}

## What This Project Does
<!-- 1–2 sentences. -->

## Public API Surface
<!-- Subsections for: CLI entry points, Library public API (modules/classes/functions meant for external
     import), HTTP/network surface if any. Cite evidence (README, docs, __all__, usage examples, 
    user-facing tests). Exhaustive public list; treat everything else as internal. Call out specific 
    internal modules by name only if their names would mislead a judge. -->

## Supported Scope
<!-- Language/format-processing projects: input subset supported, target runtime/model, explicit 
    non-goals. Other projects: supported use cases, platform assumptions, stated non-goals. Cite 
    sources. -->

## Error Handling
<!-- 2 sentences. Custom hierarchy vs. bare exceptions vs. result types, plus any documented semantics 
    distinguishing user-input errors from invariant violations. Omit section if conventions are uniform 
    and already implicit in Supported Scope. -->

## Intentional Deviations — Do NOT Report as Bugs
<!-- Behaviors that look wrong but are by design within scope. One entry per line with source citation. 
    Omit section if none. -->

## Deliberately Unsupported Features
<!-- Constructs the project detects and refuses at parse/compile/load time. One entry per line with 
    source citation. Must be consistent with Supported Scope. Omit if none. -->
````

## Rules

- Do not modify source, tests, or configuration.
- Do not install packages or alter the environment.
- Do not report bugs.
- Be factual and specific: exact file paths, class names, function names.
- Omit sections that don't apply — do not write "N/A" or placeholder text.
  \end{lstlisting}
\end{promptbox}

\begin{promptbox}{User}{userborder}
  \begin{lstlisting}[basicstyle=\ttfamily\fontsize{7}{7.6}\selectfont,breaklines=false,showstringspaces=false,language=,mathescape=false,escapeinside={(*@}{@*)}]
Produce a structured context report for this repository following
the instructions in CLAUDE.md. Explore the codebase at /app/ systematically.
Output your report as a single Markdown document in a fenced code block,
exactly as specified in CLAUDE.md.
  \end{lstlisting}
\end{promptbox}

\promptheading{Panel Judge}{prompt:panel-validate}{\textsc{Panel Judges}}
Panel judge prompt, with the verdict schema appended. Every judge receives this text: \luna{} and \geminiflash{} in tier 1, \opus{} and \terra{} in tier 2. The finding under judgment is mounted as \ttt{/app/input/finding.json} rather than inlined, and the judge works in its own container on the snapshot image.

\begin{promptbox}{System}{systemborder}
  \begin{lstlisting}[basicstyle=\ttfamily\fontsize{7}{7.6}\selectfont,breaklines=false,showstringspaces=false,language=,mathescape=false,escapeinside={(*@}{@*)}]
# AGENTS_PANEL.md — Bug-Finding Audit: Single-Finding Panel Validation

## CRITICAL — How to deliver your verdict

Your verdict is consumed by a downstream pipeline that parses **structured JSON only**. Prose, bullet 
    lists, markdown headers, and "Verdict: confirmed" text are **silently discarded** and your entire 
    run is thrown away as a failure.

There is exactly one accepted output. You must do **both** of the following — do not pick one:

1. **Write the JSON object to `/app/output/verdict.json`.** Use a heredoc so the JSON is unambiguous:

   ```bash
   cat > /app/output/verdict.json << 'EOF'
   {
     "verdict": "confirmed",
     "rejection_reason": null,
     "confidence": "medium",
     "file": "path/to/file",
     "line_start": 42,
     "line_end": 45,
     "severity": "high",
     "title": "...",
     "description": "...",
     "code_snippet": "...",
     "poc": "...",
     "poc_result": "...",
     "suggested_fix": "...",
     "realistic_scenario": "...",
     "validation_notes": "..."
   }
   EOF
   ```

2. **Also emit the same JSON object as the final message of your run, inside a single fenced code block 
    — and nothing else.** No prose before the fences, no prose after. Just:

   ````
   ```json
   { ...the same object... }
   ```
   ````

   This is a redundant fallback so that if the file write silently fails, the parser can still recover 
    the verdict from your message stream.

---

## Instructions

> **Read this entire document before starting work.** Your single finding is in 
    `/app/input/finding.json`.

### Task

A prior automated tool has reported a potential bug in a repository. Your job is to **validate this 
    single finding** by exploring the repository yourself, reviewing the code in context, and writing a 
    proof-of-concept script.

You are one judge in a panel of independent LLM judges. **Report your honest, calibrated belief — do not
     bias your verdict toward rejection to be "safe."** A correctly-flagged real bug at Low confidence 
    is more useful to the panel than the same bug rejected for caution. Use the `confidence` field, not 
    the `verdict` field, to express uncertainty.

Your finding is in `/app/input/finding.json`. Read it now. It contains at minimum a title, description, 
    and file path. Other fields — `code_snippet`, `poc_*`, `poc_result_*`, `validation_notes_*`, 
    `rejection_reason_*` — may be `null` or absent. **Treat all pre-filled fields as hints, not ground 
    truth.** The originating tool may be wrong about the snippet, the line numbers, or the reasoning. 
    You must independently verify every claim.

A curated repository overview may be available at `/app/repo_context.md`. If it exists, read it 
    **before** exploring the repository — it covers the public API surface, supported scope, intentional
     deviations, and architecture. It does not replace code-level verification: G1 still requires 
    `grep`/`rg` searches with file:line citations.

**Spend your time on verification, not on finding new bugs.**

---

### What You May and May Not Do

**Allowed:**
- Read any file in the repository at `/app`.
- Run the project's code, import its modules, execute its entry points.
- Write and execute PoC scripts in `/app/.scratch/`.
- Run individual tests from the test suite.
- Use `grep`, `find`, `rg` and the language's own tools (Python: `pip show`, `python -c "..."`; 
    TypeScript/JavaScript: `npm ls`, `node -e "..."`) to inspect the environment and search for callers.
- Read `/app/input/finding.json` and any pre-filled fields as context.

**Not allowed:**
- Modify any existing source file, test file, or configuration file.
- Install additional packages or change dependencies.
- Make network requests or access external services, including localhost — PoC scripts must not connect 
    to any port unless a listener was explicitly started in the same script first. Prefer in-process 
    test clients (Python: FastAPI `TestClient`, `unittest.mock.patch`; TypeScript: Hono `app.request()`)
     over real sockets.
- Run the full test suite (e.g. Python: `pytest` with no arguments; TypeScript: `npm test` / `bun run 
    test`).
- Search for bugs not reported in `/app/input/finding.json`.

---

### Gathering Context

You have full access to the repository at `/app`. Gather whatever context you need:
1. **Read the referenced file.** Open the file and lines named in the finding. Confirm the snippet 
    matches the actual source. If the finding's `code_snippet` field was non-null, compare it to what 
    you actually find at those lines — they may differ.
2. **Find callers and callees.** Use `grep -r`, `find`, or `rg` to search for callers of the suspect 
    function. Read those caller files to understand what inputs actually reach the buggy code. Do not 
    infer callers from the function signature alone.
3. **Understand project scope.** Read `/app/repo_context.md` if present; otherwise look for a README, 
    CONTRIBUTING guide, or scope documentation. This determines G2 and shapes whether divergence from an
     external reference is in scope.
4. **Use pre-filled fields as a starting point, not a conclusion.** Any `poc_*`, `validation_notes_*`, 
    or `rejection_reason_*` in the finding JSON may be the result of a prior automated pass that made 
    mistakes. Read them, consider them, but verify every claim independently.

---

### The Reachability-and-Behavior Gate

This is the central concept of the validation pass. A PoC that triggers misbehavior is **necessary but 
    not sufficient** for confirmation. Before confirming any finding, walk through both gates below. 
    Failing either means `rejected`.

#### (G1) Reachable in normal use

Reachability is a property of the call path the PoC *actually takes*, not one that could in principle 
    exist. The PoC must enter through a public entry point and reach the buggy code by the same path a 
    real caller would. Importing the suspect function directly and feeding it crafted input only proves 
    the function misbehaves on that input — a weaker claim than reachability.

The triggering input must be one that:
- an internal caller actually passes, **or**
- a documented external user of the function would plausibly supply in non-adversarial use (or, for 
    security bugs, a documented attacker surface).

"Documented external user" is defined by the project's stated public-API surface — not by export 
    visibility, not by the absence of a leading underscore, not by the function being importable. If 
    documentation doesn't identify a function as part of the public surface, treat it as internal and 
    check whether any internal caller actually passes the problematic input.

**How to check G1:** Search for callers using `grep -r` or `rg` across the repository at `/app`. Do not 
    assert G1 from the function's type signature or the fact that the function is callable — a signature
     that permits an input is not evidence that any caller passes it. For each cross-boundary claim, 
    cite specific file paths and line numbers where you found (or did not find) the triggering input 
    being produced. If a repo-wide search finds no internal caller that passes the problematic input, 
    that is the conclusion — state it explicitly.

If no realistic caller passes the triggering input — repo-wide search finds nothing, an upstream guard 
    blocks it, or the specific caller cited by the finding does not actually pass that input — and the 
    function is not part of the public-API surface, reject as `unreachable`. Justify the evidence in 
    `validation_notes` (which queries you ran, which guards you read, etc.).

#### (G2) Expected behavior on supported input

This gate combines scope and outcome into one question: *given the input is in scope, does the system 
    produce wrong output, crash, hang, corrupt state, or breach security?*

Classify the observable outcome:
- **Wrong result, crash, hang, data corruption, or security breach on supported input** → bug.
- **Input is out of the project's supported scope, or the project explicitly refuses it (at 
    compile/setup/load time, or via a documented limitation)** → not a bug. Reject as `out-of-scope`.
- **System correctly rejects in-scope-but-invalid input AND the only complaint is cosmetic AND no 
    sibling code path handles the same case more cleanly** → not a bug. Reject as `error-mode-only`. See
     the narrowed definition in the rejection table — error-handling code that itself crashes, recurses 
    infinitely, or swallows information is *not* error-mode-only.

**Tricky cases.** If the project deliberately implements a restricted subset (a DSL, a compiler 
    targeting a constrained runtime, a tool with explicit non-goals), divergence from a fuller 
    language's or library's semantics is only a bug when the diverging behavior produces wrong results 
    **within the supported subset**. If the finding contrasts the project against an external reference 
    (stdlib, a specification, another library), **verify that reference** — read the actual source or 
    documentation before accepting the prior tool's characterization of it.

---

### PoC Construction

**Enter through a public entry point** — the CLI, top-level API function, request handler, or pipeline 
    `run()` a real user invokes. Importing the suspect function and feeding it crafted input 
    demonstrates the *mechanism* but not reachability, and does not pass the gate.

**Design the PoC to falsify, not just to reproduce.** Before writing, name (mentally — this doesn't go 
    in output) the single most plausible reason this could be a false positive: intentional behavior in 
    the repo context, an upstream guard, a precondition no caller passes, a misread external reference. 
    Then design the PoC to *go through* that suspected guard or caller and observe whether the bug fires
     anyway. A PoC that bypasses the guard via mocking or direct calls into private helpers shows the 
    mechanism but not reachability. If the PoC needs to traverse a guard whose semantics you assumed but
     didn't verify, read the guard's code first — the High-confidence bar requires it.

**Keep it minimal.** The shortest script that exercises the public entry point with inputs sufficient to
     trigger the bug. Place PoCs in `/app/.scratch/` (e.g., `/app/.scratch/poc.py` or 
    `/app/.scratch/poc.ts`). Do not copy-paste source from the repo — import from it directly.

**If you cannot construct inputs that survive the pipeline and reach the buggy code**, that is evidence 
    the finding fails G1 — reject as `unreachable`. A localized PoC calling the leaf function is not a 
    fallback.

**Drop to an internal entry point only** if the high-level path requires infrastructure you can't 
    simulate (database, network, specific OS state), and explain why in `validation_notes`. 
    Inconvenience of pipeline setup does not qualify.

---

### Verdict Criteria

There are only two verdicts: `confirmed` and `rejected`.

#### `confirmed`

A finding is `confirmed` when **both** the mechanism evidence and the Reachability-and-Behavior Gate are
     satisfied.

**Mechanism evidence.** One of the following:
- **Preferred:** a runnable PoC entering through a public path demonstrates the bug via observable 
    misbehavior — wrong output, exception, crash, wrong state.
- **Reasoning fallback:** the PoC is genuinely infeasible — requires network access the environment 
    forbids, production runtime state, disk-full / OOM / mid-operation timeouts, or concurrency timing 
    the validator cannot reproduce. In this case, confirm on rigorous code reasoning, **cap `confidence`
     at `Low`**, and explain the blocker in `validation_notes`. Inconvenience of pipeline setup does not
     qualify — try harder.

Difficulty getting input past upstream validation is evidence *against* confirmation (G1 likely fails), 
    not grounds for the reasoning fallback.

**Reachability-and-Behavior Gate.** Both G1 and G2 above must hold.

When confirming, the **`realistic_scenario`** field must contain a one-sentence concrete scenario in 
    which the bug fires:

- **For correctness bugs:** a non-adversarial usage scenario — a plausible thing a real user does that 
    produces the triggering input.
  - *Not sufficient:* "a user calls `parse()` with a malformed string."
  - *Sufficient:* "a user pastes a CSV exported from Excel (which uses CRLF line endings) into 
    `parse()`, which is documented for CSV input."
- **For security bugs:** a concrete deployment path in which an attacker reaches the vulnerable code — 
    name the attack surface, not just "an attacker."
  - *Not sufficient:* "an attacker exploits the SQL injection."
  - *Sufficient:* "the `search` query parameter on the public `/api/users` endpoint is interpolated 
    unescaped into the WHERE clause, reachable by any unauthenticated request."

In either case the difference is concreteness about **why** the triggering input arises in real 
    deployment.

#### `rejected`

The finding is not a real bug for the benchmark. Pick the single best `rejection_reason`:

| `rejection_reason` | When to use |
|---|---|
| `code-mismatch` | The bug itself is wrong — hallucinated function, snippet not in the file, wrong file
     with no analogous bug. Off-by-N line numbers, paraphrased snippets, and other surface details that 
    are wrong while the bug is clearly present nearby are **corrections, not rejections** — fix them in 
    your output and continue validating. |
| `unreachable` | No realistic caller passes the triggering input, and the function is not part of the 
    public-API surface. The function may permit the input by signature, but the call graph doesn't 
    produce it. Justify the evidence in `validation_notes` (which queries you ran, which guards you 
    read, etc.). |
| `intentional` | Behavior is intentional — documented, tested, listed as an intentional behavior in the
     repo context, or clearly consistent with project design. |
| `error-mode-only` | Input is correctly rejected by a clean error path AND no sibling code path handles
     the same case more cleanly. **Applies only when the only complaint is cosmetic** (exception 
    type/message/trace). Does *not* apply when: (i) the error-handling code itself crashes, hangs, or 
    recurses infinitely; (ii) wrong-type input causes stack overflow, memory blowup, or data corruption;
     (iii) an analogous sibling function handles the case cleanly — the inconsistency itself is a real 
    bug (typically Low severity); (iv) the exception swallows information the caller needed to act on. |
| `out-of-scope` | The project does not claim to handle this input or construct. Covers: divergence from
     a reference language/library that doesn't affect the project's supported use cases; constructs the 
    project detects and refuses cleanly at compile/setup/load time (known limitations, not runtime 
    bugs); inputs documented as not yet supported (TODO, issue reference, deliberately unsupported). |
| `style-only` | Style, naming, formatting, or documentation issue with no runtime impact. |

A rejection can be supported by a PoC showing the code behaves correctly on the claimed triggering input
     plus an explanation of why that generalizes, or by code reasoning alone when the flaw in the prior 
    tool's argument is clear. If rejecting on reasoning alone, `validation_notes` must be specific and 
    cite concrete code.

**Do not reject a finding solely because a PoC could not be built.** Use the reasoning fallback (with 
    capped confidence) instead.

#### When in doubt

Report your honest belief and use the `confidence` field to express uncertainty. If you genuinely cannot
     decide between `confirmed` and `rejected`, pick the verdict your reasoning best supports and set 
    `confidence` to `Low`, with `validation_notes` stating what evidence would have resolved the 
    question.

---

### Severity and Confidence

Severity and confidence are **independent axes**. A bug can be Critical severity with Medium confidence.
     Do not let one bleed into the other.

#### Severity — about impact, assuming the bug is real

Assigned only on confirmed findings. Set to `null` on rejections.

- **Critical**: Data loss, data corruption, security breach (RCE, auth bypass, privilege escalation, 
    secret exfiltration), or process crash triggered by inputs that occur in normal usage. Critical 
    should be rare — before assigning it, state in `validation_notes` the concrete sequence of events in
     a realistic deployment where the data loss, breach, or crash occurs. A crash on a valid but 
    adversarially-constructed input is High, not Critical. A security issue that requires an attacker to
     already have write access to the codebase is not Critical.
- **High**: Incorrect results, wrong state transitions, or denial of service that a developer following 
    documented workflows would encounter without assembling an unusual combination of inputs, 
    configurations, or call sequences. The triggering condition must arise in straightforward use of the
     feature. No data loss or security breach.
- **Medium**: Incorrect behavior that requires a specific combination of inputs, configurations, or call
     sequences a typical user would not assemble without deliberate effort. Includes recoverable errors 
    surfaced wrong, meaningful performance degradation without breakage, and concurrency bugs that 
    require contention or timing conditions not present under light or single-threaded use.
- **Low**: Real defect with minimal practical impact: defensive issues, bugs behind preconditions that 
    are unlikely to be met, misleading error messages that don't cause wrong behavior, error-handling 
    inconsistencies between sibling functions where one handles a case cleanly and another doesn't.

**When in doubt between two adjacent levels, assign the lower one.** Over-inflation is a worse 
    calibration error than under-inflation.

#### Confidence — about the verdict

Assigned on every finding. The meaning depends on the verdict:

- On `confirmed`: confidence that the bug is real and meets the gate.
- On `rejected`: confidence that the rejection reason correctly applies.

There are three levels: **Low**, **Medium**, **High**. Assign the level whose anchor your evidence 
    actually meets — do not inflate because a prior tool sounded confident, and do not deflate a 
    well-supported verdict out of caution.

- **Low** — A material part of the verdict rests on something you could not verify or resolve. Use Low 
    when **any** of the following holds:
  - A supporting link rests on an unverified assumption — a caller you didn't read, a library behavior 
    you inferred from docs but didn't test, a guard whose semantics you assumed.
  - The gate turns on a judgment call you can't decisively resolve: is this input "supported" or 
    "out-of-scope"? does the sibling function really count as analogous? is the documented contract 
    precise enough to call this a violation?
  - The bug is real but the realistic-scenario you constructed sits at the edge of "concrete enough" — a
     maintainer could go either way on whether it's a bug worth fixing.

- **Medium** — A solid, well-supported verdict with no material gap. Your reasoning is end-to-end and, 
    for confirmations, the PoC runs **through the public entry point** with no mocking or direct 
    internal calls; no link in the argument rests on an unverified assumption. You have not necessarily 
    constructed an input designed specifically to falsify the verdict.

- **High** — You actively tried to falsify the verdict and it held, or the code mechanically forces it. 
    Uncommon.
  - *On a confirmation:* everything Medium requires, **plus** you read every caller in the relevant grep
     results that could plausibly produce or block the triggering input, you cite file/line for each 
    link in G1, and you named a specific input shape that *would* invalidate the bug if the code weren't
     actually broken — and the code produced the bug anyway. You're certain a maintainer would actually 
    consider this a bug. Alternatively, the verdict is mechanically forced — alternative outcomes are 
    excluded by lines you cite.
  - *On a rejection:* the rejection reason is mechanically forced by code you cite. For `unreachable`, 
    this means a repo-wide search whose queries you can name and whose hits you've enumerated, **or** a 
    specific guard whose code you quote and whose semantics you verified.

**Downgrade triggers — each one caps the confidence at Low:**
- The PoC required mocking, monkey-patching, or direct calls into private functions.
- Reachability rests on "the function is exported / has no leading underscore" rather than a caller you 
    read.
- The finding contrasts against a stdlib/spec/library and you didn't open the actual reference.
- The triggering input survives an upstream layer whose code you didn't read.
- You're rejecting on `unreachable` based only on the immediate file's context, without a repo-wide 
    search.
- You're not certain a maintainer would actually consider this a bug.

**Hard cap at Low:** verdicts confirmed via the reasoning fallback — i.e., when a PoC was genuinely 
    infeasible — cannot exceed Low confidence regardless of how strong the reasoning is.

**Calibration check.** Before finalizing, re-check the level against its anchor: claim **High** only if 
    you actively falsified the verdict (or code mechanically forces it); claim **Medium** only if no 
    link in the argument is unverified; otherwise use **Low**. Do not inflate because the prior tool 
    sounded confident.

---

### Workflow

#### Phase 1 — Orient

1. Check whether `/app/repo_context.md` exists. If it does, read it in full before opening any source 
    file. Key sections to absorb:
   - **Public API Surface** — determines what counts as a public entry point for G1.
   - **Supported Scope** — determines G2 (`out-of-scope` rejections).
   - **Intentional Deviations / Deliberately Unsupported Features** — prime sources of `intentional` and
     `out-of-scope` rejections; check these before confirming.

   Treat the file as a strong prior, but verify any specific claim that bears on your verdict against 
    the actual source.

2. Read `/app/input/finding.json`.
3. Read the file and lines named in the finding. Verify the code snippet against the actual source.

#### Phase 2 — Validate

1. **Verify the source.** Confirm the snippet matches the actual source. Reject as `code-mismatch` 
    **only when the bug itself does not exist at the referenced location and is not present nearby** — 
    for example, a hallucinated function or a snippet that's not in the file. If line numbers are 
    slightly off, the snippet is paraphrased, or other surface details are wrong but the bug is clearly 
    present, **correct the details in your output and continue validating**. Note any corrections in 
    `validation_notes`.

2. **Review the reasoning.** Does the prior tool's explanation hold up? For cross-boundary claims, 
    search the repo with `grep -r` or `rg`. For logic errors, trace the code path. If the finding 
    contrasts against an external reference, read the actual reference and verify the comparison.

3. **Write a PoC entering through a public entry point.** Design it to falsify (per the PoC Construction
     section), not just reproduce. Place it in `/app/.scratch/`. If the inputs cannot survive the 
    pipeline, that is evidence for `unreachable` — do not work around it with a localized PoC.

4. **If a PoC is genuinely infeasible** (network, OOM, disk full, concurrency timing, production runtime
     state), confirm on reasoning and cap confidence at Low. Do not reject solely because a PoC could 
    not be built.

5. **Apply the Reachability-and-Behavior Gate.** Walk through G1 and G2 explicitly in 
    `validation_notes`. For G1, cite specific file paths and line numbers where you found (or searched 
    for and did not find) the triggering input being passed by actual callers. Any gate failure means 
    `rejected`.

6. **For confirmed findings**, assign severity. Write the realistic-scenario sentence in the 
    `realistic_scenario` field. Apply confidence anchors and the automatic downgrade triggers honestly.

#### Phase 3 — Output

Deliver your verdict on **both** channels described in §CRITICAL — How to deliver your verdict at the 
    top of this document:

1. Write the JSON object to `/app/output/verdict.json` (heredoc shown above).
2. Emit the same JSON object as the final message of your run, inside one fenced ```json … ``` block, 
    with no surrounding prose.

Prose verdicts (`**Verdict** - \`confirmed\``, `Confirmed.` followed by bullets, "Verdict: confirmed" 
    inline, etc.) are silently discarded. Use the schema below for the JSON object.

---

### Output Format

Write a single JSON object to `/app/output/verdict.json` (see the heredoc in §CRITICAL for the exact 
    shape). Do not wrap it in a list or add commentary outside the JSON. The object must match the JSON 
    Schema appended to this prompt; its field descriptions are part of the output contract, and it 
    defines the lowercase enum values for `verdict`, `rejection_reason`, `confidence`, and `severity`.

#### What goes in each field

The structured fields and `validation_notes` have non-overlapping roles. Do not duplicate content 
    between them.

- **`rejection_reason`** — One enum value from the rejection table when `verdict` is `rejected`; `null` 
    when `confirmed`. The *justification* for the rejection (which code supports it, why the alternative
     reasons don't apply) goes in `validation_notes`, not here.
- **`suggested_fix`** — A brief description of how to fix the bug (a sentence or two — not a full 
    patch). Only meaningful on `confirmed`; set to `null` on `rejected`.
- **`realistic_scenario`** — A single concrete sentence describing how the bug fires in real use, per 
    the rules in §Verdict Criteria. Required and non-null on `confirmed`; `null` on `rejected`. Do not 
    duplicate this sentence into `validation_notes`.
- **`validation_notes`** — Free-text reasoning, required on every finding. This is the catch-all for 
    everything not captured by a structured field. Include:
  - G1/G2 analysis with file:line citations (G1 must cite specific paths and line numbers).
  - Any corrections you made to the original finding (line numbers, snippet text, attribution).
  - Severity rationale, on confirmed findings.
  - Confidence rationale, including which automatic downgrades you applied and why.
  - The justification for the chosen `rejection_reason`, on rejected findings.
  - For findings confirmed via the reasoning fallback, why a PoC was infeasible.

  Do **not** repeat the realistic scenario or the suggested fix here — they have their own fields.

#### Field constraints

- `verdict` ∈ {`confirmed`, `rejected`}.
- `rejection_reason` is one of the enum values when `verdict` is `rejected`; `null` when `confirmed`.
- `confidence` ∈ {`low`, `medium`, `high`} on every finding.
- `realistic_scenario` is a non-null single sentence when `verdict` is `confirmed`; `null` when 
    `rejected`.
- `severity` ∈ {`critical`, `high`, `medium`, `low`} when `verdict` is `confirmed`; `null` when 
    `rejected`.
- `suggested_fix` is a brief sentence or two when `verdict` is `confirmed`; `null` when `rejected`.
- `validation_notes` is required for every finding. G1 analysis must cite specific file paths and line 
    numbers.
- `poc` and `poc_result` are both non-null together when a PoC was run; both `null` otherwise (including
     when you used the reasoning fallback).
- If `poc` and `poc_result` are both `null` on a `confirmed` finding, `confidence` must be `low`.

---

### Rules

- Validate the **single finding** in `/app/input/finding.json` — do not report additional bugs.
- Do NOT modify any existing source code, tests, or configuration.
- Do NOT install packages or alter the environment.
- Be specific: include exact file paths and corrected line numbers.
- Include the actual code snippet verified against the repository.
- Include both PoC code AND its output for every finding where a PoC was run.
- PoCs must enter through a public entry point; calling the suspect function directly does not satisfy 
    the mechanism-evidence requirement unless the project documentation lists it as part of the 
    public-API surface.
- For confirmed findings, include a `realistic_scenario` sentence and explicitly walk through G1 and G2 
    in `validation_notes` with file/line citations for G1.
- For rejected findings, set `rejection_reason` and point to concrete code supporting it in 
    `validation_notes`.
- Apply severity and confidence anchors mechanically and independently; honor the downgrade triggers and
     the Low cap on reasoning-fallback confirmations.
- A successful PoC alone is **not** sufficient for `confirmed`. The Reachability-and-Behavior Gate must 
    also pass.
- Report your honest verdict and use `confidence` to express uncertainty — do not bias toward rejection.
- `code-mismatch` is reserved for findings where the bug itself is wrong (hallucinated, snippet not in 
    file). Off-by-N line numbers and other surface details are corrections, not rejections.
- Do not reject solely because a PoC could not be built — use the reasoning fallback at Low confidence 
    instead.
- Write your output as a JSON object to `/app/output/verdict.json` **and** emit the same JSON in your 
    final message inside a single fenced ```json … ``` block. Bulleted summaries, prose verdicts, and 
    "Verdict: confirmed"-style text are silently discarded — see §CRITICAL — How to deliver your 
    verdict.
  \end{lstlisting}
\end{promptbox}

\begin{promptbox}{User}{userborder}
  \begin{lstlisting}[basicstyle=\ttfamily\fontsize{7}{7.6}\selectfont,breaklines=false,showstringspaces=false,language=,mathescape=false,escapeinside={(*@}{@*)}]
Read /app/AGENTS_PANEL.md carefully, then follow its instructions exactly.
  \end{lstlisting}
\end{promptbox}

\needspace{8\baselineskip}
\subsection{Bug-Type Taxonomy}

\promptheading{Category Induction}{prompt:taxonomy-induce}{\opus}
Induction of the bug-type categories. One call over a sample of the taxonomy corpus (human-fixed historical bugs and agent-found bugs), stratified by repository and source, devises the categories that \cref{prompt:taxonomy-merge} consolidates and \cref{prompt:taxonomy-attribute} assigns.

\begin{promptbox}{Prompt}{userborder}
  \begin{lstlisting}[basicstyle=\ttfamily\fontsize{7}{7.6}\selectfont,breaklines=false,showstringspaces=false,language=,mathescape=false,escapeinside={(*@}{@*)}]
You are a software-reliability expert. Below is a JSON array of (*@\textit{sample size}@*) real bugs, each with a 
    title and a description. They are a representative, stratified sample of a corpus of (*@\textit{corpus size}@*) 
    bugs drawn from 20 open-source Python and TypeScript/JavaScript repositories.

Devise a flat taxonomy of between (*@\textit{minimum category count}@*) and (*@\textit{maximum category count}@*) mutually 
    exclusive fault-type CATEGORIES that classify bugs by the kind of programming mistake, never by 
    repository, programming language, severity, or affected feature.

Rules:
- Classify by ROOT CAUSE, not symptom: the mistake a fix would correct, not the failure the user saw. A 
    crash, a wrong result and a hang belong together when the same kind of mistake caused them.
- Granularity: no category should end up holding more than about (*@\textit{largest category share}@*)% of the corpus
     ((*@\textit{largest category size}@*) bugs), and do not create one you expect to hold fewer than about 
    (*@\textit{smallest category size}@*) bugs unless it is conceptually irreducible. Split the common kinds of 
    mistake into their distinct sub-kinds and merge the rare ones until both hold.
- Every bug must have exactly one obvious category. No residual or "other" category.
- These often-conflated mistakes must be separable: a missing conditional branch for a supported input 
    variant; a missing entry in a dispatch or lookup table; an unhandled degenerate value (empty, zero, 
    null); missing validation of an input that should be rejected; an unimplemented feature.
- Each definition is one or two sentences: what belongs, plus a clause naming the adjacent category a 
    borderline bug belongs to instead ("not X, which belongs under Y"). Use commas, colons or brackets, 
    never dashes.
- Names are short noun phrases in sentence case (only the first word and proper nouns capitalised), in 
    British spelling, and unique.

Respond with ONLY a JSON object of this exact shape, no prose:
{"categories": [{"name": "...", "definition": "..."}]}

BUGS:
(*@\textit{JSON array of the sampled bugs, title and description only}@*)
  \end{lstlisting}
\end{promptbox}

\promptheading{Thin-Category Merge}{prompt:taxonomy-merge}{\opus}
Merge of thin categories. After every corpus bug is assigned, the model sees each category with its measured count and either merges one below the floor into a survivor or keeps it as irreducible. A merge remaps assignments without re-attributing them.

\begin{promptbox}{Prompt}{userborder}
  \begin{lstlisting}[basicstyle=\ttfamily\fontsize{7}{7.6}\selectfont,breaklines=false,showstringspaces=false,language=,mathescape=false,escapeinside={(*@}{@*)}]
You are enforcing the granularity criterion of a flat bug fault-type taxonomy after a full 
    classification pass over a (*@\textit{corpus size}@*)-bug corpus. Below is every category: id, name, definition, 
    measured bug count, and example bug titles.

The criterion: a category holding fewer than (*@\textit{smallest category size}@*) bugs merges into the closest 
    existing category unless it is conceptually irreducible, that is a distinct kind of programming 
    mistake no other category covers. Near-duplicate categories must merge regardless of size. Never 
    merge genuinely distinct mistakes just to reach the floor.

For every category below (*@\textit{smallest category size}@*) bugs, decide: absorb it into a survivor, or keep it 
    with a one-sentence reason why it is irreducible. When a merge changes what a survivor covers, 
    restate its definition in one or two sentences, keeping a boundary clause, in British spelling and 
    without dashes.

Respond with ONLY a JSON object of this exact shape, no prose:
{"merges": [{"into": 1, "absorb": [2, 3], "definition": "..."}], "keep": [{"id": 4, "reason": "..."}]}

- "into" is the surviving category's id and "absorb" the ids merged into it; "definition" optionally 
    restates the survivor's definition.
- "keep" lists every below-floor category you leave standing, each with its reason.

CATEGORIES:
(*@\textit{JSON array of every category: id, name, definition, measured count and example titles}@*)
  \end{lstlisting}
\end{promptbox}

\promptheading{Bug-Type Attribution}{prompt:taxonomy-attribute}{\sonnet}
Attribution of bug types, in batches. The same prompt assigns the taxonomy corpus and every evaluated configuration's findings, so the distributions of \cref{fig:fault-type-jsd} are measured on one taxonomy.

\begin{promptbox}{Prompt}{userborder}
  \begin{lstlisting}[basicstyle=\ttfamily\fontsize{7}{7.6}\selectfont,breaklines=false,showstringspaces=false,language=,mathescape=false,escapeinside={(*@}{@*)}]
You are classifying software bugs. Each bug gets one fault-type category and one level on each of the 
    fixed dimensions below.

CATEGORIES, the fault type, that is the kind of programming mistake a fix would correct:
(*@\textit{JSON array of the categories: id, name and definition}@*)

DIMENSIONS, each a question about the bug as described, with numbered levels:
(*@\textit{JSON array of the fixed dimensions and their levels}@*)

For every bug below, output its key, the single best-fitting category_id, and one level id per 
    dimension. Include every key exactly once; invent no ids.

Rules:
- Category: classify by root cause, not symptom. When two categories fit, choose the one naming the more
     specific mistake. Answer 0 only when no listed category describes the mistake.
- Dimensions: judge from the bug as described, not from how hard the fix is. When two levels both fit, 
    choose the lower-numbered one. Answer 0 only when the record does not say enough to decide, never 
    because two levels fit.
- The description states the defect; the code snippet is context, not the defect itself.

Respond with ONLY a JSON object of this exact shape, no prose:
{"assignments": [{"key": "...", "category_id": 0, (*@\textit{one level field per dimension}@*)}]}

BUGS:
(*@\textit{JSON array of the batch's bugs: key, title, description and truncated code snippet}@*)
  \end{lstlisting}
\end{promptbox}

}

\end{document}